\documentclass[%
reprint,onecolumn,
preprint,
amsmath,amssymb,
aps,
longbibliography,
]{revtex4-2}

\usepackage{graphicx}
\usepackage{bm}
\usepackage{amsmath}
\usepackage{float}
\usepackage{natbib}

\usepackage{color}

\usepackage[whole]{bxcjkjatype}

\begin{document}

\preprint{APS/123-QED}

\title{Two-time Lagrangian velocity correlation function for particle pairs in two-dimensional inverse energy-cascade turbulence}

\author{Tatsuro Kishi}
\email{tatsuro@kyoryu.scphys.kyoto-u.ac.jp}
\author{Takeshi Matsumoto}
\author{Sadayoshi Toh}
\affiliation{%
  Division of Physics and Astronomy,
  Graduate School of Science,
  Kyoto University,
  Kitashirakawa Oiwaketyo Sakyoku, Kyoto 606-8502, Japan
}%
\date{\today}

\begin{abstract}
  We numerically investigate a two-time Lagrangian velocity correlation function (TTLVCF)
  for particle pairs in two-dimensional energy inverse-cascade turbulence.
  We consider self similarity of the correlation function by means of incomplete similarity.
  In this framework, we propose a self-similar form of the correlation function,
  whose scaling exponents cannot be determined by only using the dimensional analysis
  based on the Kolmogorov's phenomenology.
  As a result, the scaling laws of the correlation function can depend on the initial separation.
  This initial-separation dependency is frequently observed in
  laboratory experiments and direct numerical simulations of the relative dispersion,
  which is directly related to the correlation function,
  at moderate Reynolds numbers.
  We numerically verify the self-similar form
  by direct numerical simulations of two-dimensional energy inverse-cascade turbulence.
  The involved scaling exponents and the dependencies on
  finite Reynolds-number effects are determined empirically.
  Then, we consider implication of the scaling laws of
  the correlation function on the relative dispersion,
  i.e. the Richardson--Obukhov $t^3$ law.
  Our results suggest a possibility not to recover the Richardson--Obukhov $t^3$ law at infinite Reynolds number.
\end{abstract}

\maketitle

\section{Introduction\label{sec:introduction}}

Velocity correlation function is fundamental to characterize turbulence.
We can understand dynamical couplings between
two points and two times in turbulence through the correlation function \cite{He2017a}.
Eulerian and Lagrangian velocity correlation functions have different characteristic
time scales from each other, which are essential to develop two-point closure approximations
to the Navier-Stokes equations without ad-hoc parameters \cite{Kraichnan1959, Kraichnan1965, Kaneda1981}.
These direct-interaction approximations or Lagrangian renormalized approximations
provide consistent results with Kolmogorov's 1941 (K41) phenomenology \cite{Kolmogorov1941} and also with
the 2D analog \cite{Kraichnan1967,Leith1968a,Batchelor1969}.
In particular, the success of these closures lies in applying the approximation not to
the Eulerian velocity correlation but to the Lagrangian velocity correlation.

The most general form of the second-order Lagrangian velocity correlation function
can be written in terms of Kraichnan's generalized velocity notation \cite{Kraichnan1965} as
\begin{equation}
  Q^L_{ij}(\bm{a},s_1|t_1; \bm{b},s_2|t_2) \equiv \langle v_i(\bm{a},s_1|t_1) v_j(\bm{b},s_2|t_2) \rangle.
  \label{gc}
\end{equation}
Here $v_i(\bm{a},s|t)$ is the $i$-th component of the Lagrangian velocity measured at time $t$
of a Lagrangian particle passed through a point $\bm{a}$ at time $s$
and $\langle \cdot \rangle$ denotes an ensemble average.
Since the general form, Eq.(\ref{gc}), is too hard to tackle,
a majority of theoretical, numerical and experimental investigations on the Lagrangian velocity correlation
are devoted to the abridged form of Eq.(\ref{gc})
by setting $s_1 = s_2$ and $t_2 = s_2$ \cite{Kraichnan1966, Kaneda1991, Gotoh1992,Kaneda1998,Kaneda1999a,He2009},
namely,
\begin{equation}
  Q^L_{ij}(\bm{a},s|t; \bm{b},s|s) = \langle v_i(\bm{a},s|t) v_j(\bm{b},s|s) \rangle.
  \label{ac}
\end{equation}
It should be noticed that $v_j(\bm{b},s|s)$ coincides with the
Eulerian velocity at a point $\bm{b}$ at time $s$.
Thus, the correlation Eq.(\ref{ac}) is between the Lagrangian velocity
and the Eulerian velocity.

An abridged form of Eq.(\ref{gc}) but involving only the Lagrangian velocity
can be
\begin{equation}
  Q^L_{ij}(\bm{a},s|t_1; \bm{b},s|t_2) = \langle v_i(\bm{a},s|t_1) v_j(\bm{b},s|t_2) \rangle,
  \label{lc}
\end{equation}
where the measuring times $t_1$ and $t_2$ are different from the labeling time $s$.
There are few studies on the Lagrangian correlation function Eq.(\ref{lc})
that is an essential ingredient to solve unsteady problems of turbulence
such as turbulent diffusion and mixing \cite{MoninYaglom1975}.
In a notable study of the correlation Eq.(\ref{lc}) \cite{Ishihara2002},
the authors performed a direct numerical simulation (DNS) to
obtain the Lagrangian velocity correlation function Eq.(\ref{lc}).
However, to analyze the computed correlation function, they had to resort
to the theory developed for the Lagrangian-Eulerian correlation Eq.(\ref{ac}).
This may be not only because a theory for the Lagrangian correlation Eq.(\ref{lc})
is not developed, but also because its simple characterization remains
to be done.
By the simple characterization, we mean answers to the following cascading questions:
does the Lagrangian correlation function Eq.(\ref{lc}) have a self-similar form?;
if this is so, what is the self-similar form?;
if the self-similar form is a power-law function, what are scaling exponents?
In this paper, we address these questions with phenomenological theory beyond the
dimensional analysis and direct numerical simulations. Our simulation here is limited to
two-dimensional (2D) energy inverse cascade turbulence, but the theory is applicable both to
two and three dimensions.

\begin{figure}[htbp]
  \centering{
    \includegraphics[scale=0.8,clip]{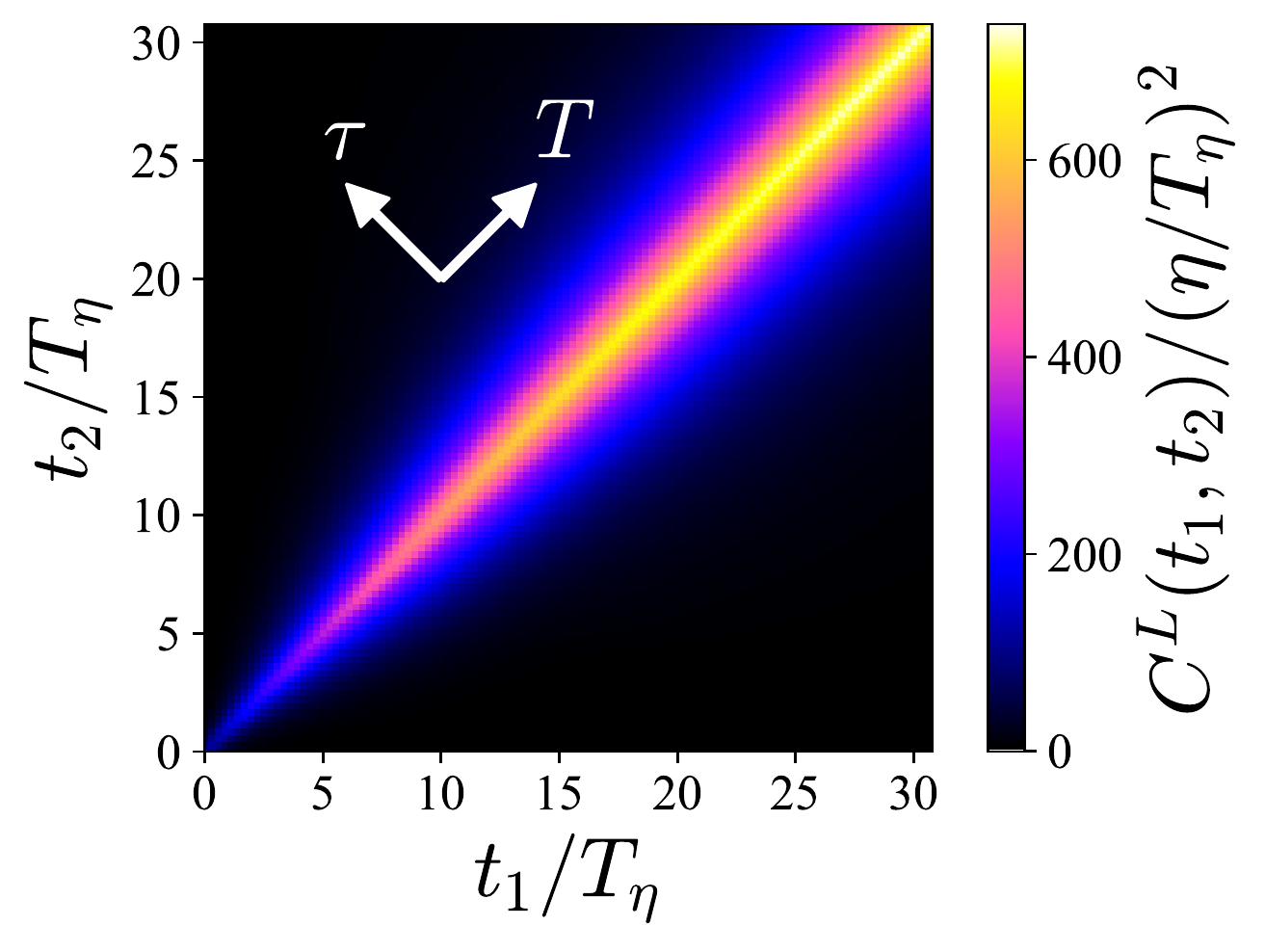}
  }
  \caption{
    Color map of the TTLVCF, Eq.(\ref{vcdef}),
    with $T_B=3.4T_\eta$ for $\mathrm{Re}_\alpha = 160$, where
    $T_B$ and $T_\eta$ are the Batchelor time and the Kolmogorov dissipation time scale, respectively.
    The infrared Reynolds number $\mathrm{Re}_\alpha$ is defined in the main text.
    See Sec.~\ref{sec:dns} for details.
  }
  \label{fig:intro}
\end{figure}

One of the difficulties in these questions is
that the Lagrangian velocity correlation function Eq.(\ref{lc})
is intrinsically dependent on both two times, $t_1$ and $t_2$.
To illustrate our approach, let us show
a color map of the Lagrangian correlation function as a function of $t_1$
and $t_2$ in Fig.\ref{fig:intro}.
For reasons described shortly below, we do not consider the Lagrangian correlation Eq.(\ref{lc}).
Instead, we study the Lagrangian velocity increment or, equivalently, the relative velocity between
two Lagrangian particles whose Lagrangian labels are $\bm{a}$ and $\bm{a} + \bm{r}_0$ at time $s$,
\begin{equation}
  \delta v_i(\bm{a}, \bm{r}_0, s|t) \equiv v_i(\bm{a}+\bm{r_0},s|t) - v_i(\bm{a},s|t)
  \label{relvel}
\end{equation}
and its correlation
\begin{equation}
  C^{L}_{ij}(r_0, t_1, t_2) = \langle  \delta v_i(\bm{a}, \bm{r}_0, s|t)   \delta v_j(\bm{a}, \bm{r}_0, s|t)   \rangle,
  \label{vcdef}
\end{equation}
where $r_0 = |\bm{r}_0|$.
We call the Lagrangian correlation Eq.(\ref{vcdef}) two-time Lagrangian
velocity correlation function (TTLVCF) in this paper
(on the left hand side (lhs) of Eq.(\ref{vcdef}), we omit the
dependence on $\bm{a}, s$ for simplicity).
The TTLVCF shown in Fig.\ref{fig:intro}
is numerically computed in 2D
energy inverse-cascade turbulence. The details will be explained in
Sec.\ref{sec:dns} in this paper.

To characterize the TTLVCF shown in Fig.\ref{fig:intro}, one way
is to look at it along the diagonal line through the origin, $t_1 = t_2$.
The line is parallel to the ``$T$'' axis written in Fig.\ref{fig:intro}.
The other way is obviously to study it along the lines perpendicular
to the diagonal line, i.e, $t_1 + t_2 = c_p$, where $c_p$ is a positive constant.
These lines are parallel to the ``$\tau$'' axis written in Fig.\ref{fig:intro}.
Accordingly, a correlation time can be defined for each line.
From Fig.\ref{fig:intro}, it can be observed that a correlation time along
a perpendicular line grows as we increase the constant $c_p$.
This sort of non-stationary behavior is not present in the two-point Eulerian velocity
correlation, whose correlation time is constant due to the statistical stationarity.
Unlike the Eulerian one, the TTLVCF has more than one degree of freedom.
It implies that the scaling law of the TTLVCF
cannot be obtained by dimensional analysis.
For this kind of problems, the incomplete similarity \cite{Barenblatt2003} provides
a framework to specify possible self-similar forms.
In this study, by using both the incomplete similarity and DNS,
we propose a self-similar form of the TTLVCF
shown in Fig.\ref{fig:intro}

Now let us explain why we consider the correlation of the Lagrangian velocity increment,
Eq.(\ref{relvel}).
The TTLVCF is directly related to the relative dispersion as follows.
The relative separation of a particle pair, whose
Lagrangian labels are $\bm{a}$ and $\bm{r}_0$ at time $0$, is written as
\begin{equation}
  \bm{r}(t)  = \bm{r}_0 + \int_0^t \delta \bm{v}(t') ~dt'.
  \label{eq:r_intro}
\end{equation}
The squared separation can then be written in terms of the TTLVCF as
\begin{equation}
  \langle r^2(t) \rangle = r_0^2
  + 2 \int_0^t \bm{r_0} \cdot \langle \delta \bm{v}(t_1) \rangle ~dt_1
  +  \int_0^t \int_0^t \langle \delta \bm{v}(t_1) \cdot \delta \bm{v}(t_2) \rangle ~dt_1 dt_2,
  \label{eq:r2_intro}
\end{equation}
where $r = |\bm{r}|$.
The turbulent relative dispersion has been widely investigated
since the pioneering work by Richardson \cite{LewisFRichardson1926},
who first predicted that the lhs of Eq.(\ref{eq:r2_intro})
grows as $\langle r^2(t) \rangle \propto t^3$ over an appropriate time interval,
see Ref.\cite{Salazar2008} as a review.
The $t^3$ prediction can be also derived by naively using the K41 dimensional analysis
in the inertial subrange \cite{Obukhov1941, Batchelor1950} and therefore,
the $t^3$ law is referred to as the Richardson--Obukhov law.

In the K41 framework,
statistics on the relative separation, whose best studied quantity is the second
moment $\langle r^2(t) \rangle$,
is considered to be independent of the initial separation, $r_0$, as long as $r(t)$ is
in the sufficiently wide inertial subrange.
This implies that, when we plot $\langle r^2(t) \rangle$ starting from various $r_0$'s,
the curves become independent of $r_0$ and eventually
collapses to the only one curve independent of $r_0$, which is proportional to $t^3$.
A tendency toward such an asymptotic state is indicated by DNS in three dimensions
at high Reynolds numbers \cite{Sawford2008,D.Buaria2015,Buaria2016}.
By contrast, one has never clearly observed the $r_0$-independence
in laboratory experiments in three dimensions \cite{Ouellette2006a, OTT2000}
and in two dimensions \cite{Jullien1999,Rivera2005,VonKameke2011,Rivera2016}.
The situation in numerical simulations
of the 2D inverse energy-cascade turbulence \cite{Boffetta2002,Boffetta2002c, Kishi2020}
is similar to that of the 3D turbulence
\cite{Yeung2004,Biferale2005,Sawford2008,Bitane2013b,D.Buaria2015}.
In this sense, the Richardson--Obukhov law is not verified satisfactorily by observations.
Of course, with much higher Reynolds number, a cleaner $t^3$ law irrespective of $r_0$ may be
observed.
In this paper, we parametrize finite Reynolds-number effect on the Richardson--Obukhov law
by analyzing the TTLVCF Eq.(\ref{vcdef}).
Specifically, we will study Reynolds-number dependence of the correlation function.
Then, through Eq.(\ref{eq:r2_intro}), we argue that $\langle r^2(t) \rangle$ depends on $r_0$
at finite Reynolds numbers and infer its asymptotic form at infinite Reynolds number.

In particular, at moderate Reynolds numbers it is known that
the $t^3$ scaling behavior of $\langle r^2(t) \rangle$ is observed
for a certain selected initial separation, see \cite{Boffetta2002,Kellay2002, Kishi2020} for further discussion.
This special initial separation is around the Kolmogorov dissipation length
for both 2D and 3D and thus dependent on the Reynolds number.
Our argument on the $r_0$-dependence reveals the nature of this special initial separation.

We organize the paper as follows.
In Sec. \ref{sec:scaling},
we make ansatz for the scaling laws of the TTLVCF shown in Fig.\ref{fig:intro}
by adapting the incomplete similarity \cite{Barenblatt2003}.
The method enables us to explore qualitatively scaling laws that deviate
from the K41 dimensional analysis. In particular, we take the finite Reynolds number effect
and the dependence on $r_0$ into consideration.
It should be noticed that this method does not depend on specific dimensions.
We can apply the method to both 2D and 3D turbulence.

Next, in Sec.\ref{sec:dns},
we verify the ansatz and determine quantitatively the involved scaling exponents
by comparing with DNS data of the 2D inverse energy-cascade turbulence.
We use a modified Navier-Stokes equations which have a superviscosity and
a hypodrag.
We have already estimated the artificial effects from these numerical terms
and verified that these effects can be ignored for the statistics of particle pairs in turbulence
in Ref. \cite{Kishi2020}.
We estimate the scaling exponents as a function of Reynolds number by using
DNSs with four different Reynolds numbers.
Subsequently, we infer asymptotic values of the scaling exponents at infinite
Reynolds number by extrapolating them from those at finite Reynolds numbers.
There are several reasons for selecting the 2D system;
detailed numerical studies are more feasible;
the Eulerian velocity is intermittency free \cite{Paret1997,Boffetta2000} and therefore
we factor out the intermittency effects on the Lagrangian statistics.
Of course, careful discussion and further investigation are required when one
applies the method used here to the 3D system.
We discuss a possibility to justify its validity to 3D turbulence in Sec. \ref{sec:conclusion}.

In Sec.\ref{sec:richardson},
we discuss implications of the self-similar ansatz of the TTLVCF
on the Richardson--Obukhov law.
Furthermore, we explain why the $t^3$ scaling law of $\langle r^2(t) \rangle$
is observed for a selected initial separation at moderate Reynolds numbers.
Concluding remarks are made in Sec.\ref{sec:conclusion}.

\section{Incomplete self-similarity of the Lagrangian correlation and scaling exponents\label{sec:scaling}}

In this section, we present a scaling ansatz of the TTLVCF for particle pairs.
It should be noted that the argument below
is independent of specific spatial dimensions.
Therefore, we expect that the ansatz is meaningful for both 2D and 3D turbulence.

As we discussed in Sec.\ref{sec:introduction},
we consider the Lagrangian correlation function, $C^L_{ii}(r_0, t_1,t_2)$,
of the velocity difference Eq.(\ref{relvel}), where we use the Einstein summation
convention of the repeated indices. We write $C^L_{ii}(r_0, t_1, t_2)$ as $C^L(r_0, t_1, t_2)$
below and term it the TTLVCF as well.

We consider a statistically steady, homogeneous, and isotropic
turbulent Eulerian velocity field, and therefore we deal with an external forcing which leads to
such a statistical state.
In the following scaling argument, we ignore effects of the external forcing on the TTLVCF.
Later, in Sec. \ref{sec:conclusion}, we discuss the effects
when we analyze the DNS data of the TTLVCF.

Let us first change time variables from $t_1$ and $t_2$
to the average time, $T$, and the relative time, $\tau$ as
\begin{equation}
  T \equiv \frac{t_1+t_2}{2}, \quad
  \tau \equiv t_1-t_2.
\end{equation}
The new variables are useful because the TTLVCF is symmetric with respect
to the $T$-axis as shown in Fig.\ref{fig:intro}.

Now we present the scaling ansatz for the TTLVCF.
Our arguments are given thereafter.
The TTLVCF can be this form:
\begin{equation}
  C^L(r_0, T,\tau) =  \varepsilon T \Phi(r_0, T, \tau),
  \label{eq:CL_scaling_law}
\end{equation}
which we assume to be valid at an appropriate time interval. Here $\varepsilon$
is the energy dissipation rate.
The non-dimensional function $\Phi(r_0, \tau)$ includes the deviation from the dimensional analysis.
By using the idea of incomplete similarity \cite{Barenblatt2003},
we argue that $\Phi(r_0, T, \tau)$ can be written as
\begin{equation}
  \Phi(r_0, T, \tau)
  = G  \left(\frac{T_B}{T}\right)^\gamma  g^L\left(\frac{\tau}{T_B^{\beta} T^{1-\beta}}\right),
  \label{Phig}
\end{equation}
where $G$ is a non-dimensional, non-zero constant
and $g^L(X)$ is a non-dimensional self-similar function with $g^{L}(0) = 1$.
Here the two scaling exponents, $\beta$ and $\gamma$, appear. They are not determined by dimensional analysis.
With the ansatz Eq.(\ref{Phig}), the width of the ridge along the diagonal line $t_1 = t_2$ shown
in Fig.\ref{fig:intro} is given by $T (T_B / T)^\beta$.
Later in Sec.\ref{sec:dns}, by comparing with DNS data, we will show that the exponents take
the following functional forms
\begin{align}
  \beta  & = \beta_0 + \tilde{\beta}\left(\frac{T_\eta}{T_B}, \frac{T_L}{T_B}\right),\label{eq:beta_scaling}    \\
  \gamma & = \gamma_0 + \tilde{\gamma}\left(\frac{T_\eta}{T_B}, \frac{T_L}{T_B}\right).\label{eq:gamma_scaling}
\end{align}
Here, we introduce three time scales, $T_B$, $T_\eta$, and $T_L$, which are
given by
\begin{equation}
  T_B  \equiv \left( \frac{r_0^2}{\varepsilon}\right)^{1/3},
  T_\eta  \equiv \left( \frac{\eta^2}{\varepsilon}\right)^{1/3},
  T_L  \equiv \left( \frac{L^2}{\varepsilon}\right)^{1/3},
  \label{eq:time_scale}
\end{equation}
where $T_B$ is the Batchelor time associated with the initial separation $r_0$ \cite{Batchelor1950},
$T_\eta$ is the smallest time scale of turbulence associated with the smallest length scale,
$\eta$, such as Kolmogorov dissipation length, and
$T_L$ is the largest time scale of turbulence associated with the largest length scale, $L$,
such as the integral scale.
Moreover, $\beta_0$ and $\gamma_0$ are the asymptotic exponents at infinite Reynolds number, and
therefore their values are independent of $T_\eta$ and $T_L$.

Now let us explain how we reach the scaling ansatz, (\ref{eq:CL_scaling_law})--(\ref{eq:gamma_scaling}),
of the TTLVCF.
Our argument here follows the self-similar analysis of Ref.\cite{Barenblatt2003}.
First, we specify the system of all the governing parameters of the correlation function, $C^L(r_0, T, \tau)$.
It depends on the average time, $T$, the relative time, $\tau$,
the initial separation of particle pairs, $r_0$, the energy dissipation rate or the average energy flux in the inertial subrange, $\varepsilon$,
the smallest length scale of turbulence such as the Kolmogorov length, $\eta$, and
the largest length scale of turbulence such as the integral scale, $L$.
Taking them into account, we rewrite the arguments of the TTLVCF as
\begin{equation}\label{eq:corr}
  C^L(r_0, T,\tau, \varepsilon, \eta, L)
  = \langle  \delta v_i(\bm{a}, \bm{r}_0, s = 0|t_1)   \delta v_i(\bm{a}, \bm{r}_0, s = 0|t_2)   \rangle,
\end{equation}
Here, we take average over the Lagrangian label $\bm{a}$. Hence we omit the dependence on $\bm{a}$.
We set the labeling time to zero, i.e.,  $s = 0$ and, $t_1$ and $t_2$ are measured from this time
origin. We also omit dependence on $s$ on the lhs of Eq.(\ref{eq:corr}).
For the 2D inverse energy-cascade turbulence, we can use the characteristic length of the drag
as $L$ instead of the integral scale and use the energy flux cascading inversely in the inertial subrange
as $\varepsilon$.  In this case, we can explicitly write down $L$ dimensionally by using the drag
coefficient and the energy flux. This may be an advantage of the 2D inverse energy-cascade turbulence.

Second, we apply the Buckingham $\Pi$-theorem \cite{Barenblatt2003} to Eq.(\ref{eq:corr})
by assuming that the independent dimensions are $\varepsilon$ and $T$. This means that
all the other governing parameters are non-dimensionalized by $\varepsilon$ and $T$.
This leads to an expression with the dimensionless function, $C_*^L$, as
\begin{equation}
  C^L(r_0, T,\tau, \varepsilon, \eta, L)
  =
  \varepsilon T ~
  C^L_*\left(
  \frac{\tau}{T},
  \frac{r_0}{\varepsilon^{1/2}T^{3/2}},
  \frac{\eta}{\varepsilon^{1/2}T^{3/2}},
  \frac{L}{\varepsilon^{1/2}T^{3/2}}
  \right).
  \label{eq:Pi_thm}
\end{equation}
This is rewritten by using the time scales (\ref{eq:time_scale}) as
\begin{equation}
  C^L(r_0, T,\tau, \varepsilon, \eta, L)
  = \varepsilon T
  C^L_*\left(
  \frac{\tau}{T}, \left(\frac{T_B}{T}\right)^{3/2},
  \left(\frac{T_\eta}{T}\right)^{3/2},
  \left(\frac{T_L}{T}\right)^{3/2}\right).
  \label{CLall}
\end{equation}

Third, we consider intermediate asympotics of the time scales, $T_\eta$, $T$, and $T_L$,
and reduce the number of the arguments on the rhs of Eq.(\ref{CLall}).
As is clear from our choice of the independent dimensions,
we assume that $T_\eta$ and $T_L$ are sufficiently separated and that
\begin{equation}
  T_\eta \ll T
  \ll T_L .
  \label{ineq:inertial}
\end{equation}
This intermediate time interval for $T$ is a Lagrangian counterpart of the inertial subrange
of the Eulerian velocity statistics.
We call this time interval the inertial subrange in this paper.
In this inertial subrange, we assume that the correlation function becomes independent
of $T_\eta$ and $T_L$. This implies that the complete similarity holds for $T_\eta$ and $T_L$.
Hence, with a dimensionless function $C^L_{**}(\zeta, \xi)$,
Eq.(\ref{CLall}) is simplified as
\begin{equation}
  C^L(r_0, T,\tau, \varepsilon)
  = \varepsilon T C^L_{**}\left( \frac{\tau}{T}, \frac{T_B}{T}\right).
\end{equation}

Fourth, let us also assume that the initial separation, $r_0$, is sufficiently small.
Namely, we consider that the average time is much larger than $T_B ~(\propto r_0^{2/3})$:
\begin{equation}
  (T_\eta \ll) \quad T_B \ll T  \quad (\ll T_L)
  \label{ineq:Batchelor}
\end{equation}
To discuss behavior of $C^L$ in this time range, for simplicity,
we rewrite the dimensionless times as
\begin{equation}
  \zeta \equiv \frac{\tau}{T}, \quad \xi \equiv \frac{T_B}{T}.
\end{equation}
The additional asymptotics (\ref{ineq:Batchelor}) implies $\xi \to 0$.
Now there are two possibilities for
the asymptotic behavior of $C^L_{**}(\zeta, \xi)$ as $\xi \to 0$
\cite{Barenblatt2003}:
\begin{center}
  \begin{enumerate}
    \renewcommand{\labelenumi}{(\roman{enumi})}
    \item the limit of $C^L_{**}(\zeta, 0)$ exists and is finite and non-zero,
    \item no finite limit of $C^L_{**}(\zeta, 0)$ exists, or the limit is zero if it exists.
  \end{enumerate}
\end{center}
We do not know a priori which case holds
unless the full functional dependence of $C^L(r_0, T, \tau, \varepsilon)$
was obtained theoretically from the Navier-Stokes equations.
It is necessary to study data for small $\xi$, which is obtained from DNS or
laboratory experiment in order to conclude which case is valid \cite{Barenblatt2003}.

Now let us discuss implications of each case.
In the case (i),
a scaling law for $C^L(\zeta, 0)$ is consistent with the K41 phenomenology.
In other words, we can determine the scaling relations for $C^L(\zeta, 0)$
by dimensional analysis: complete similarity.
In this case, we can estimate $\xi$-dependence by the Taylor series
\begin{equation}
  C^L_{**}(\zeta, \xi) = C^L_{**}(\zeta,0) + A_1(\zeta) \xi + O(\xi^2),
  \label{eq:taylor}
\end{equation}
where $A_1(\zeta) = (\partial C^L_{**}/\partial \xi)_{\xi=0}$.
Therefore, in the case (i), the scaling law for $C^L(r_0, T, \tau)$ is as follows:
\begin{equation}
  C^L(r_0, T, \tau) = \varepsilon T A_0\left(\frac{\tau}{T}\right) + \varepsilon T_B A_1\left(\frac{\tau}{T}\right) + O( \xi^2 ),
  \label{eq:complete}
\end{equation}
where $A_0(\tau/T) = C^L_{**}(\tau/T, 0)$.

Furthermore, when the intermediate asymptotics (\ref{ineq:inertial}) is insufficient,
we assume that complete similarity for $C^L(\zeta, 0)$ still holds but the constants
$A_i$ depend on $T_\eta$ and $T_L$ \cite{Barenblatt2014}.
It should be noted that at this situation,
$\xi$ has a lower bound and therefore does not approach zero.
Under this assumption, in the inertial subrange, Eq.(\ref{eq:complete}) may be modified as
\begin{equation}
  C^L(r_0, T, \tau) = \varepsilon T A_0\left(\frac{\tau}{T}, \frac{T_\eta}{T_L} \right) + \varepsilon T_B A_1\left(\frac{\tau}{T}, \frac{T_\eta}{T_L} \right) + O( \xi^2 ).
  \label{eq:complete_mod}
\end{equation}
It should be noted that the width at $T$ of the ridge along the diagonal line $t_1 = t_2$
shown in Fig.\ref{fig:intro} is given by $T$ in the inertial subrange for any Reynolds number.

On the other hand, in the case (ii), a scaling law for $C^L(\zeta, \xi)$ has non-trivial
scaling exponents which cannot be determined by dimensional analysis: incomplete similarity.
When $\xi$ is sufficiently small, as a natural self-similar form suggested in \cite{Barenblatt2003}, we propose
\begin{equation}
  C^L_{**}(\zeta, \xi) = G \xi^\gamma g^L\left(\frac{\zeta}{\xi^\beta}\right),
  \label{ssform}
\end{equation}
where $g^L(X)$ is a dimensionless function and $g^L(0) = 1$.
Here, $G$ is a non-zero constant factor and independent of $T_B$. It should be noticed
that Eq.(\ref{ssform}) is consistent with the case (ii) since the function $g^L(X)$ is bounded.
The scaling exponents, $\beta$ and $\gamma$, are determined either
by the Navier-Stokes equations (or, more precisely, closure equations for the TTLVCF)
or by comparison with experimental data.
As well as the case (i), when the intermediate asymptotics (\ref{ineq:inertial})
is insufficient and then $\xi$ is finite,
we expect that the incomplete self-similar form (\ref{ssform}) still holds
but the scaling exponents $\beta$ and $\gamma$ may depend on $T_\eta$, $T_L$ and $T_B$.
The functional dependency of $\beta$ and $\gamma$ on $T_B$
are not obvious like Eq.(\ref{eq:complete_mod}) because
the Taylor expansion for $\xi$ does not exist.

As we will show in Sec.\ref{sec:dns}, the case (ii) yields a better
agreement with DNS data of the 2D inverse energy-cascade turbulence
than for the case (i).
Therefore, we conclude that the case (ii) holds for the 2D inverse energy-cascade turbulence.
Finally, we arrive at Eqs (\ref{eq:CL_scaling_law}) and (\ref{Phig}),
which is supposed to hold under the conditions (\ref{ineq:inertial}) and (\ref{ineq:Batchelor}).

Having obtained the ansatz for the temporal inertial subrange, we now
consider finite-Reynolds number effects.
The argument below is heuristic and should be justified experimentally.
At finite Reynolds numbers, we assume that the above self-similar form (\ref{eq:CL_scaling_law})
is applicable. However, we assume that the scaling exponents, $\beta$ and $\gamma$,
are dependent on $T_\eta$ and $T_L$ as in the Eqs (\ref{eq:beta_scaling}) and (\ref{eq:gamma_scaling}).
In the next section we show that these hypothetical formulae of the exponents are useful
to fit the DNS data obtained at finite Reynolds numbers and to infer the asymptotic behavior of $C^L$.

It should be noted that undetermined scaling exponents such as $\beta$ and $\gamma$
do not appear in scaling relations for the two-time Eulerian correlation function $C^E(r, t_1, t_2)$
of the velocity increments, which is defined by
\begin{equation}
  C^E(r, t_1, t_2)  = \langle \delta u_i(\bm{x}, \bm{r}, t_1) \delta u_i(\bm{x}, \bm{r}, t_2) \rangle,
  \label{defce}
\end{equation}
where $r = |\bm{r}|$ and the Eulerian velocity increment is given by
$\delta u_i(\bm{x}, \bm{r}, t_1) = u_i(\bm{x} + \bm{r}, t_1) - u_i(\bm{x}, t_1)$.
We used spatial homogeneity and isotropy.

Because of the statistically steady state,
$C^E(r, t_1, t_2)$ does not depend on $T$ and can be written with all the governing parameters by
\begin{equation}
  C^E(r, t_1, t_2) = C^E(r, \tau, \varepsilon, \eta, L).
\end{equation}
According to the $\Pi$-theorem,
there exists a dimensionless function, $C^E_{*}$, by regarding $\varepsilon$ and $r$ as the independent parameters,
such that $C^E$ has the form,
\begin{equation}
  C^E(r, \tau, \varepsilon, \eta, L) =
  \varepsilon^{2/3}r^{2/3} C^E_*\left(\frac{\tau}{\varepsilon^{-1/3}r^{2/3}},\frac{\eta}{r}, \frac{L}{r} \right).
\end{equation}
Furthermore, when we consider that $r$ is in the inertial subrange,
\begin{equation}
  \eta \ll r \ll L,
  \label{ineq:euler_inertial}
\end{equation}
we assume that, as $\eta \to 0$ and $L \to \infty$, the dependence on $\eta$ and $L$
can be ignored. Then $C^E$ has a reduced form,
\begin{equation}
  C^E(r, \tau, \varepsilon) = C_2 \varepsilon^{2/3}r^{2/3} g^E\left(\frac{\tau}{\varepsilon^{-1/3}r^{2/3}}\right),
  \label{eq:CE_scaling_law}
\end{equation}
where $C_2$ is a universal constant related to the Kolmogorov constant and the function $g^E(X)$ satisfies
$g^E(0) = 1$. This is consistent with Kolmogorov's phenomenology.
In this way, the scaling law of the Eulerian velocity correlation function can be determined
by the dimensional analysis thanks to the statistical stationarity.
This is different from the TTLVCF.
However, it should be noted that the sweeping effect by large-scale advection of eddies
\cite{Kraichnan1964f,Tennekes1975} may be more dominant than the Kolmogorov time scale
$\varepsilon^{-1/3} r^{2/3}$. If that is the case, the Eulerian correlation function
may be different from the the scaling law given in Eq.(\ref{eq:CE_scaling_law}),
see Ref.\cite{Wallace2014a,He2017a} for review.

Before we leave this section,
we summarize the assumptions that we made to arrive at the scaling law (\ref{Phig}),
(\ref{eq:beta_scaling}), and (\ref{eq:gamma_scaling}) with their physical meaning
and how to validate them.
\begin{description}
  \item[(Assumption 1) Governing parameters of the TTLVCF]
    They appear in the arguments of the TTLVCF on the lhs of Eq.(\ref{eq:corr}).
    It should be noticed that we do not include parameters related to the external force.
    We assumed here that they are not relevant for the sake of argument.
    However, this may not be valid for the small-scale forcing in 2D turbulence.
    We will later discuss this point in Sec. \ref{sec:conclusion}.

  \item[(Assumption 2) Scale separations of time, Eqs.(\ref{ineq:inertial}) and (\ref{ineq:Batchelor})]
    We assume that the Batchelor time $T_B$ is in the inertial range to arrive at
    main result (\ref{Phig}).

  \item[(Assumption 3) Vanishing dependence on $\bm{T_\eta}$ and $\bm{T_L}$ in the intermediate asymptotics]
    This is related to Assumption 2 above.
    It is a natural assumption in the inertial range as made in the K41 theory.
    However, it is not obvious
    whether the assumption can be applied to the Lagrangian statistics.
    To validate this point, we will perform DNS with different Reynolds numbers.

  \item[(Assumption 4) Persistent dependence of $\bm{T_B}$ in the intermediate asymptotics]
    This is also related to Assumption 2 above.
    In our argument from Eq.(\ref{eq:corr}) to Eq.(\ref{ineq:Batchelor}),
    we assume that the effects of the initial separations remain
    for a long time and therefore that the statistics of particle pairs
    keep being dependent on their histories.
    In the following sections, we will investigate this memory effects of
    the initial separations by means of DNS of 2D inverse energy-cascade turbulence.

  \item[(Assumption 5) Functional form of the TTLVCF at short times, Eq.(\ref{ssform})]
    Equation (\ref{ssform}) is a standard candidate of the functional forms
    in the incomplete self-similarity framework.
    The validation of this assumption with DNS is our principal goal of the next section
    \ref{sec:dns}.

  \item[(Assumption 6) Finite Reynolds-number effects on the exponents $\bm{\beta}$ and $\bm{\gamma}$]
    In Eqs.(\ref{eq:beta_scaling}) and (\ref{eq:gamma_scaling}),
    we proposed that specific functional forms of $\beta$ and $\gamma$ where
    $T_\eta$ and $T_L$ re-appear, in spite of Assumptions 2 and 3.
    This is an empirical form we find with DNS to parameterize finite Reynolds number effects,
    as we will see in the next section.

\end{description}
\section{Numerical experiments\label{sec:dns}}

\subsection{Numerical details}
We perform DNS of the 2D inverse energy-cascade turbulence
in order to numerically verify the ansatz made in Sec.\ref{sec:scaling}.
We suppose that the Eulerian velocity field, $\bm{u}(\bm{x},t)$, follows
the 2D Navier-Stokes equations with forcing, hyperviscous, and hypodrag terms.
We numerically solve the equations in terms of the vorticity,
\begin{equation}\label{eq:NS}
  \frac{\partial \omega}{\partial t} +
  \left(\bm{u}\cdot\nabla\right)\omega =
  (-1)^{h + 1}\nu\Delta^h\omega + \alpha\Delta^{-q}\omega + f,
\end{equation}
where $\omega$ is vorticity filed, $\omega(\bm{x},t) = \partial_x u_y(\bm{x},t) - \partial_y u_x(\bm{x},t)$.
The hyperviscous, and hypodrag terms are the first and second ones on the right hand side (rhs)
of Eq.(\ref{eq:NS}) and $f$ is an external forcing term.
For the 2D inverse energy-cascade turbulence, the smallest and largest time scales
can be explicitly described by the viscous coefficient, $\nu$, and
the drag coefficient, $\alpha$, respectively as below:
\begin{equation*}
  T_\eta  \equiv \left(\frac{\nu}{\varepsilon^{h}}\right)^{\frac{1}{3h-1}}, \quad
  T_L  \equiv \left(\frac{1}{\alpha \varepsilon^q}\right)^{\frac{1}{3q+1}}.
\end{equation*}
This is an advantage of the 2D energy inverse-cascade turbulence,
because the integral time scale of the 3D turbulence
cannot be explicitly described.

The forcing term, $f({\bm x}, t)$, is
given in terms of the Fourier coefficients,
$\hat{f}(\bm{k},t) = k^2\varepsilon_{in}
  /[n_f \hat{\omega}^{*}(\bm{k},t)]$,
where
\, $\hat{f}$ \, denotes the Fourier transform of the function
$f(\bm{x}, t)$.
The energy input rate is denoted by $\varepsilon_{in}$, and
$n_f$ denotes the number of the Fourier modes
in the following forcing wavenumber range.
We select the coefficients, $\hat{f}({\bm k}, t)$,
as non-zero only in high wave numbers, $\bm{k}$,
satisfying $k_f - 1 < |\bm{k}| < k_f + 1$.
Thus, the energy input rate is maintained as a constant in time.
Numerical integration of Eq.~(\ref{eq:NS})
is performed via the pseudospectral method
with the 2/3 dealiasing rule in space and
the 4-th order Runge--Kutta method in time.
The setting and our numerical method are identical to those used
in \cite{XIAO2009,Mizuta2013}.
The typical wavenumber of the hypodrag is dimensionally estimated as
$(\alpha^3/\varepsilon)^{1/(6q+2)}$, which is termed as the frictional wave number,
$k_\alpha$.
Here we use the infrared Reynolds number,
$\mathrm{Re}_\alpha \equiv k_f/k_\alpha$,
as proposed by Vallgren \cite{Vallgren2011}
in order to quantify the span of the inertial subrange.
In Table \ref{tab:parameter} we list the parameters of
simulations used in the study.
\begin{table}
  \begin{center}
    \begin{tabular*}{\textwidth}{@{\extracolsep{\fill} } ccccccccccccccccc}\\
      $N^2$  & $\delta x$ & $\delta t$ & $\nu$ & $h$ &$\alpha$ & $q$ & $k_f$ & $\varepsilon_{in}$ & $N_p^2$ & $\varepsilon$ & $\sigma_\varepsilon$ & $L$ & $u_{\mathrm{rms}}$ & $\mathrm{Re}_\alpha$ & $T_\eta$ & $T_L$ \\
      $1024^2$  & $0.006$ & $0.002$ & $1.82 \times 10^{-38}$ & 8 & $35$ & $1$ & $249$ & $0.1$ & $2048^2$ & $0.019$ & $2.9 \times 10^{-4}$ & $0.38$ & $0.5$ & $40$ & $0.091$ & $1.1$\\
      $2048^2$ & $0.003$ & $0.001$ & $4.664 \times 10^{-43}$ & 8 & $35$ & $1$ & $496$ & $0.1$ & $2048^2$ & $0.019$ & $2.9 \times 10^{-4}$ & $0.37$ & $0.5$ & $80$ & $0.057$ & $1.1$\\
      $4096^2$ & $0.0015$ & $0.001$ & $1.05 \times 10^{-47}$ & 8 & $35$ & $1$ & $997$ & $0.1$ & $4096^2$ & $0.019$ & $2.6 \times 10^{-4}$ & $0.36$ & $0.5$ & $160$ & $0.036$ & $1.1$\\
    \end{tabular*}
    \caption{Parameters of numerical simulations:
      $N^2$, $\delta x = 2\pi / N$, $\delta t$,
      $\nu$, $h$, $\alpha$, $q$, $k_f$, $\varepsilon_{in}$, and $N_p^2$
      denote the number of grid points, grid spacing,
      size of the time step,
      hyperviscosity coefficient,
      order of the Laplacian of the hyperviscosity,
      hypodrag coefficient,
      order of the inverse Laplacian of the hypodrag,
      forcing wavenumber,
      energy input rate of the forcing, and,
      number of the Lagrangian particles, respectively.
      Turbulent characteristics:
      $\varepsilon$, $\sigma_\varepsilon$,
      $L$, $u_{\mathrm{rms}}$, $Re_\alpha$, $T_\eta$, $T_L$, and $N_p^2$
      denote
      mean of the resultant energy flux in the inertial subrange,
      standard deviation of the resultant energy flux,
      integral scale,
      root-mean-square velocity,
      infrared Reynolds number,
      viscous time scale, and,
      drag time scale,
      respectively.}
    \label{tab:parameter}
  \end{center}
  \hrulefill
\end{table}

To obtain the Lagrangian statistics,
we employ a standard particle tracking method.
The flow is seeded with a
large number of tracer particles.
The number of particles, $N_p^2$, for each simulation is described
in Table \ref{tab:parameter}.
The particles are tracked in time
via integrating the advection equation,
\begin{equation}\label{eq:particle}
  \frac{\mathrm{d}}{\mathrm{d} t}\bm{x}_p(t)
  = \bm{u}(\bm{x}_p(t),t),
\end{equation}
where $\bm{x}_p(t)$ denotes the particle position vector.
The numerical integration of Eq.~(\ref{eq:particle})
is performed using the Euler method.
The velocity value at an off-grid particle position is
estimated by the fourth-order Lagrangian interpolation
of the velocity calculated on the grid points.

In Eq.(\ref{eq:NS}),
we use the hyperviscosity, $h=8$, rather than the normal viscosity, $h=1$, for DNSs.
This is because the hyperviscosity extends
the inertial subrange for a given spatial resolution.
We confirmed that the hyperviscosity does not affect
the particle-pair statistics in Ref. \cite{Kishi2020}.

First of all, let us consider to what extent the assumptions on the time separations,
$T_\eta \ll T \ll T_L$ (Eq.(\ref{ineq:inertial}))
and $T_B \ll T$  (Eq.(\ref{ineq:Batchelor})), made in Sec.\ref{sec:scaling}
hold in our DNS.
In the DNS,  $T_L/T_\eta \lesssim 10^2$.
Certainly, this poses limitations on studying whether the asymptotic behavior of the TTLVCF,
Eq.(\ref{eq:CL_scaling_law}), is valid.
In theory, if $T_\eta \ll T_B \ll T_L$, then the particle pairs may be hardly influenced by
neither the viscosity nor the large scale drag from the beginning of the relative diffusion.
However, in practice, due to the limited scale separation,
$\xi = T_B / T$ may not become sufficiently small in our DNS, as $T = (t_1 + t_2) / 2$ increases
while satisfying $T_\eta < T_B < T < T_L$.
Therefore, it is inevitable to consider that
the numerically obtained TTLVCF, $C^L(r_0, T,\tau, \varepsilon)$, depends on $T_\eta$ and $T_L$
even if the large $T_B$ condition, $T_\eta <  T_B < T_L$, is satisfied.

Given these practical limitations, it is useful to relax the large $T_B$ condition and
to consider the case $T_B < T_\eta$, which we call the small $T_B$ condition.
Obviously under the small $T_B$ condition, we cannot ignore viscous effects on particle-pair statistics.
However, the value of $\xi = T_B / T$ can become smaller as the average time $T$ increases in $T_\eta \ll T \ll T_L$
than under the large $T_B$ condition.
Some previous studies investigate a particle-pair statistics under the small $T_B$
condition \cite{Jullien1999,Boffetta2002c,Scatamacchia2012,Biferale2014}.
Of course, it is not obvious that the two different conditions
give the same asymptotic behavior of $C^L(r_0, T,\tau, \varepsilon)$ as $\xi = T_B/ T \to 0$.
Therefore, we investigate dependencies on $T_\eta, T_L$ and $T_B$ for both conditions
in the following subsections.

More specifically,
we investigate the two scaling exponents, $\beta$ and $\gamma$, appeared
in our ansatz (\ref{eq:CL_scaling_law}).
For this purpose, we decompose the TTLVCF
$C^L(r_0, T,\tau, \varepsilon)$ into two parts:
\begin{equation}
  C^L(r_0, T,\tau, \varepsilon) = C_d^L(T, T_B) C_p^L(T, \tau, T_B),
  \label{defcd}
\end{equation}
where $C_d^L(T, T_B)$ corresponds to the TTLVCF along the diagonal line $t_1 = t_2$,
that is, $C_d^L(T, T_B) \equiv C^L(r_0, T, \tau = 0, \varepsilon)$.
The other part $C_p^L(T, \tau, T_B)$ corresponds
to the TTLVCF along a line $t_1 + t_2 = 2T$,
which is perpendicular to the diagonal line. Its value at $\tau = 0$ is normalized:
$C_p^L(T, \tau = 0, T_B) = 1$.
If the ansatz is correct, $C_p^L(T, \tau, T_B) = g^{L}(\tau / [T^{1- \beta} T_B^\beta] )$.

In what follows, the values of the exponents, $\gamma$ and $\beta$, are estimated
from numerically calculated $C_d^L(T, T_B)$ and $C_p^L(T, \tau, T_B)$, respectively,
as we vary $T_B$ and $T_\eta$.
We consider first the large $T_B$ condition ($T_B > T_\eta$)
and then the small $T_B$ condition.

\subsection{Large $\bm{T_B}$ condition: $\bm{T_\eta \ll T_B \ll T \ll T_L}$}

\begin{figure}[htbp]
  \centering{
    \includegraphics[clip, scale=0.42]{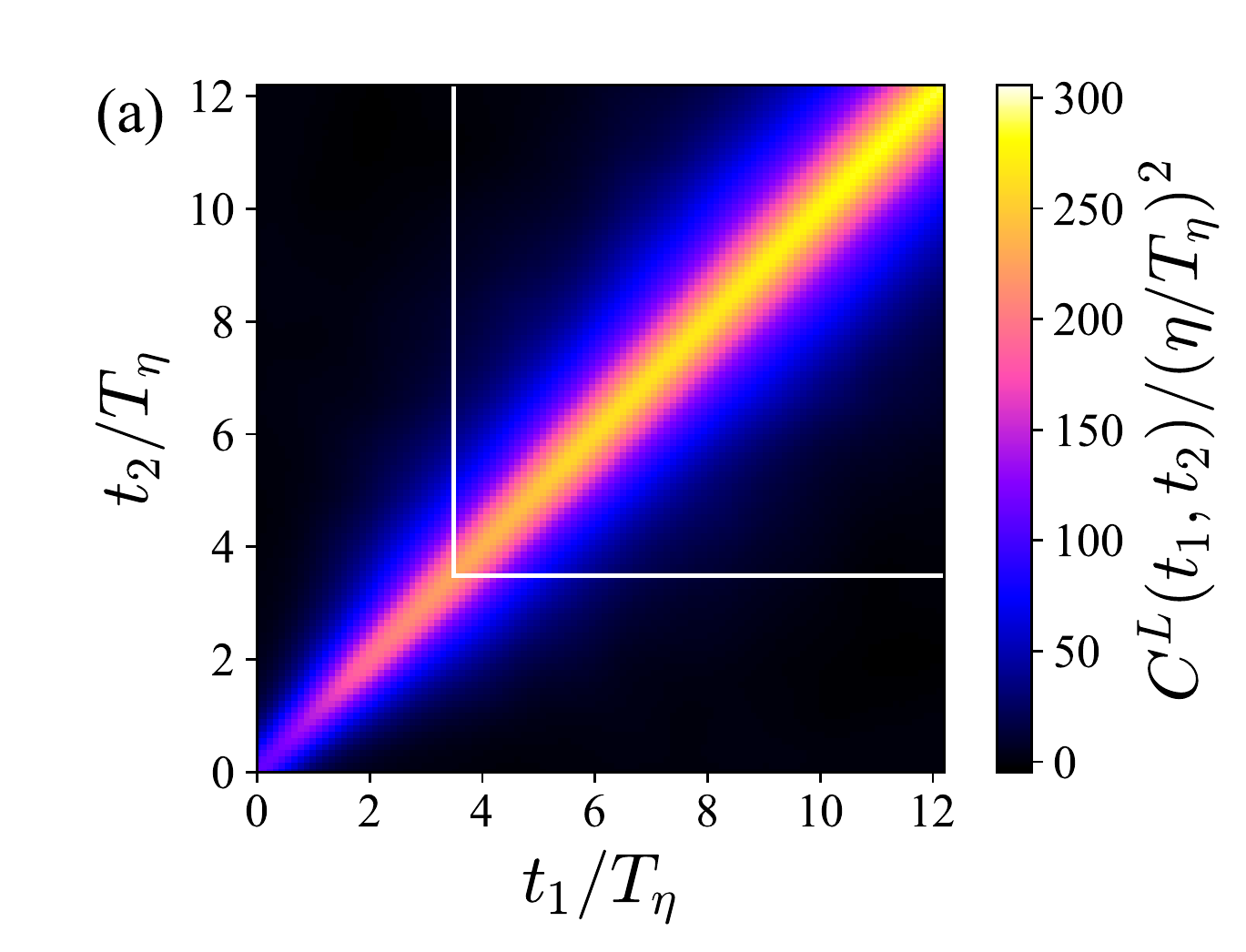}
    \includegraphics[clip, scale=0.42]{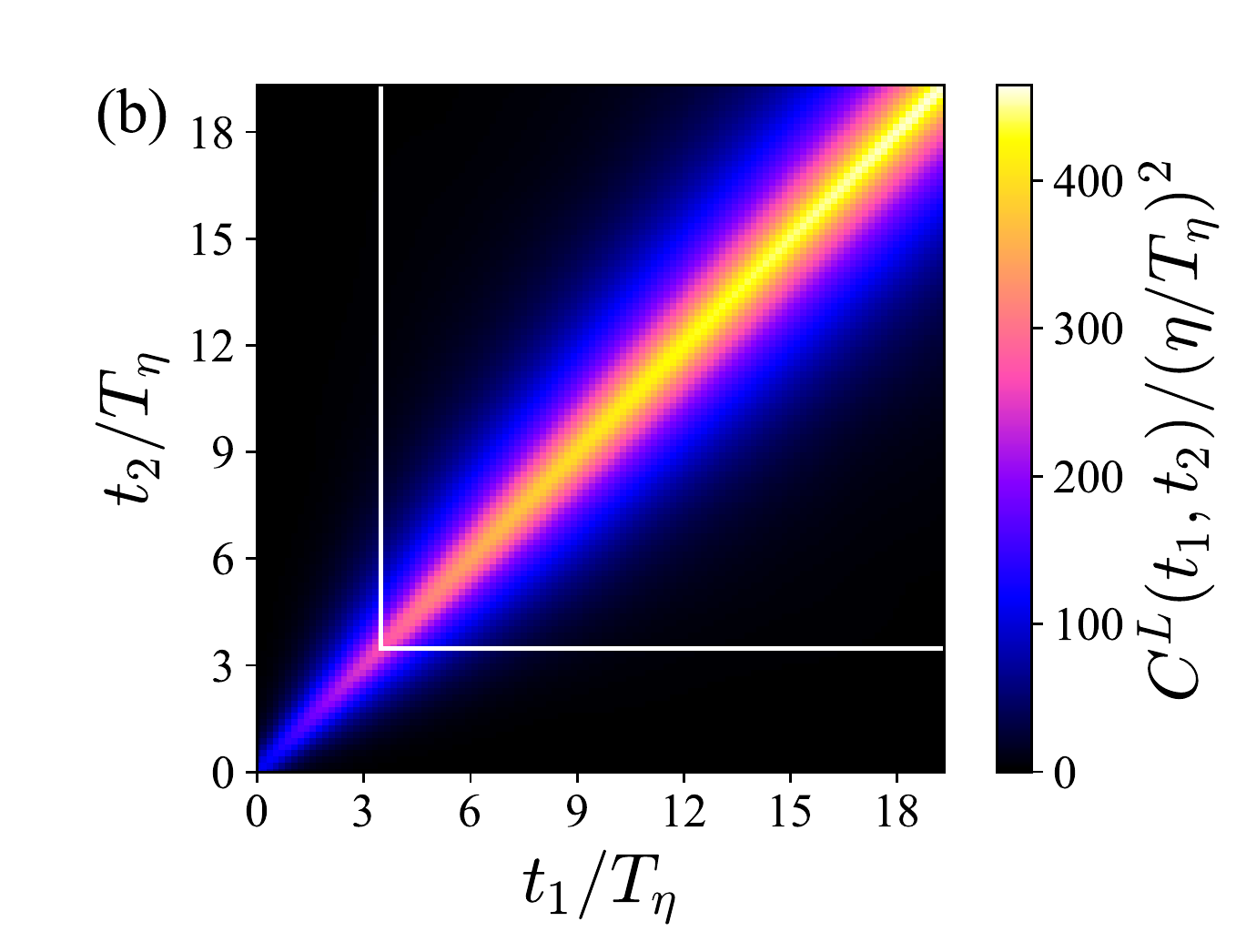}
    \includegraphics[clip, scale=0.42]{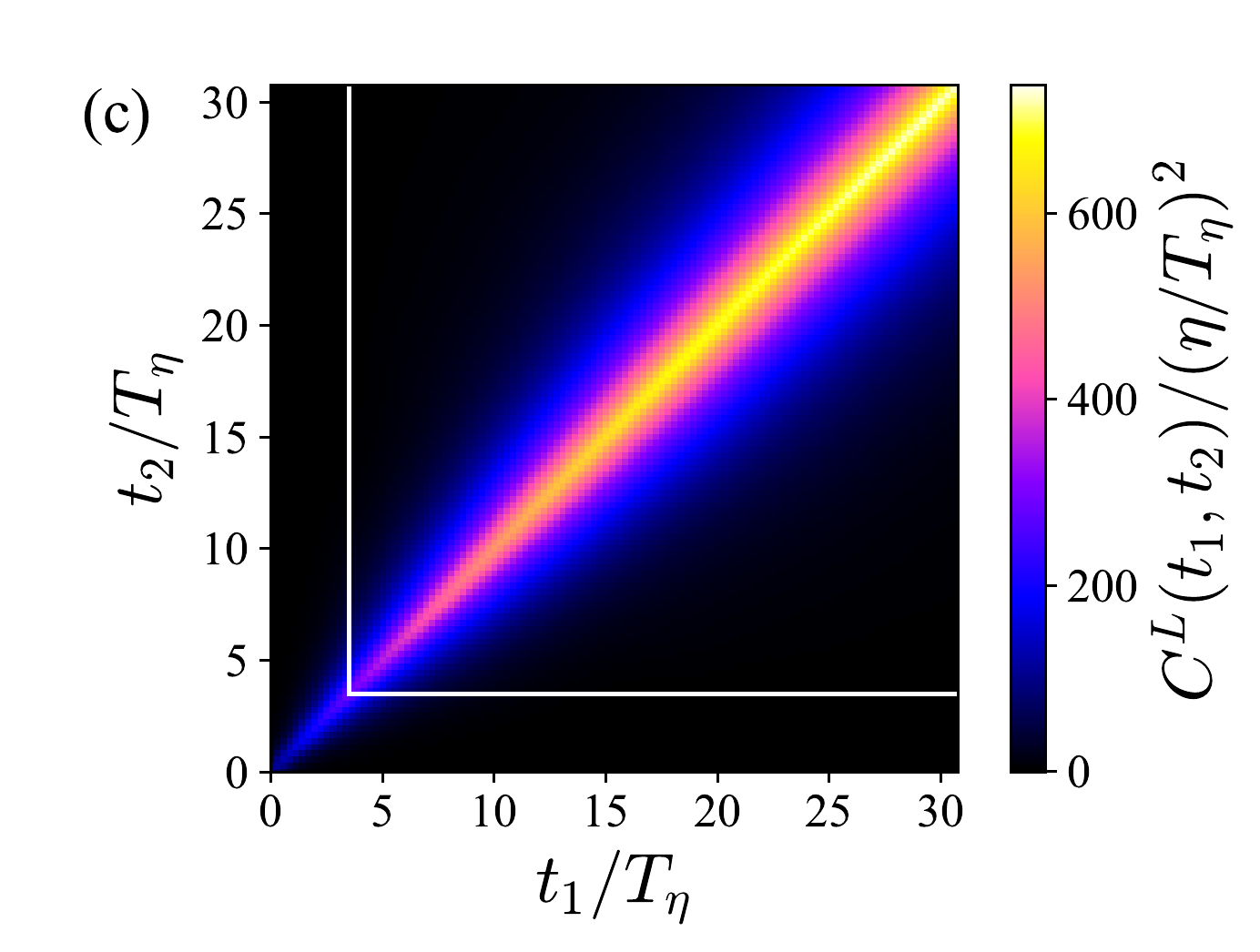}
  }
  \caption{
  Color maps of time evolution of the TTLVCF
  $C^L_{ii}(r_0, t_1, t_2)$ defined in Eq.(\ref{vcdef})
  with $T_B = (r_0^2 / \varepsilon)^{1/3} = 3.5T_\eta$.
  (a) $\mathrm{Re}_\alpha = 40$, (b) $\mathrm{Re}_\alpha = 80$, (c) $\mathrm{Re}_\alpha = 160$.
  The white lines indicate $t_i = T_B$ ($i=1,2$).
  Here the time axes, $t_i/T_\eta$ ($i=1,2$), span from $0$ to $T_L/T_\eta$.
  }
  \label{fig:colormap}
\end{figure}

Let us consider the scaling laws of $C^L(r_0, T,\tau, \varepsilon)$ under the large $T_B$ condition.
In DNS, although this condition, $T_\eta < T_B < T_L$ is satisfied,
Assumption 2 is not sufficient even in our largest simulation with $\mathrm{Re}_\alpha = 160$.
Figure \ref{fig:colormap} shows color maps of $C^L(r_0, T,\tau, \varepsilon)$ in terms of
the original time variables $t_1$ and $t_2$ with $T_B = 3.5 T_\eta$ for the three values of $\mathrm{Re}_\alpha$.
We observe that the width at $T$ of the ridge along the diagonal line (the region where
$C^L(r_0, T,\tau, \varepsilon)$ remains large) becomes wider
as the average time $T = (t_1 + t_2) / 2$ increases.
We also observe that qualitatively this tendency appears to be independent of $\mathrm{Re}_\alpha$.

\begin{figure}[htbp]
  \includegraphics[clip,scale=0.4]{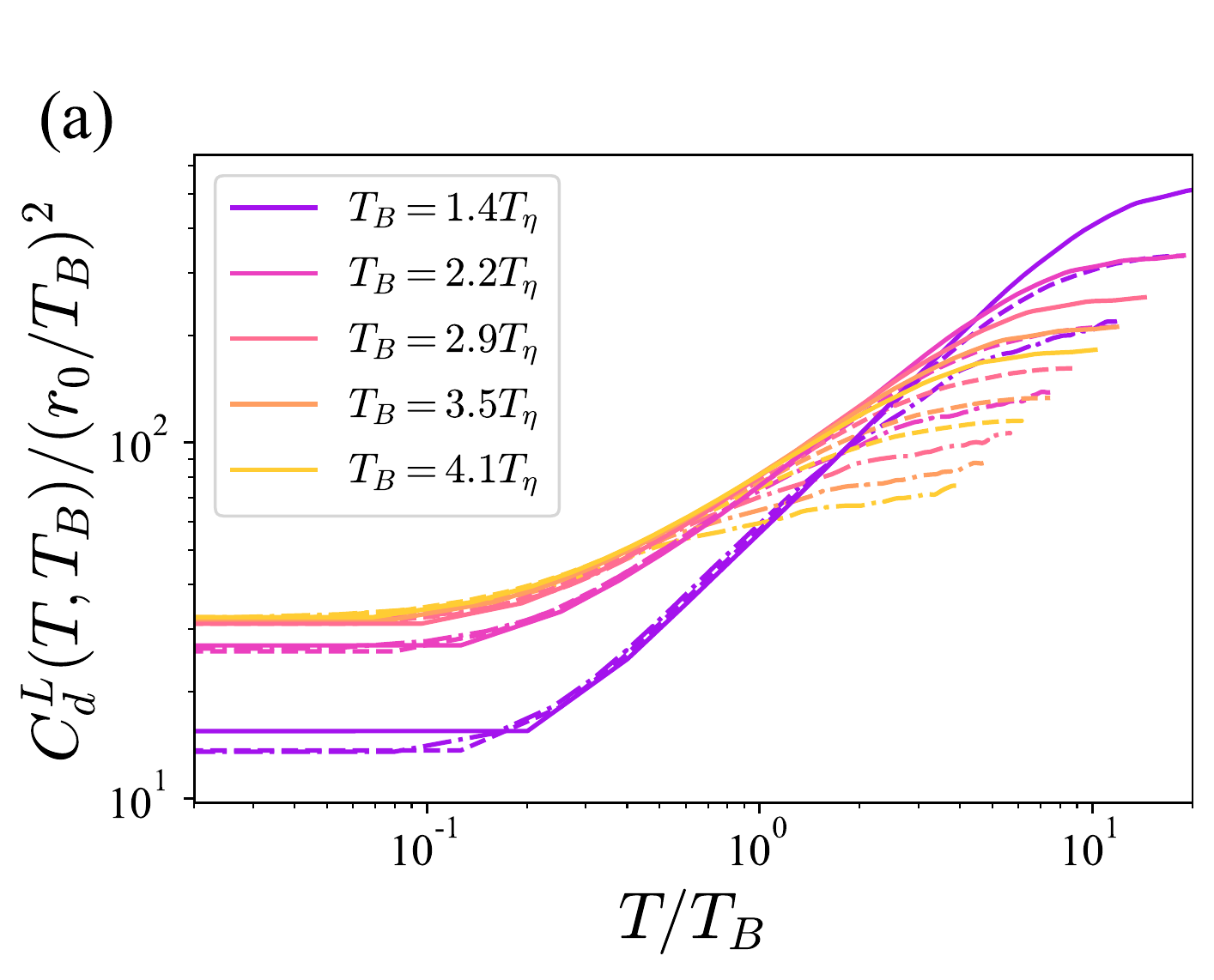}
  \includegraphics[clip,scale=0.4]{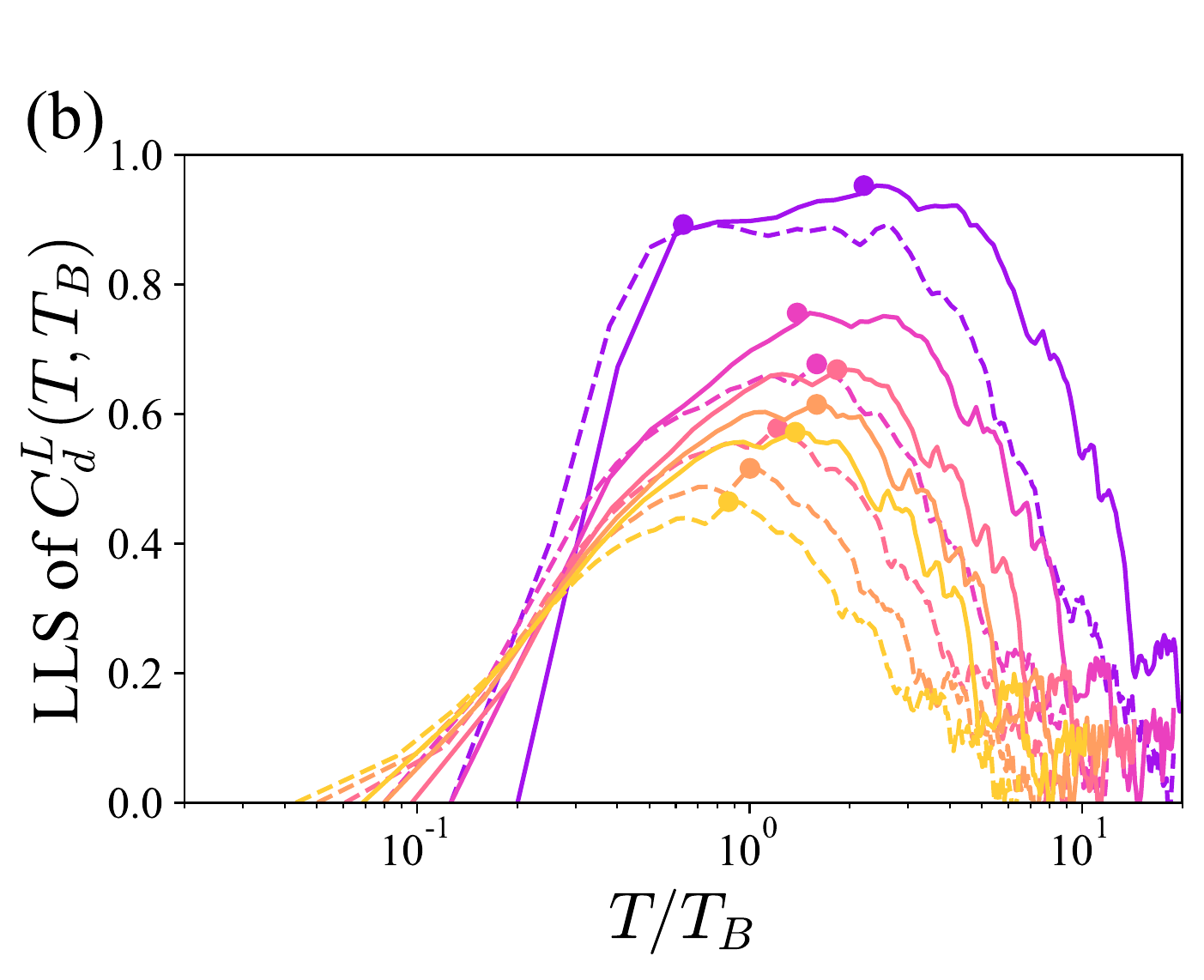}
  \includegraphics[clip,scale=0.4]{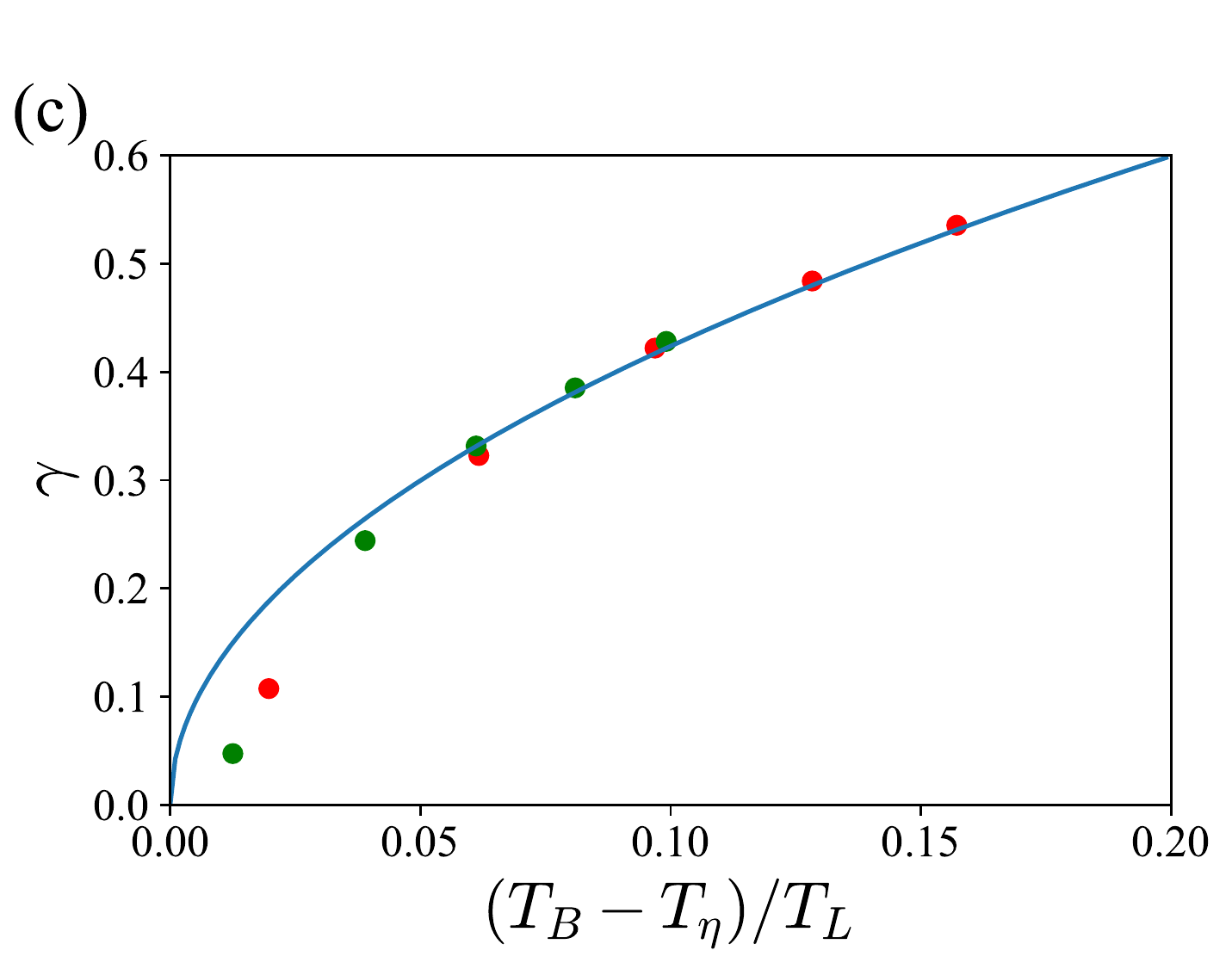}
  \caption{
    (a) Time evolution of $C_d^L(T, T_B)$ normalized by $(r_0/T_B)^2$ for various $T_B$'s at $\mathrm{Re}_\alpha = 40$ (dashed dotted), $80$ (dashed), and $160$ (solid).
    (b) Logarithmic local slope (LLS) of $C_d^L(T, T_B)$ at $\mathrm{Re}_\alpha = 40$ (dashed) and $160$ (solid).
    The filled circle on each line indicates the position of the maximum.
    (c) Value of the exponent $\gamma$ suggested by the maximum value of the LLS
    plotted as a function of $(T_B - T_\eta)/T_L$
    at $\mathrm{Re}_\alpha = 40$ (red) and $160$ (green).
    The blue solid line corresponds to  $1.34 [(T_B-T_\eta)/T_L]^{0.5}$,
    which is determined by the least square fit in the range $0.039 \le (T_B - T_\eta)/T_L \le 0.16$.
  }
  \label{fig:CL_large}
\end{figure}

Now we focus on the behavior of the correlation function along the diagonal line.
Figure \ref{fig:CL_large}(a) shows time evolution of $C_d^L(T, T_B) = C^{L}(r_0, T, \tau = 0, \varepsilon)$
for various $T_B$'s.
Obviously, it indicates that a scaling exponent, if it exists, depends on $T_B$.
Figure \ref{fig:CL_large}(b) shows logarithmic local slopes (LLSs) of $C_d^L(T, T_B)$.
If the ansatz (\ref{eq:CL_scaling_law}) is valid, the LLS becomes $1-\gamma$ (constant).
We see that a narrow plateau region for each LLS.
As $T_B$ approaches $T_\eta$ from above, we observe that
it becomes wider and that the value of the plateau region
becomes closer to $1$ which corresponds to $\gamma=0$.
For further quantification, we infer the value of $\gamma$ for each curve from
the maximum value of the LLS.
The data shown here indicate the dependence on $T_B, T_\eta$ and $T_L$.
To circumvent this, we now use the empirical form for $\gamma$ given in
Eq.(\ref{eq:gamma_scaling}).

Figure \ref{fig:CL_large}(c) shows the maximum values of the LLSs, which we regard as
$\gamma$ in Eq.(\ref{eq:gamma_scaling}); $\gamma$ is dependent on $T_B, T_\eta$ and $T_L$.
The horizontal axis of Fig.\ref{fig:CL_large}(c) is set to $(T_B - T_\eta)/T_L$.
We find empirically this combination of the independent variables, $(T_B - T_\eta)/T_L$,
to make the data points collapse onto a single curve. 
The first observation concerns the behavior  as  $(T_B - T_\eta)/T_L \to 0$
(when $T_B$ approaches $T_\eta$ from above): the exponent $\gamma$ seems to approach $0$.
However, this limit $T_B \to T_\eta$ violates
the large $T_B$ condition, $T_\eta \ll T_B \ll T_\eta$.
The second observation is about the behavior under the large (but smaller than 1) $(T_B - T_\eta)/T_L$ range,
which is consistent with the large $T_B$ condition.
In this range, we observe that the master curve becomes independent of $\mathrm{Re}_\alpha$
as shown in Fig.\ref{fig:CL_large}(c).
Our best fit function to the curve for the exponent $\gamma$ is
\begin{equation}
  \gamma\left(\frac{T_\eta}{T_B}, \frac{T_L}{T_B}\right)
  = \overline{\gamma}_0 \left(\frac{T_B-T_\eta}{T_L}\right)^{1/2},
  \label{eq:gamma_large}
\end{equation}
which is plotted as a solid line in Fig.\ref{fig:CL_large}(c).
Here $\overline{\gamma}_0$ is a constant estimated about $1.34 \pm 0.01$ by the fitting.

Now we come back to the ansatz (\ref{eq:CL_scaling_law}) leading to
$C^L(r_0, T, \tau = 0, \varepsilon) \propto T^{1 - \gamma}$.
The functional form of the exponent (\ref{eq:gamma_large}) indicates that
$\gamma \to 0$ at infinite Reynolds number.
This implies that the K41 scaling,
$C^L(r_0, T, \tau = 0, \varepsilon) = C_d^{L}(T, T_B) \propto T$, is recovered
under the sufficient  scale separation.
However, for this recovery,
the exponent $1/2$ in Eq.(\ref{eq:gamma_large}) suggests
that we need an enormously large $\mathrm{Re}_\alpha$.
For example, in order to get the value of $\gamma$ valid for one effective figure,
$\gamma \sim 0.01$, we may need $T_L/T_\eta \sim 10^5$ (in our DNS here $T_L/T_\eta \sim 30$ at most), which
may correspond to $\mathrm{Re}_\alpha \sim 10^6$.

On the other hand, at small values of $(T_B-T_\eta)/T_L$,
$\gamma$ deviates from the relation (\ref{eq:gamma_large}) as shown in Fig. \ref{fig:CL_large}(c).
Let us suppose that the deviation persists at larger Reynolds numbers.
Then $\gamma$ may have a negative limit value as  $T_B \to T_\eta$ (approaching $T_\eta$ from above).
If we extrapolate the deviation to $(T_B-T_\eta)/T_L = 0$ with linear decrease,
the limit value of $\gamma$ is about $-0.25$.
We cannot conclude whether the deviation remains at sufficient large Reynolds numbers from our DNS.

In summary of the result for the TTLVCF along the diagonal line,
our simulation data suggest that the scaling law of $C_d^L(T, T_B)$
at sufficiently large $\mathrm{Re}_\alpha$ is,
\begin{equation}
  C^L(r_0, T, \tau = 0, \varepsilon) =
  C_d^L(T, T_B) =
  G \varepsilon T_B^{\overline{\gamma}_0\sqrt{\frac{T_B-T_\eta}{T_L}}}~  T^{1-\overline{\gamma}_0\sqrt{\frac{T_B-T_\eta}{T_L}}}.
\end{equation}
where $G$, which is the constant appeared in Eq.(\ref{eq:CL_scaling_law}),
is estimated as $G \sim 80$ from the compensated plot of Fig.\ref{fig:CL_large}(a)
by $T_B^{\gamma} T^{1-\gamma}$ (the compensated plot is not shown).

It is noted that $T_L$ is kept constant and $T_\eta$ is changed
when $\mathrm{Re}_\alpha$ is increased in our DNS.
Thus, the limit $(T_B-T_\eta)/T_L \to 0$ is consistent with $T_B \to T_\eta$ in this study.
On the other hand, if we can change both values of $T_\eta$ and $T_L$,
the limit $(T_B-T_\eta)/T_L \to 0$ indicates two states:
One is $T_B \to T_\eta$ and the other is $T_L \to \infty$ while keeping $T_B-T_\eta$ constant,
where the large $T_B$ condition can hold.
Hence, if we change the value of $T_L$ and fix the value of $T_B-T_\eta$,
the data points of $\gamma$ may collapse onto another curve
different from the former one at small $(T_B-T_\eta)/T_L$,
which indicates the latter state.
Nevertheless, we assume that the asymptotic value for the limit $(T_B-T_\eta)/T_L \to 0$
is the same for the two states.

\begin{figure}[htbp]
  \centering{
    \includegraphics[clip,scale=0.4]{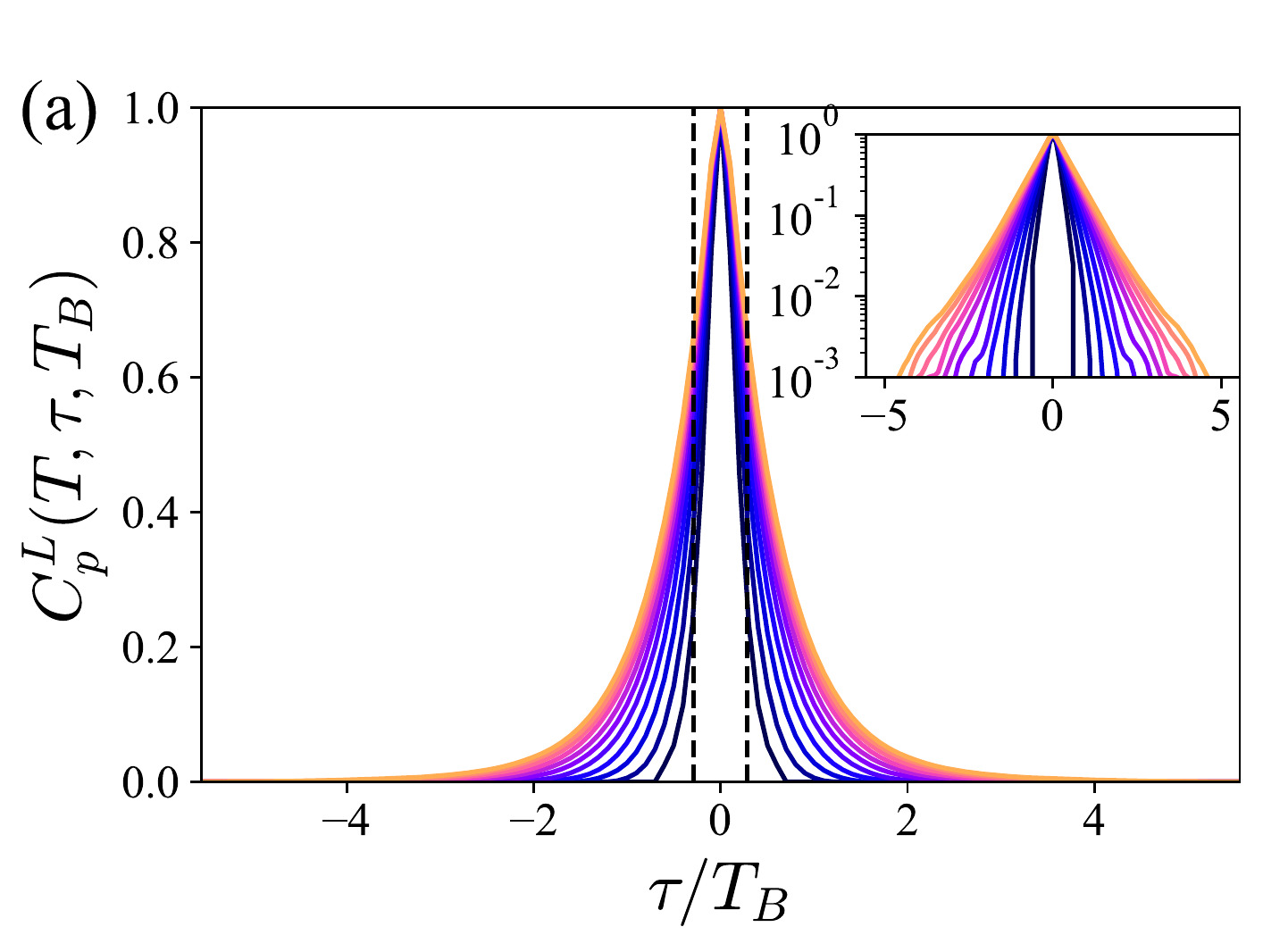}
    \includegraphics[clip,scale=0.4]{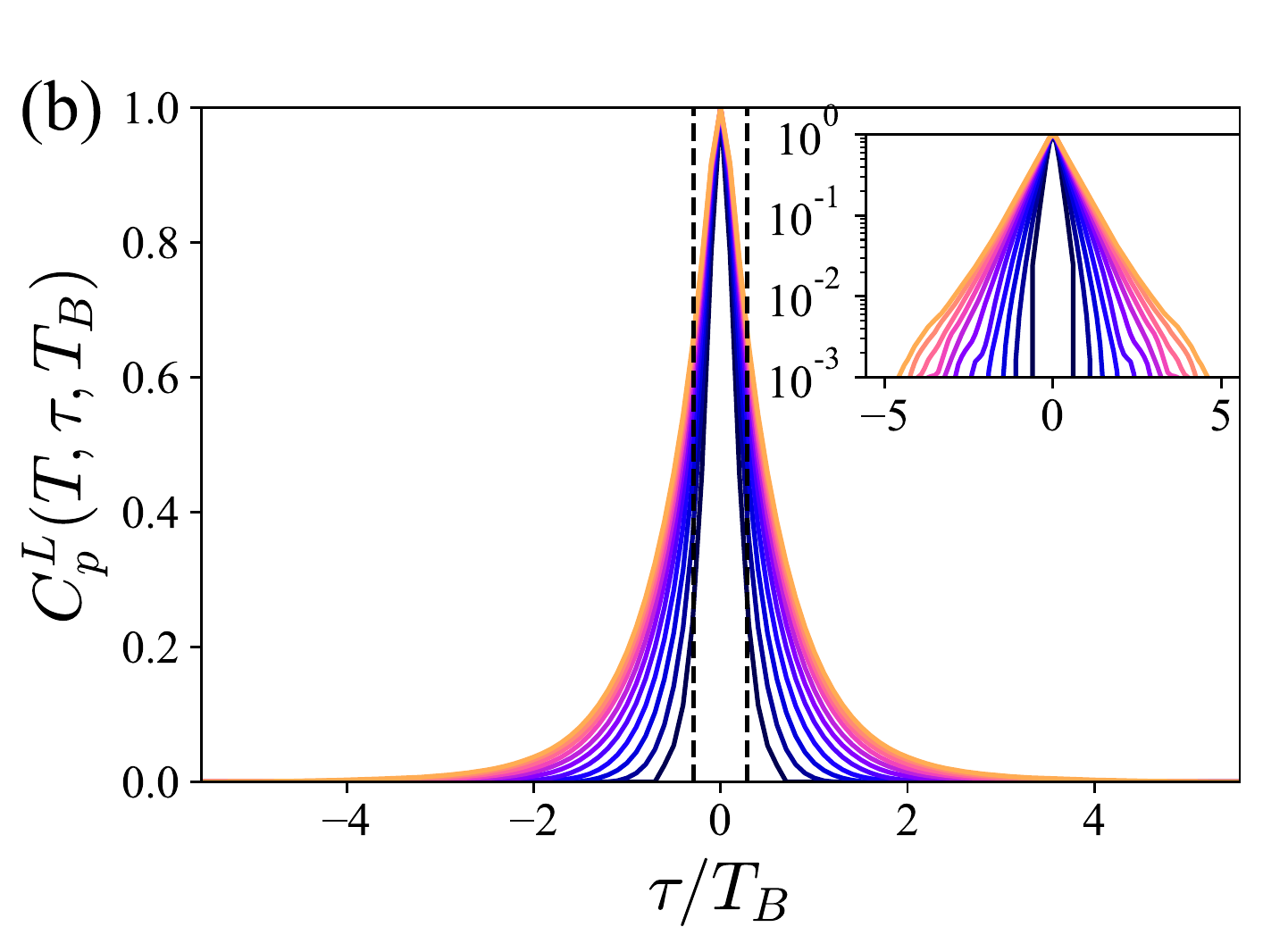}
    \includegraphics[clip,scale=0.4]{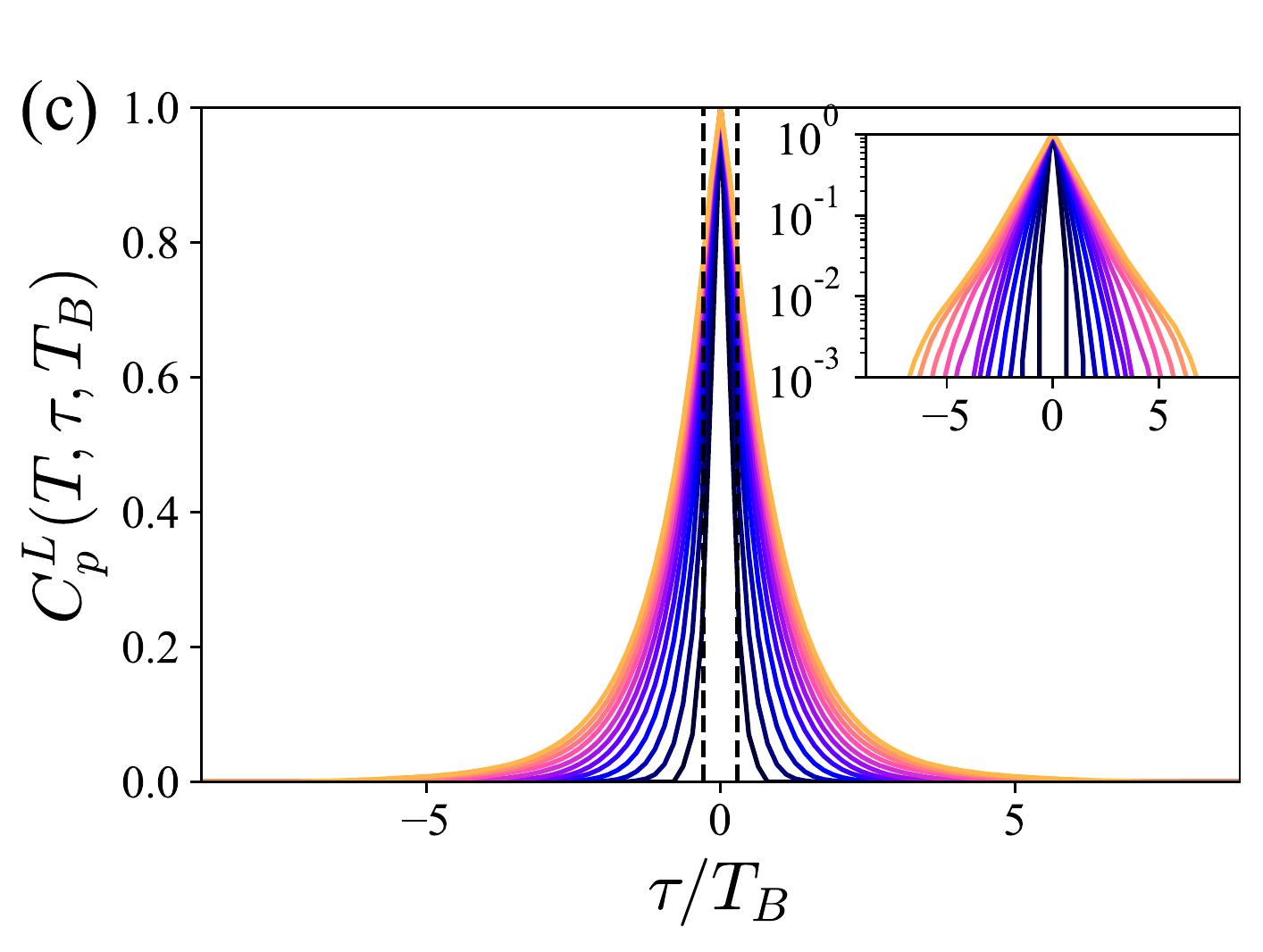}
  }
  \caption{
    Normalized correlation function, $C_p^L(T,\tau, T_B)$, defined in Eq.(\ref{defcd})
    as a function of the relative time $\tau$ with $T_B = 3.5 T_\eta$ for various $T$'s.
    The average time varies in $T_\eta < T < 0.54T_L$ and the corresponding curves are colored
    from black to yellow. The three panels correspond to
    (a) $\mathrm{Re}_\alpha=40$, (b) $\mathrm{Re}_\alpha=80$, and (c) $\mathrm{Re}_\alpha = 160$.
    Two vertical dashed lines in each panel show $\tau/T_B=\pm T_\eta/T_B$, respectively.
    The horizontal axis spans in $- T_L / T_B \le \tau /T_B  \le T_L/ T_B$.
    The insets show the same plots as the outsets but in the lin-log coordinates.
  }
  \label{fig:gL_large}
\end{figure}

Next, we consider the behavior of the TTLVCF along the lines perpendicular
to the diagonal line, which is given by $C_p^L(T,\tau, T_B)$.
Figure \ref{fig:gL_large} shows sectional views of the color map shown in Fig.\ref{fig:colormap} for
various sections given by the lines $t_1 +  t_2 = 2T$. It should be noted that
the curves shown in Fig.\ref{fig:gL_large} are normalized by $C_d^L(T, T_B)$. Hence they
are the graphs of $C_p^L(T,\tau, T_B)$ as a function of the relative time $\tau = t_2 - t_1$ at a fixed average
time $T$.

We first notice that the typical width of the peak of $C_p^L(T,\tau, T_B)$ centered at zero relative time $\tau = 0$ is
given by the dissipation time scale $T_\eta$ initially, i.e., for small average time $T$.
Then, the width becomes larger and larger as the average time increases.
At large average times, $T \lesssim T_L$,
the function $C_p^L(T,\tau, T_B)$ decreases exponentially
as shown in the insets of Fig. \ref{fig:gL_large}.
Moreover, the data indicate that $C_p^L(T,\tau, T_B)$ decreases faster than exponential
at $\tau \sim T_L$.

To quantify the decay of $C_p^{L}(T, \tau, T_B)$,
we use an $n$-th decay time scale, $\tau_{1/n}(T)$, defined as
\begin{equation}
  C_p^L(T,~\tau = \tau_{1/n}(T),~ T_B) = \frac{1}{n}.
\end{equation}
If $\tau_{1/n}(T)$ is power-law such as $\tau_{1/n}(T) \propto T^{1-\beta}$
and the scaling exponent, $\beta$, is independent of the value of $n$, then
$C_p^L(T,\tau, T_B)$ has the self-similar form of the ansatz (\ref{eq:CL_scaling_law}),
that is, $C_p^L(T,\tau, T_B) = g^L(\tau/[T_B^\beta T^{1-\beta}])$.

If $\beta$ depends on $n$,
we can still expect that $C_p^L(T,\tau, T_B)$ has a self-similar form
in a certain interval of $\tau$.
Hereafter, this $n$-dependent $\beta$ is denoted by $\beta_{1/n}$.
The interval is determined by the value of $\tau_{1/n}(T)$.

\begin{figure}[htbp]
  \centering{
    \includegraphics[clip,scale=0.4]{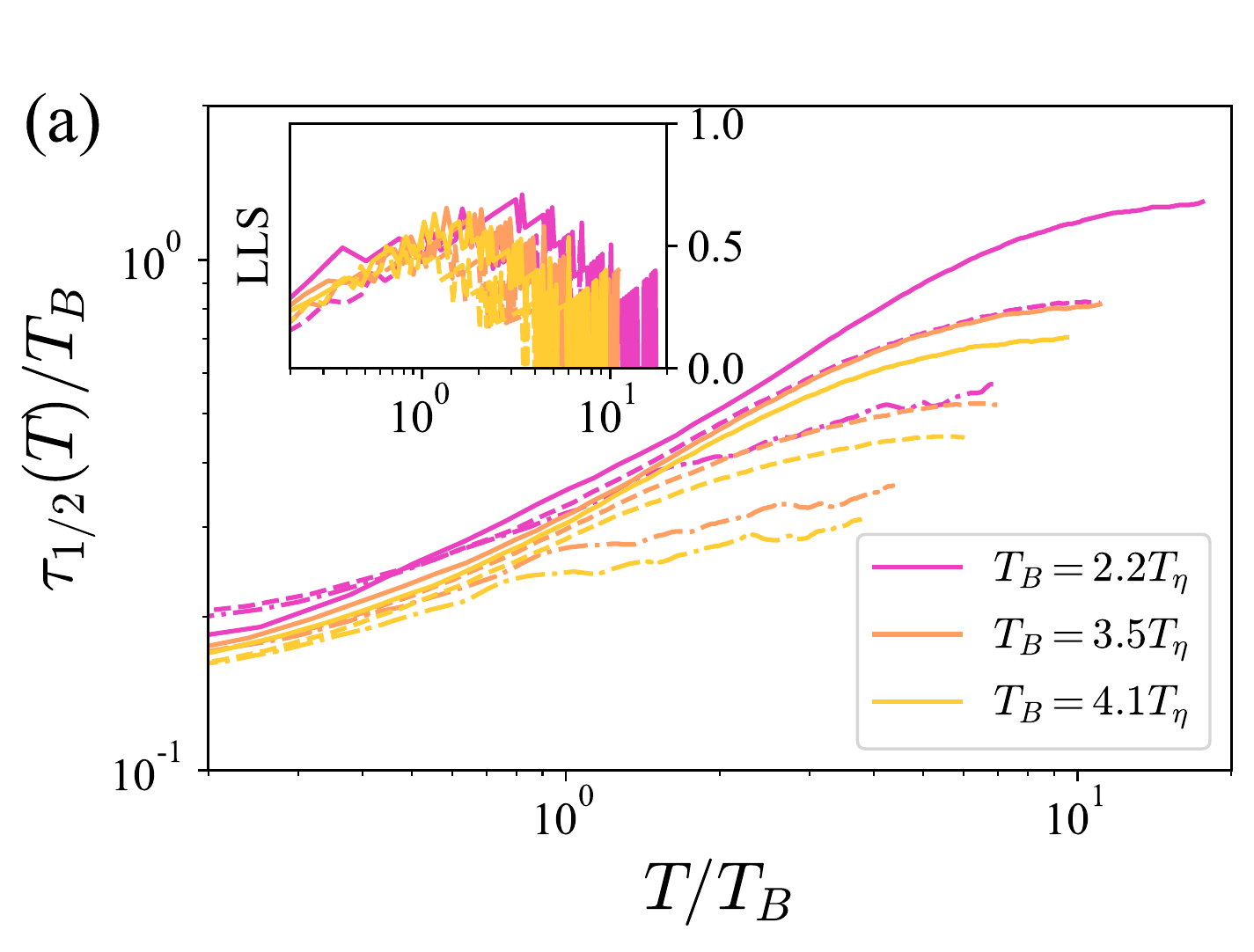}
    \includegraphics[clip,scale=0.4]{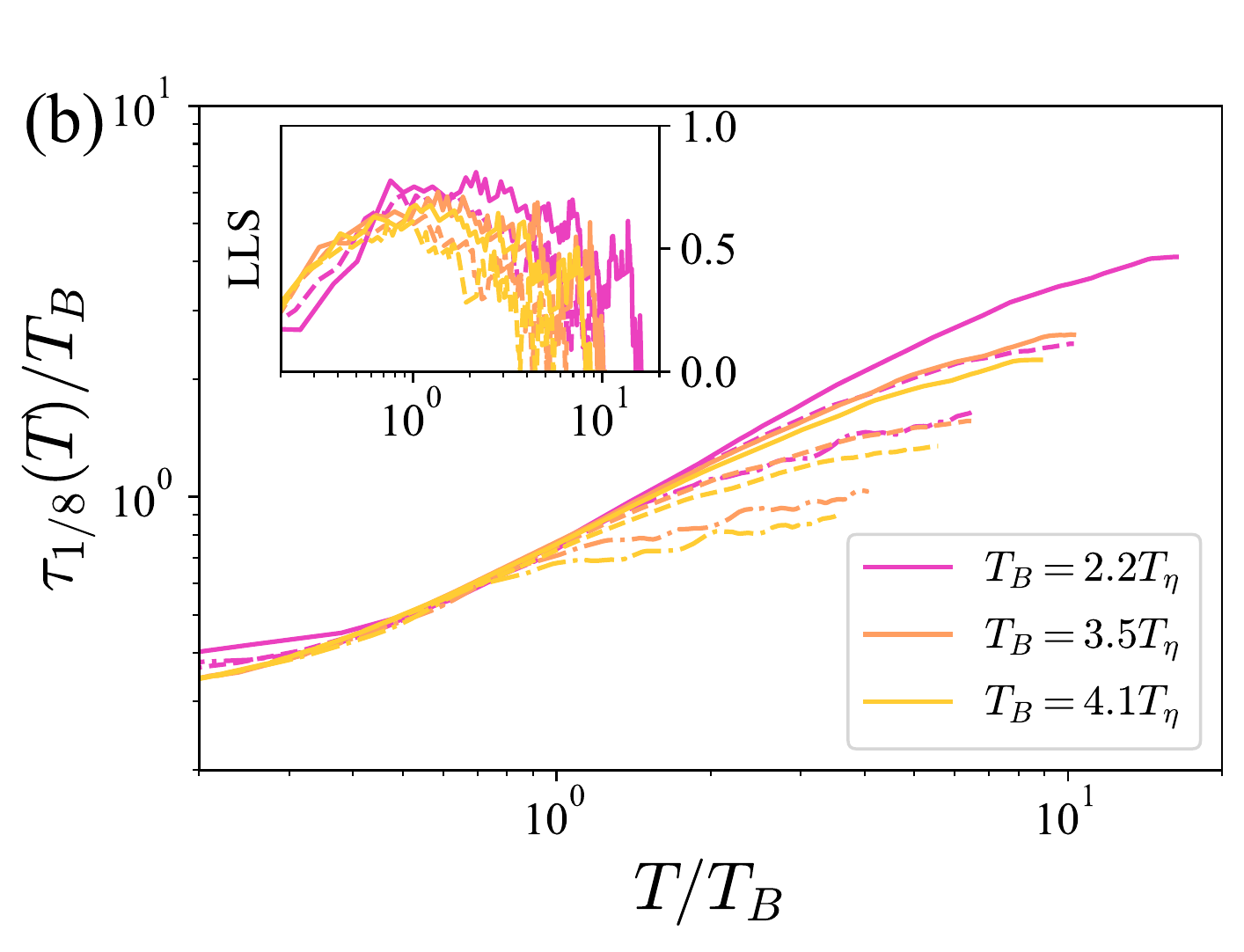}
    \includegraphics[clip,scale=0.4]{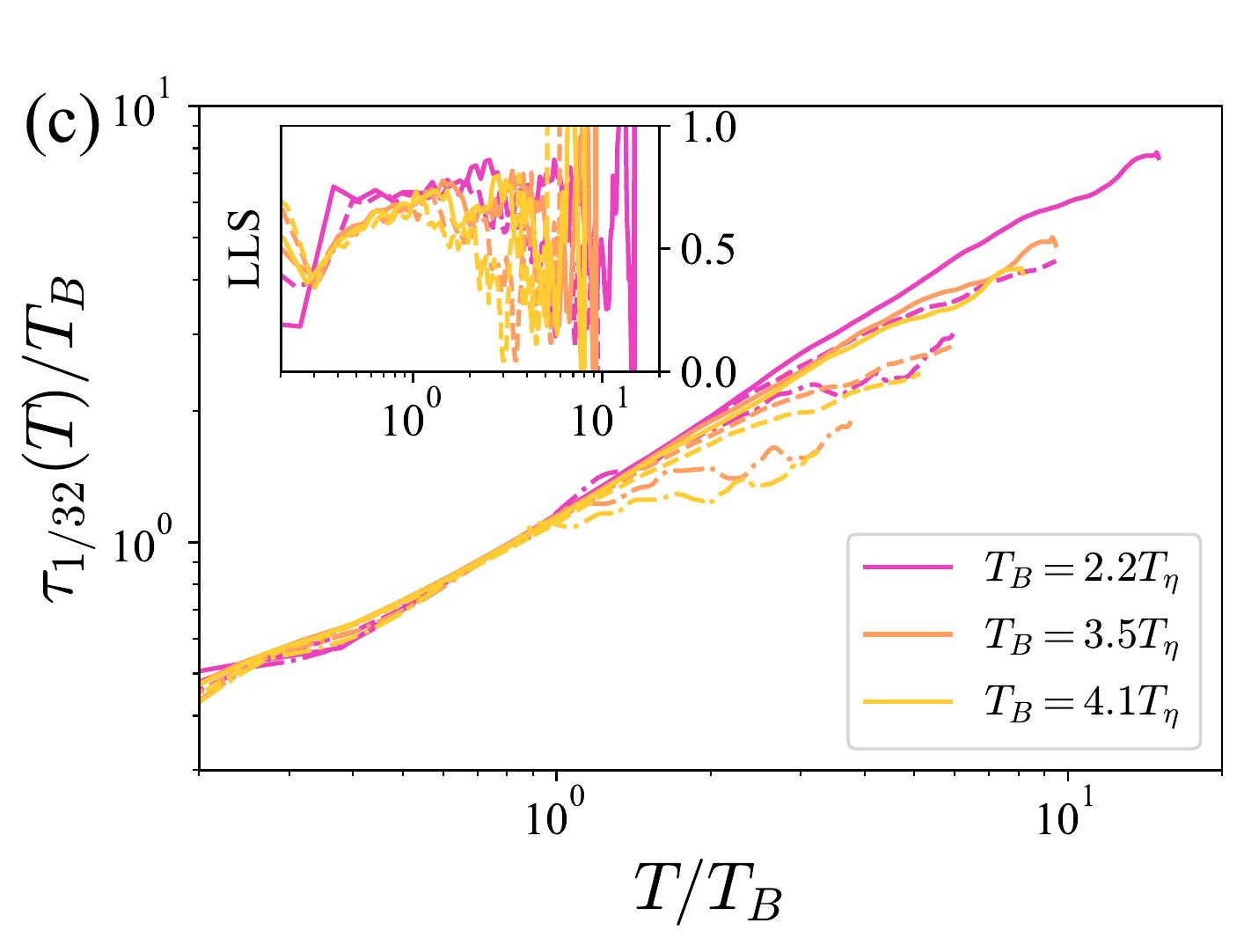}
  }
  \caption{
    $n$-th decay time scale, $\tau_{1/n} (T)$ as a function of the average time $T$
    for (a) $n=2$, (b) $n=8$, (c) $n=32$
    at $\mathrm{Re}_\alpha=40$ (dashed dotted),
    $\mathrm{Re}_\alpha=80$ (dashed),
    and $\mathrm{Re}_\alpha=160$ (solid).
    The insets show the LLS of $\tau_{1/n} (T)$ shown in the ousets.
  }
  \label{fig:tau_n_large}
\end{figure}

Figure \ref{fig:tau_n_large} shows $\tau_{1/n}(T)$ with $n=2,8$, and $32$ for various $T_B$'s.
With a small $n$ such as $n=2$,
we probe the behavior in the vicinity of the peak of $C_p^L(T,\tau, T_B)$,
and, with a large $n$ such as $n=32$,
we characterize the behavior in the tail region of $C_p^L(T,\tau, T_B)$.
For $n=2$ as shown in Fig. \ref{fig:tau_n_large}(a),
$\tau_{1/2}(T)$ strongly depends on $T_B$.
This is because $\tau_{1/2}(T)$ is smaller than $T_B$ for almost all $T$s.
Nevertheless, there may be a power-law behavior in a certain range of $T$.
On the other hand, for larger $n$ such as $n=8$ and $32$, as shown in
Fig.\ref{fig:tau_n_large}(b) and (c),
the power law behavior of $\tau_{1/n}(T)$ becomes clearer and
$\tau_{1/n}(T) \propto T^{1-\beta_{1/n}}$ holds at a certain time interval of $T$.
The scaling exponents, $\beta_{1/n}$ appear to become independent of $T_B$ and
the scaling region becomes larger as increasing $\mathrm{Re}_\alpha$.
These observations lead us to conclude that the ansatz (\ref{eq:CL_scaling_law})
is a reasonable description of the function $C_p^L(T, \tau, T_B)$.

\begin{figure}[htbp]
  \centering{
    \includegraphics[clip,scale=0.4]{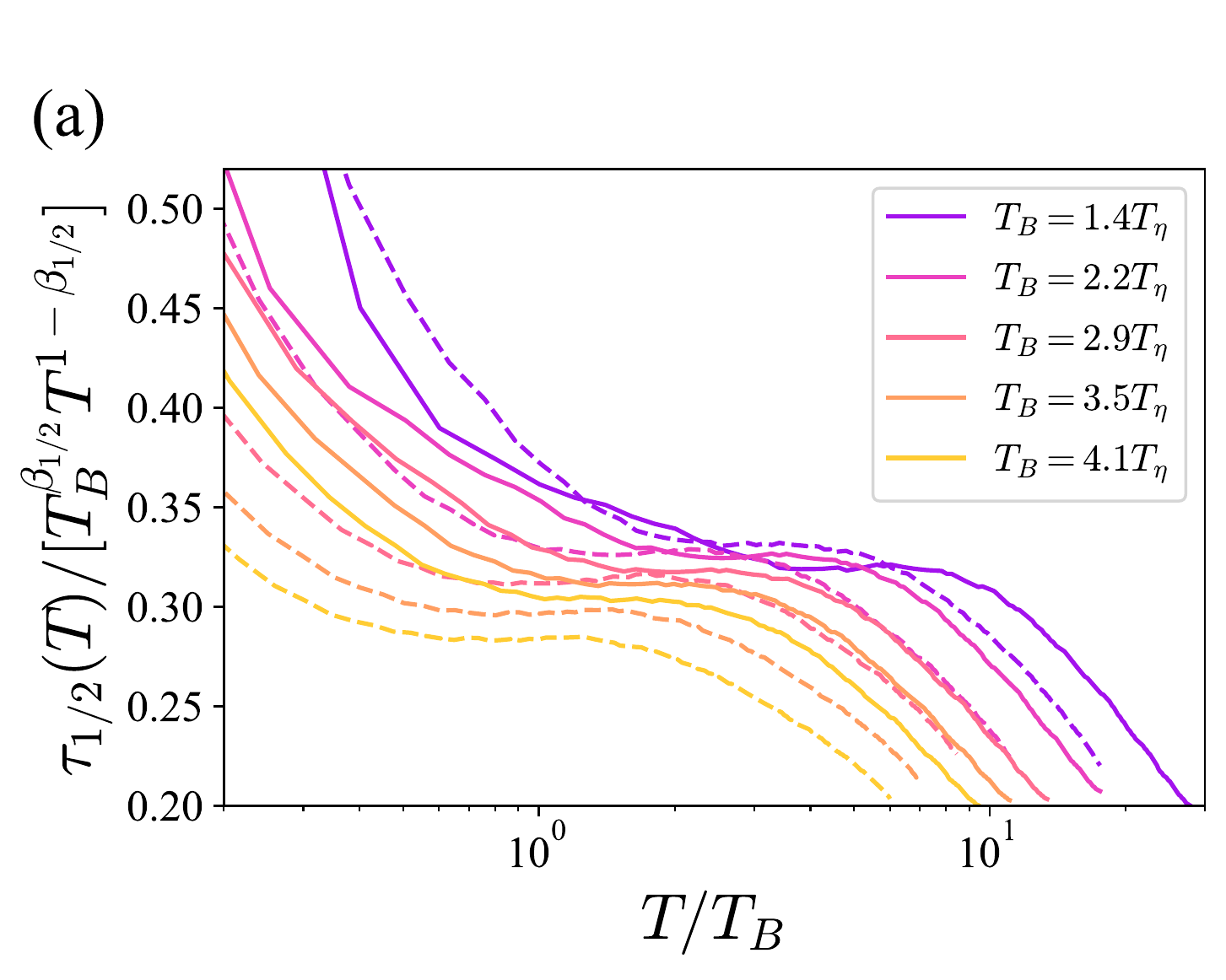}
    \includegraphics[clip,scale=0.4]{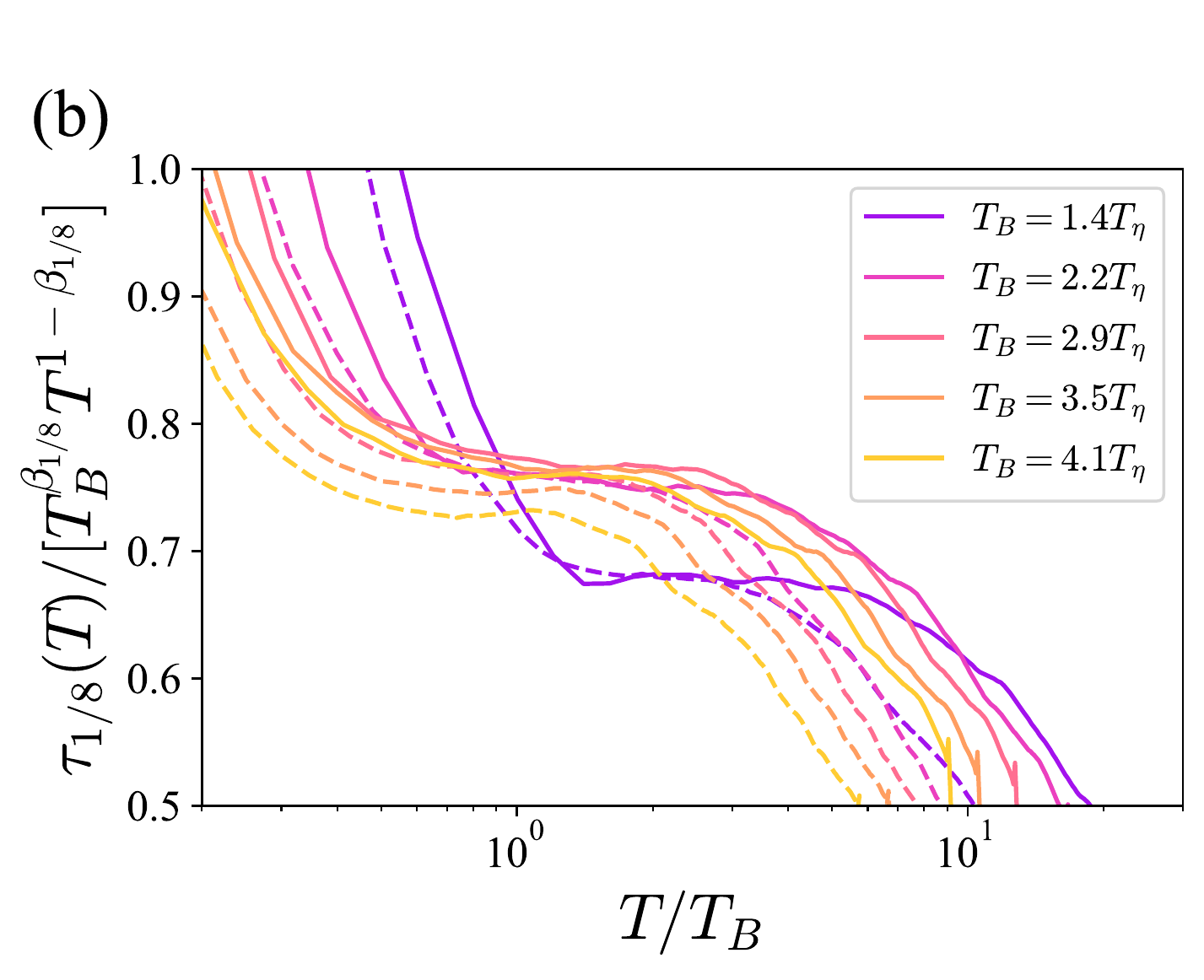}
    \includegraphics[clip,scale=0.4]{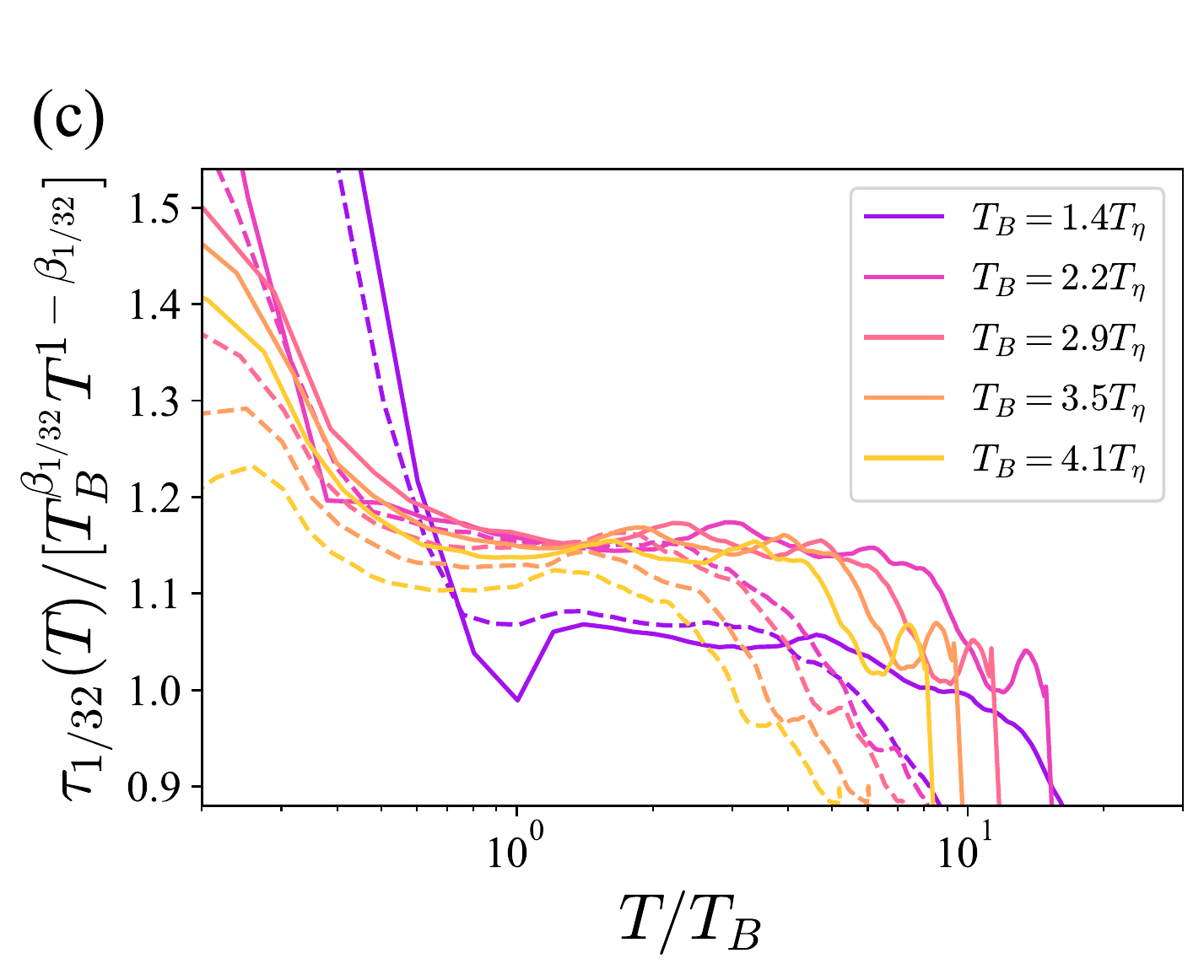}
  }
  \caption{
    Compensated plots of $\tau_{1/n}(T)$ by $T^{1-\beta_{1/n}}$
    for (a) $n=2$, (b) $n=8$, (c) $n=32$
    at $\mathrm{Re}_\alpha=80$ (dashed),
    and $\mathrm{Re}_\alpha=160$ (solid).
    The values of $\beta_{1/n}$ are determined in such a way that each of
    the compensated graphs has the widest flat region.
  }
  \label{fig:tau_n_large_compe}
\end{figure}

However, as shown in the insets of Fig. \ref{fig:tau_n_large},
the LLSs are too noisy to determine the value of $\beta_{1/n}$ accurately.
The noise may be suppressed as we increase massively the number of particle-pair samples.
Instead, here we use compensated plots of Fig. \ref{fig:tau_n_large}
to estimate the value of the scaling exponent $\beta_{1/n}$.
The compensation is based on the self-similar variable $\zeta/\xi^{\beta}$ in Eq.(\ref{ssform}), which
is the argument of the function $g^{L}$. If the self-similarity is valid at $\tau = \tau_{1/n}(T)$
with the exponent $\beta_{1/n}$, the self-similar variable
\begin{equation}
  \left.\frac{\zeta}{\xi^{\beta_{1/n}}}\right|_{\tau = \tau_{1/n}(T)}
  =
  \frac{\tau_{1/n}(T)}{T_B^{\beta_{1/n}}T^{1-\beta_{1/n}}} \equiv D_{n,\beta_{1/n}}(T)
  \label{defD}
\end{equation}
becomes constant which neither depends on $T$ nor $T_B$. For each $n$, we plot $D_{n, q}(T)$ by varying $q$
and find $q_*$ that gives the widest flat region as a function of $T$. We regard this $q_*$
as $\beta_{1/n}$.
We show $D_{n,\beta_{1/n}}(T)$ in Fig. \ref{fig:tau_n_large_compe} for $n = 2, 8$, and $32$.
These compensated plots are less noisy than the LLSs, but they still have tiny oscillations.
As increasing $\mathrm{Re}_\alpha$, we observe that $D_{n,\beta_{1/n}}(T)$ becomes independent
of $T_B$ except for $T_B=1.4T_\eta$, in particular, for $n=32$ as shown in Fig. \ref{fig:tau_n_large_compe} (c).
This indicates that the ansatz (\ref{eq:CL_scaling_law}) is reasonable for $C_p^L(T,~\tau, T_B)$,
albeit that the numerical data is noisy.

\begin{figure}[htbp]
  \centering{
    \includegraphics[clip, scale=0.4]{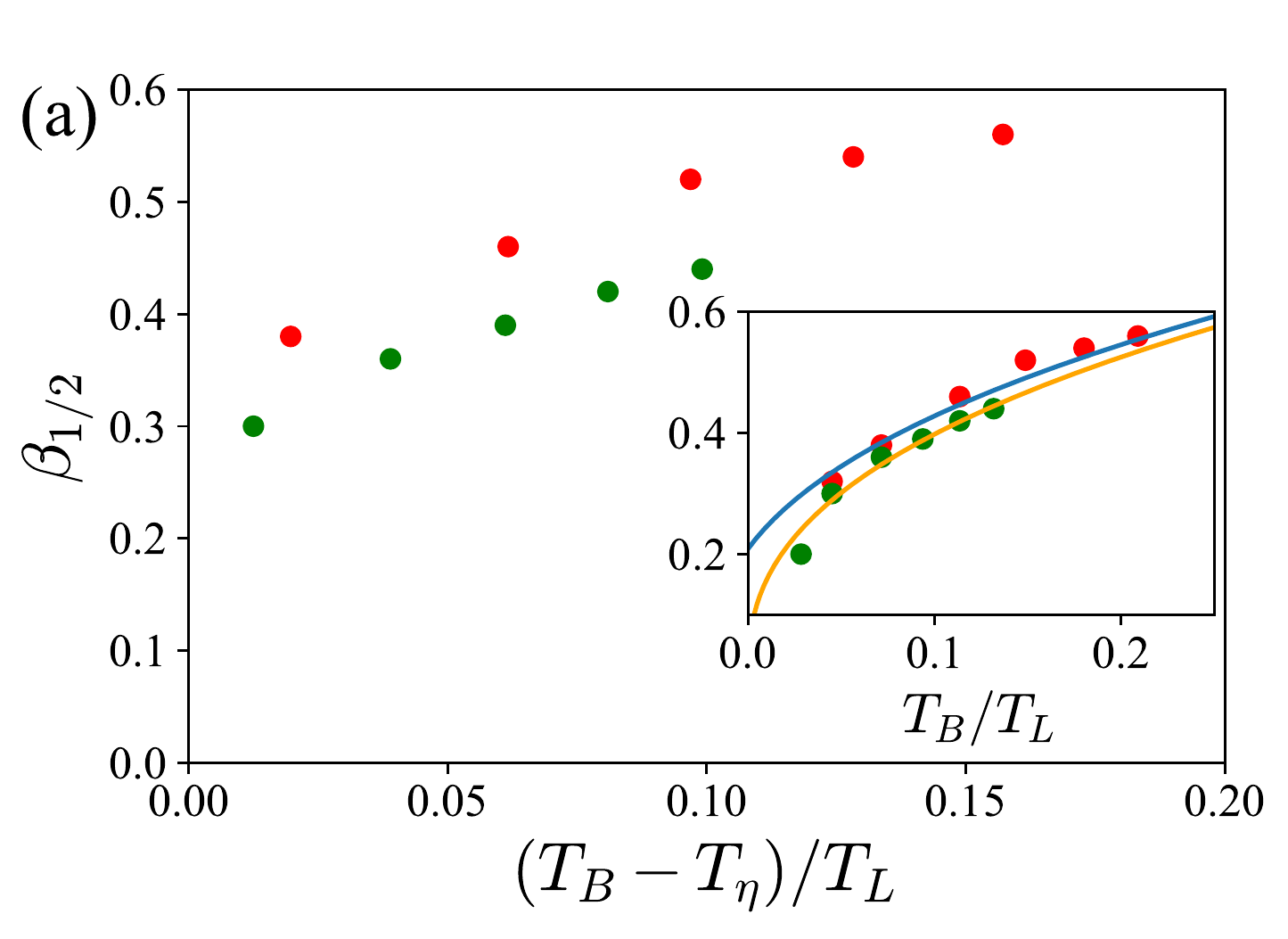}
    \includegraphics[clip, scale=0.4]{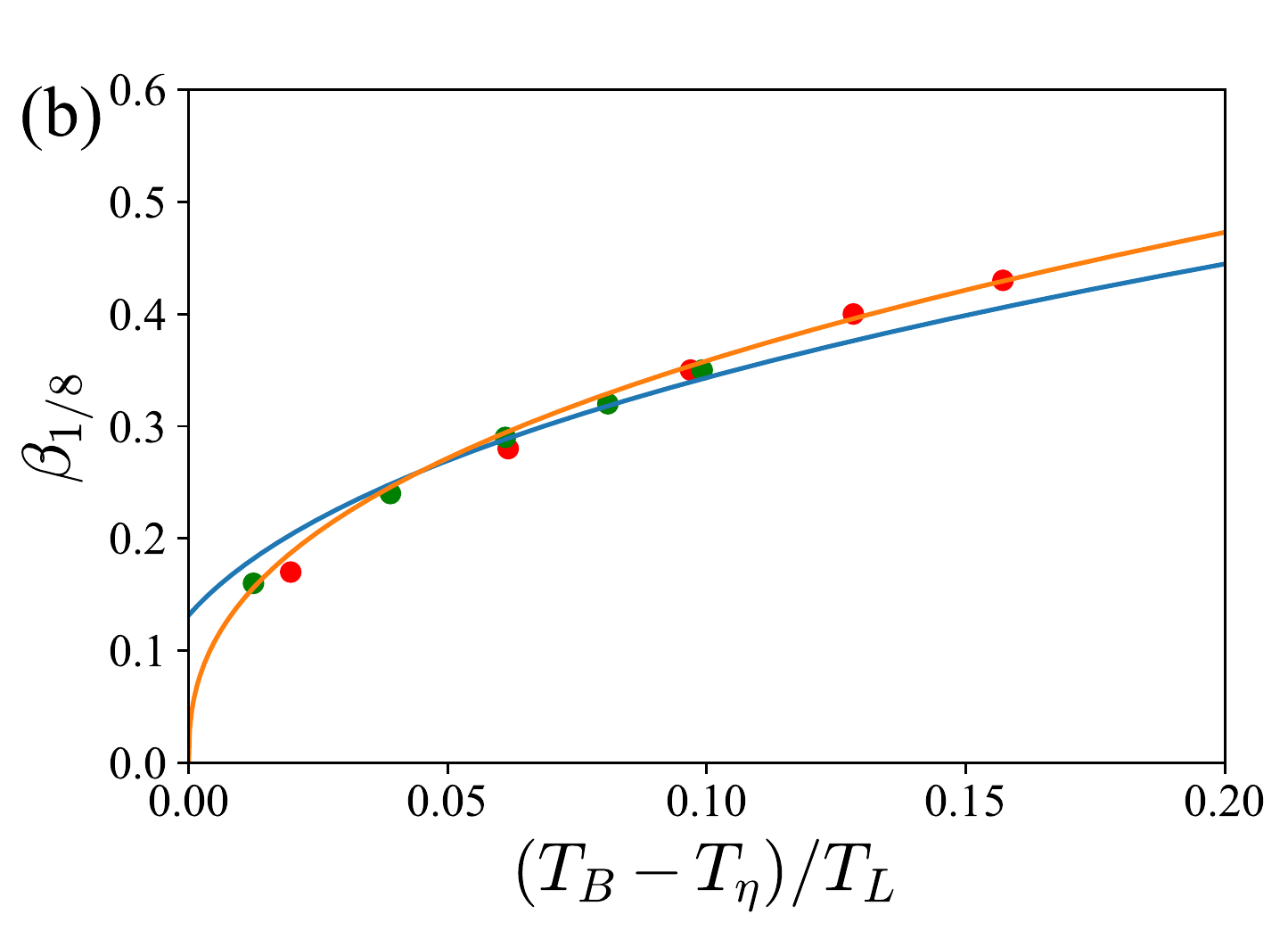}
    \includegraphics[clip, scale=0.4]{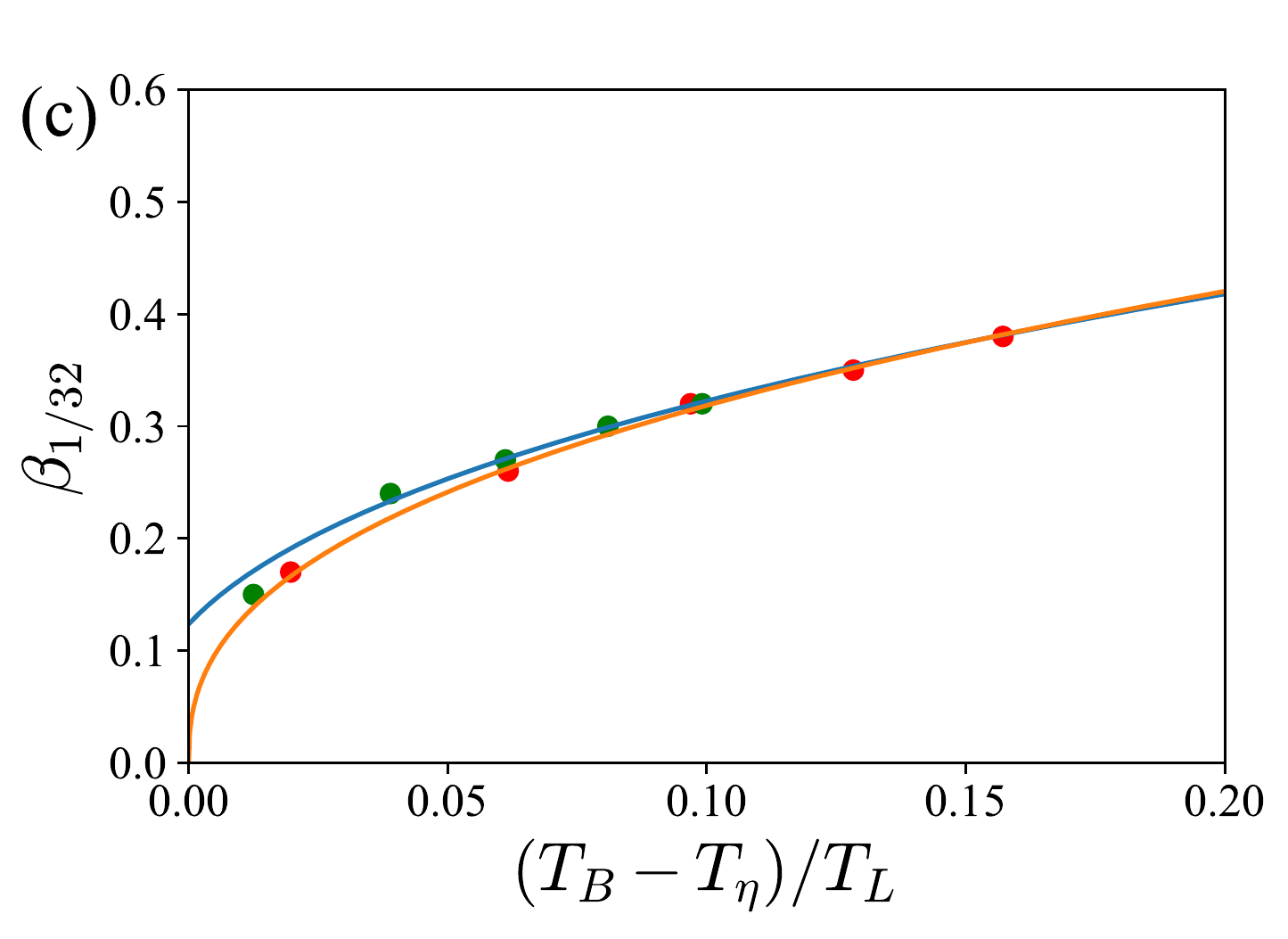}
  }
  \caption{
    Scaling exponents, $\beta_{1/n}$, evaluated from the compensated plots of Fig. \ref{fig:tau_n_large}
    for (a) $n=2$, (b) $n=8$, and (c) $n=32$ at $\mathrm{Re}_\alpha = 80$ (red) and $\mathrm{Re}_\alpha = 160$ (green).
    The inset of (a) shows the same plots in the outset, but the horizontal axis is
    changed to $T_B/T_L$ from $(T_B-T_\eta)/T_L$.
    The orange solid line shows $[T_B/T_L]^{0.4}$.
    The blue solid line shows $[T_B/T_L+0.01]^{0.4}$.
    For (b) and (c),
    the orange solid line shows $[(T_B-T_\eta)/T_L]^{0.4}$.
    The blue solid line shows $[(T_B-T_\eta)/T_L + 0.01]^{0.4}$.
  }
  \label{fig:beta_large}
\end{figure}

Now let us specify the empirical form of the exponent $\beta$ given in Eq.(\ref{eq:beta_scaling}).
Figure \ref{fig:beta_large} shows the measured values of $\beta_{1/n}$ with $n=2,8$ and $32$
as a function of $(T_B - T_\eta)/T_L$ that is the same independent variable used in Fig.\ref{fig:CL_large} (c).

For $n = 2$, the outset of Fig. \ref{fig:beta_large}(a) indicates
that $(T_B - T_\eta)/T_L$ is not appropriate since the data points are still scattered.
This leads us to search for a more suitable self-similar variable for $\beta_{1/2}$,
which is found to be $T_B/T_L$ as shown in the inset of Fig. \ref{fig:beta_large}(a).
This implies that $\beta_{1/2}$ depends only on $T_B$ and $T_L$, but not on $T_\eta$.
Empirically we now fit the collapsed curve obtained in the inset of Fig. \ref{fig:beta_large}(a)
with a function $\check{\beta}(T_B / T_L + a)^b$ with constants $\check{\beta}, a$, and $b$.
Our result is $\beta_{1/2} \propto [T_B/T_L + \omega_1]^{0.4}$,
where the constant $\omega_1$ takes zero or a non-zero small value possibly in a range,
$0 \leq \omega_1 \lesssim 0.01$. The fitted functions are shown in the inset of
Fig. \ref{fig:beta_large}(a).

On the other hand, for $n = 8$ and $32$, the exponents $\beta_{1/8}$ and $\beta_{1/{32}}$
are dependent on $T_B$, $T_\eta$, and $T_L$. Our best fit is
$\beta_{1/n} \propto [(T_B-T_\eta)/T_L + \omega_2]^{0.4}$ for $n=8$ and $32$,
as shown in Figs. \ref{fig:beta_large}(b) and (c).
Here, $\omega_2$ is a constant in a range $0 \leq \omega_2 \lesssim 0.01$.
The accurate values of $\omega_1$ and $\omega_2$ cannot be determined from the data shown
in Fig. \ref{fig:beta_large}. This is because the data are noisy and also the Reynolds numbers
are not sufficiently large for studying the behavior in $T_B/T_L \to 0$.
Nevertheless, it is obvious that the behavior of $\beta_{1/2}$ is different from the others.
On the other hand, for larger $n$'s such as $n=8$ or $32$,
the behaviors of $\beta_{1/n}$ are similar to each other.
Therefore, these results suggest that the exponent $\beta$ in the ansatz
has two different self-similar forms
depending on $\tau \lesssim T_B$ and $\tau \gg T_B$ at sufficiently large Reynolds numbers.
Specifically, we infer from the data
\begin{equation}
  \beta\left(\frac{T_\eta}{T_B}, \frac{T_L}{T_B}\right)
  =
  \begin{cases}
    \displaystyle  \left(\frac{T_B}{T_L} + \omega_1\right)^{0.4}  \equiv \beta_1        & \text{for} \quad \tau \lesssim T_B, \\
                                                                                        &                                     \\
    \displaystyle  \left(\frac{T_B-T_\eta}{T_L} + \omega_2 \right)^{0.4} \equiv \beta_2 & \text{for} \quad \tau \gg T_B,
  \end{cases}
  \label{eq:beta_large}
\end{equation}
where $\omega_1$ and $\omega_2$ are $\mathrm{Re}_\alpha$ independent constants, which
may be zero.
Accordingly, the function $g^{L}$ in the ansatz (\ref{eq:CL_scaling_law}) can be given by
\begin{equation}
  g^L\left(\frac{\tau}{T_B^{\beta} T^{1 - \beta} }\right)
  =
  \begin{cases}
    \displaystyle	g^L_1\left(\frac{\tau}{T_B^{\beta_1}T^{1-\beta_1}}\right) & \text{for} \quad \tau \lesssim T_B, \\
                                                                            &                                     \\
    \displaystyle	g^L_2\left(\frac{\tau}{T_B^{\beta_2}T^{1-\beta_2}}\right) & \text{for} \quad \tau \gg T_B,
  \end{cases}
  \label{eq:gL_scaling}
\end{equation}
where $g^L_1$ and $g^L_2$ are self-similar functions.

Now we discuss the limit of $\beta$ as $T_\eta \to 0$ and $T_L \to \infty$.
Here, $\omega_1^{0.4}$ in Eq.(\ref{eq:beta_large}) is the limit of $\beta_1$
as $T_B/T_L \to 0$. Similarly, $\omega_2^{0.4}$ is the limit of $\beta_2$ as $(T_B-T_\eta)/T_L \to 0$.
Let us suppose $\omega_2 = 0$. Then
the K41 scaling law is recovered at $\tau \gg T_B$ at sufficiently large Reynolds numbers.
It is impossible to determine the accurate value of $\omega_2$ from Fig. \ref{fig:beta_large}.
It appears that $\omega_2 = 0.01$ is the best fitted value judging from Fig. \ref{fig:beta_large}(c)
though $\omega_2=0$ is not ruled out.
Both values $\omega_1 = 0.01$ and $\omega_1 = 0$ seem equally good as in the case for $\omega_2$.
In order to determine the accurate values of $\omega_1$ and $\omega_2$,
we need to perform DNSs at much larger Reynolds number and with much larger number of
the particle pairs.

\begin{figure}[htbp]
  \centering{
    \includegraphics[clip,scale=0.5]{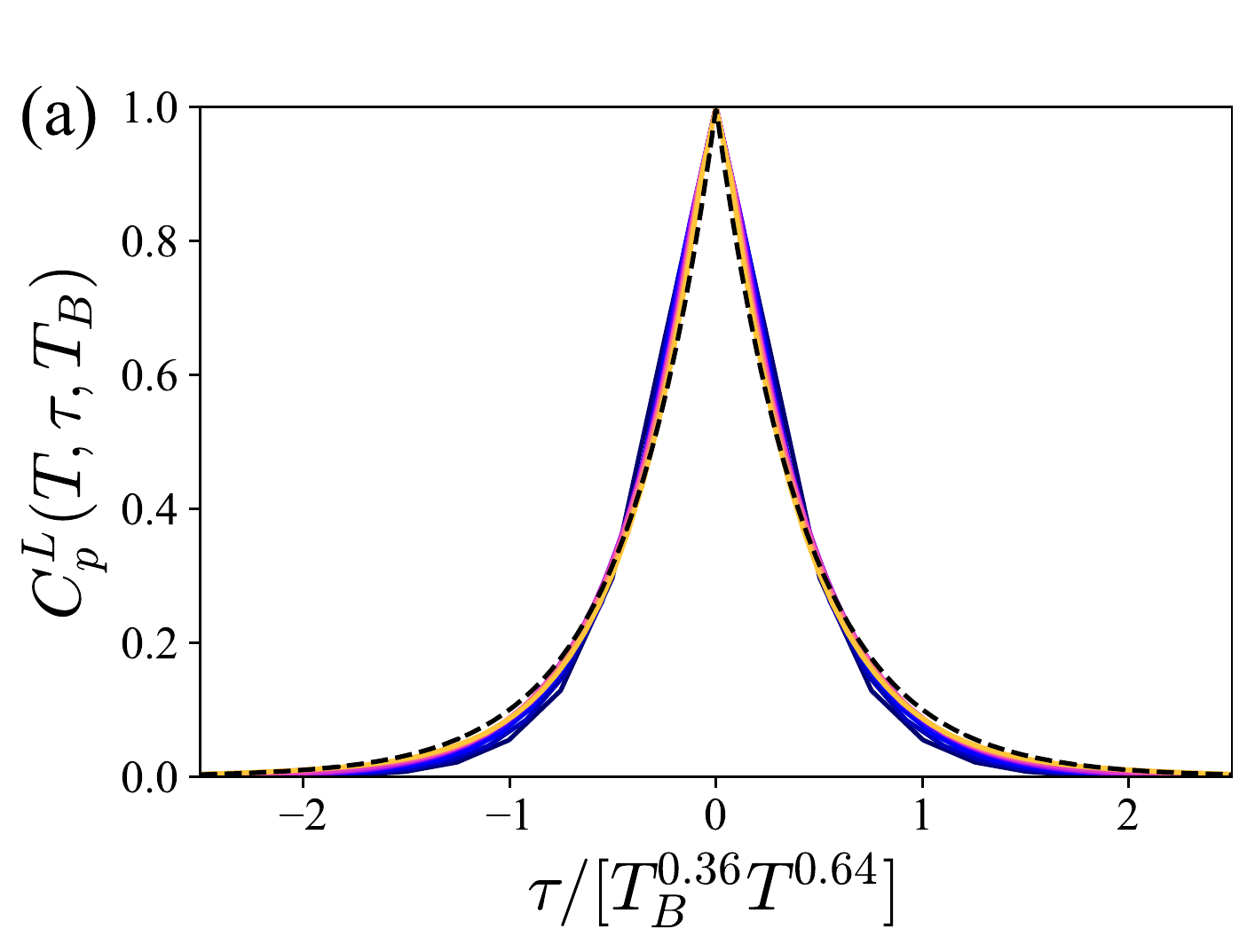}
    \includegraphics[clip,scale=0.5]{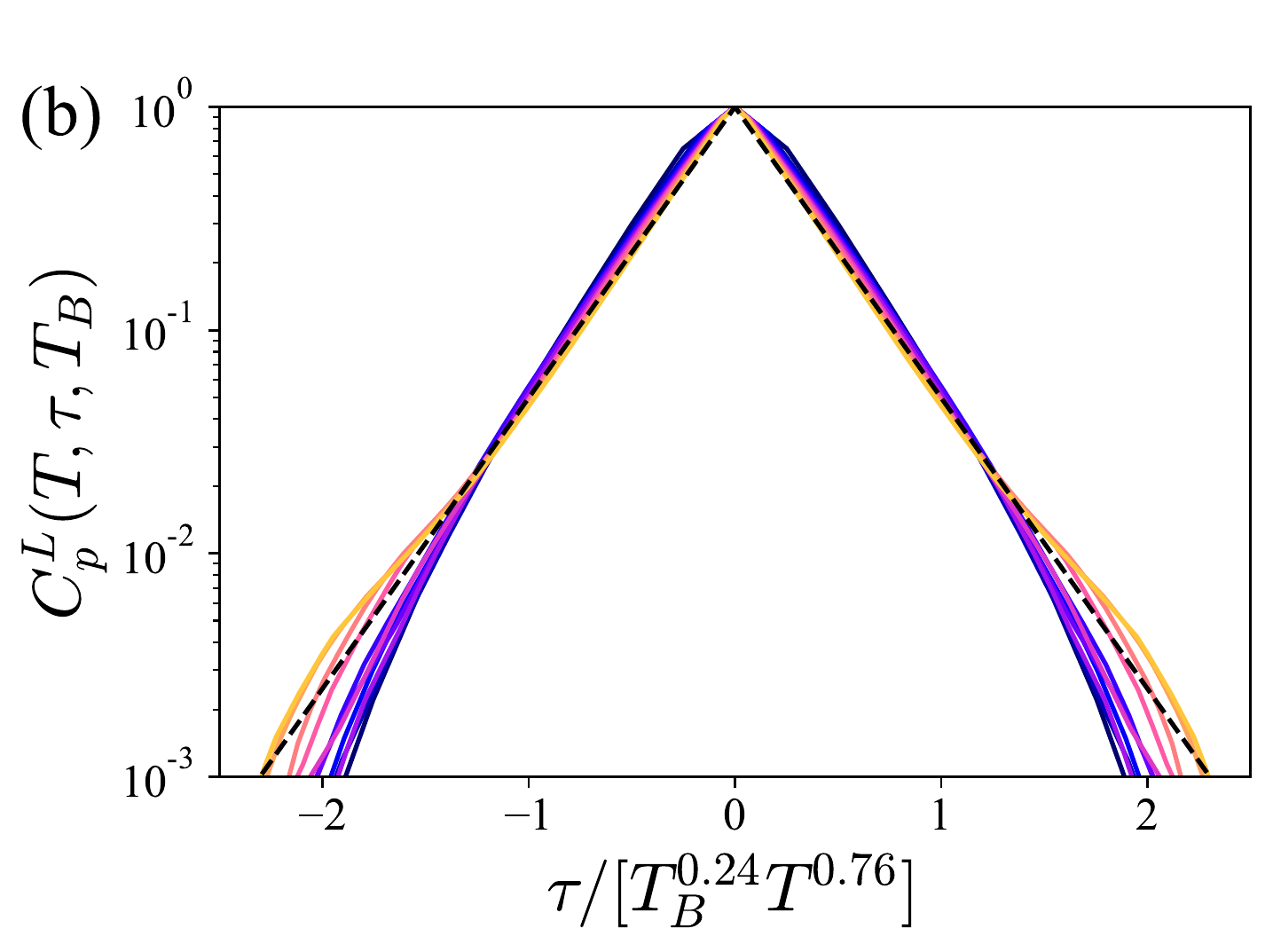}
    \includegraphics[clip,scale=0.5]{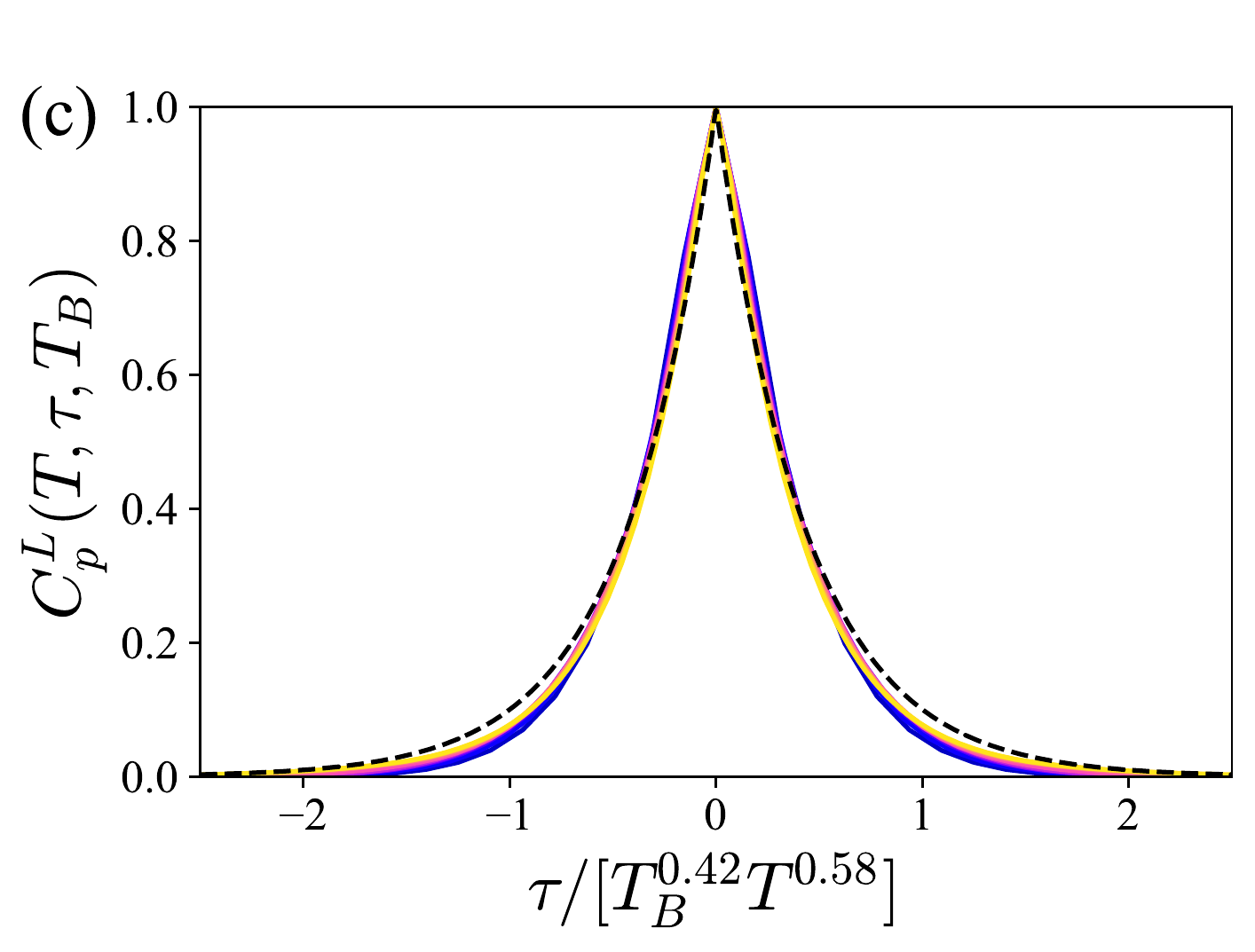}
    \includegraphics[clip,scale=0.5]{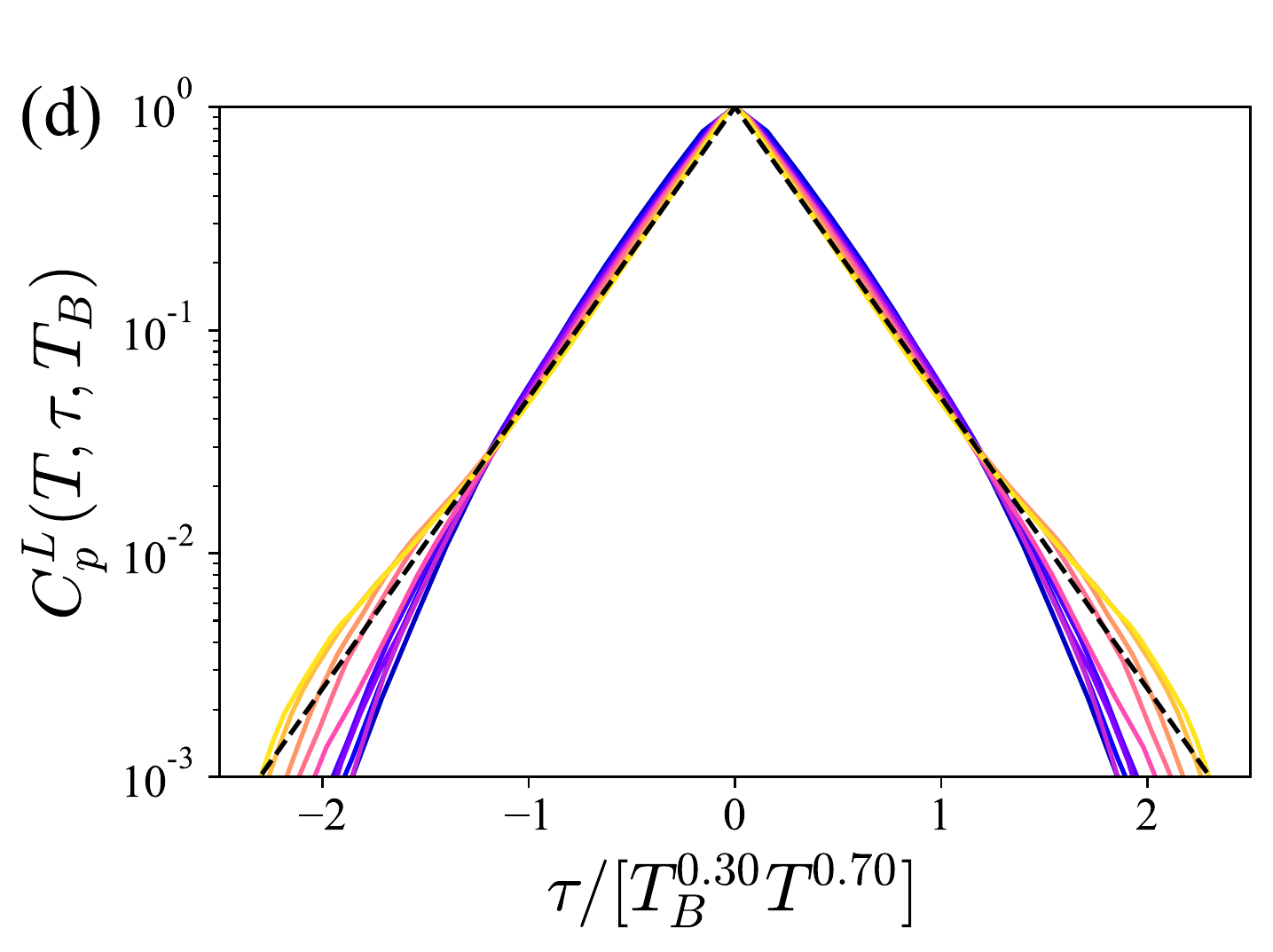}
  }
  \caption{
  Normalized correlation function, $C_p^L(T,\tau, T_B)$ for $T_B=2.2T_\eta$ rescaled as a function of
  (a) $\tau/[T_B^{\beta_{1/2}}~T^{1-\beta_{1/2}}]$ and (b) $\tau/[T_B^{\beta_{1/32}}~T^{1-\beta_{1/{32}}}]$,
  where $\beta_{1/2} = 0.64$ and $\beta_{1/{32}} = 0.76$, for $T_B < T < 0.54 T_L$ at $\mathrm{Re}_\alpha = 160$.
  The colors of the curves change from black to yellow as the average time $T$ increases, which is similar to Fig.\ref{fig:gL_large}.
  Black dashed lines show
  (a) $\exp(-2.3 |\tau|/[T_B^{\beta_{1/2}}~T^{1-\beta_{1/2}}])$ and
  (b) $\exp(-3.0 |\tau|/[T_B^{\beta_{1/32}}~T^{1-\beta_{1/32}}])$.
  (c) Same as (a) but for $T_B = 3.5 T_\eta$, where $\beta_{1/2} = 0.42$.
  (d) Same as (b) but for $T_B = 3.5 T_\eta$, where $\beta_{1/32} = 0.30$.
  }
  \label{fig:gL_norm_large}
\end{figure}

In Eq.(\ref{eq:gL_scaling}), the two different behaviors of $g^L$ are inferred
from those of $\beta$. We now demonstrate that the two forms are consistent with the DNS data.
Figure \ref{fig:gL_norm_large} shows $C_p^L(T,\tau, T_B)$ as a function of $\tau/[T^{1 - \beta_{1/n}} T_B^{\beta_{1/n}}]$
for $n=2$ in Fig.\ref{fig:gL_norm_large}(a) and $n=32$ in Fig.\ref{fig:gL_norm_large}(b).
Figure \ref{fig:gL_norm_large}(a) for $n=2$ is plotted in lin-lin coordinates,
which means that we mainly observe the regions where $C_p^L(T,\tau, T_B)$ is large.
On the other hand, Fig. \ref{fig:gL_norm_large}(b) for $n=32$ is plotted in lin-log coordinates,
which means that we mainly observe the regions where $C_p^L(T,\tau, T_B)$ is small.
The master curve in Fig.\ref{fig:gL_norm_large}(a) corresponds to $g_1^{L}$
and the one in Fig.\ref{fig:gL_norm_large}(b) corresponds to $g_2^{L}$ in Eq.(\ref{eq:gL_scaling}).
Here we assume $\beta_1 = \beta_{1/2}$ and $\beta_2 = \beta_{1/32}$.
Compare the collapsed curves in Fig.\ref{fig:gL_norm_large} to those shown in Fig.\ref{fig:gL_large}
without taking any appropriate similarity variable.
Furthermore, let us assume that the rescaled functions are exponential, namely
$g_1^L(X) = \exp(-k_1 X)$ and $g_2^L(X) = \exp(-k_2 X)$. This assumption is consistent
with Fig.\ref{fig:gL_norm_large}.
We can estimate the constants as $k_1 \sim 2.3$ and $k_2 \sim 3.0$ from Fig.\ref{fig:gL_norm_large},
though these values are also slightly dependent on $T_\eta, T_L$, and $T_B$.
The exponential forms will be used to estimate the Richardson constant in Sec.\ref{sec:richardson}.

\subsection{Small $T_B$ condition: $\bm{T_B < T_\eta}$}
Now we consider the scaling laws of $C^L(r_0, T, \tau, \varepsilon)$ under
the small initial separation condition, $T_B < T_\eta$, where
particle pairs may be strongly influenced by small-scale effects caused
by the viscosity and the forcing.
On the other hand, the condition $T_B \ll T_L$ is met more easily than
in the previous large $T_B$ condition.
Hence, we expect that $C^L(r_0, T,\tau, \varepsilon)$ is independent of
large-scale effects such as the drag.
Moreover, the $t^3$ scaling law for $\langle r^2(t) \rangle$ has been observed
under this condition in many previous studies for both 2D and 3D
as mentioned in Sec.\ref{sec:introduction}.
We also investigate the reason why the $t^3$ scaling is observed
even at moderate Reynolds numbers only for a tuned initial separation $r_0$.

\begin{figure}[htbp]
  \centering{
    \includegraphics[clip,scale=0.4]{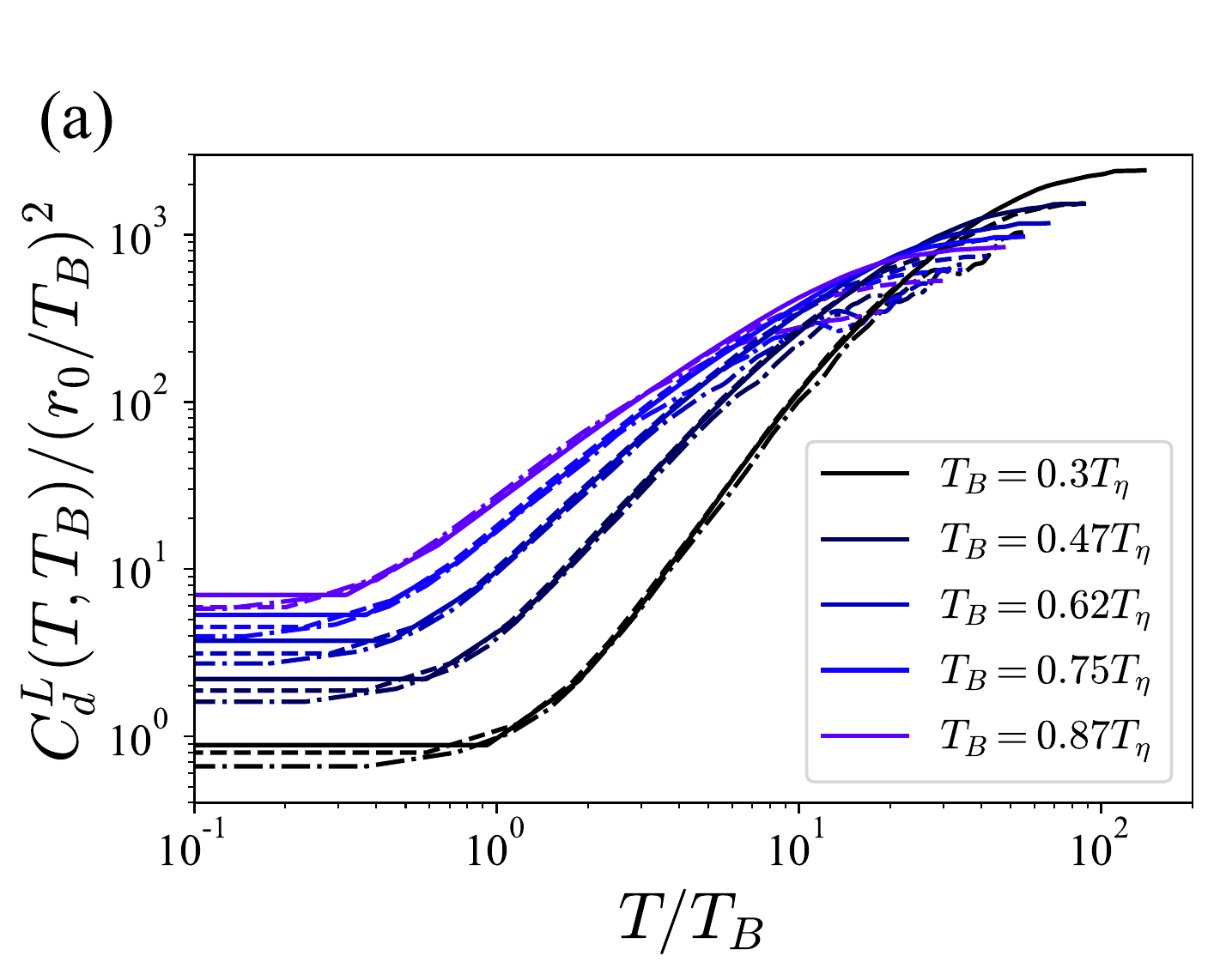}
    \includegraphics[clip,scale=0.4]{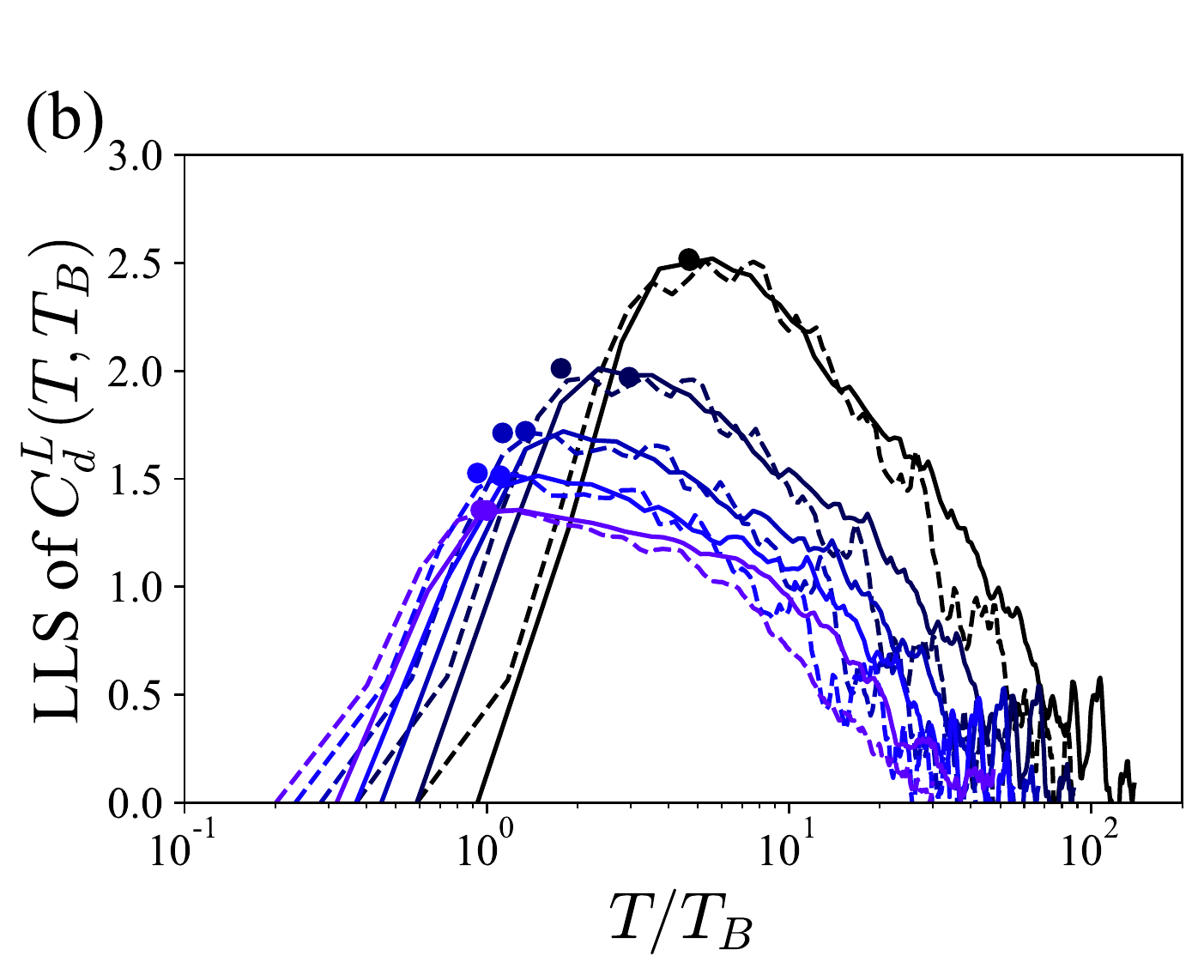}
    \includegraphics[clip,scale=0.4]{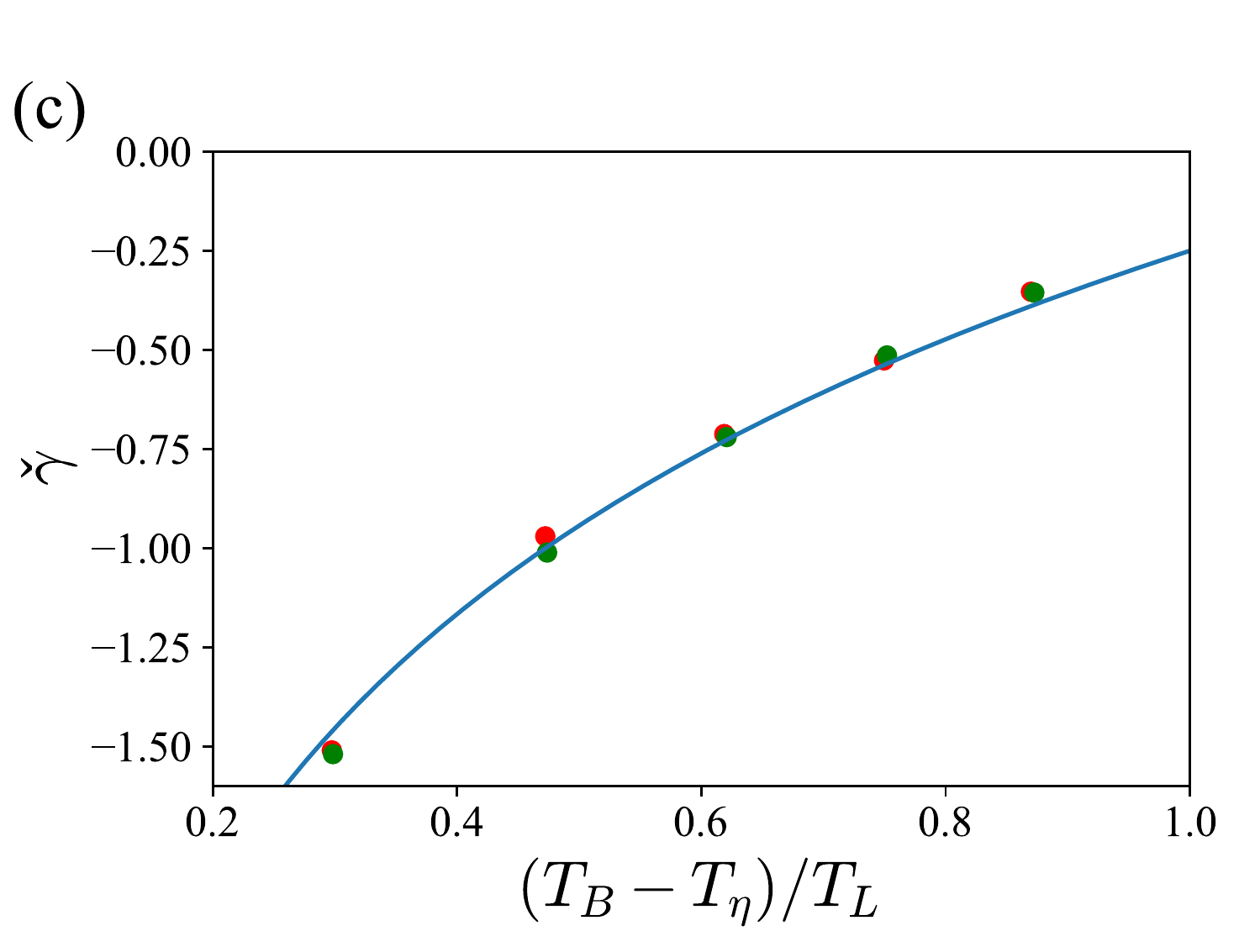}
  }
  \caption{
    (a) Time evolution of $C_d^L(T, T_B)$ for various $T_B$'s at $\mathrm{Re}_\alpha = 40$ (dashed dotted), $80$ (dashed), and $160$ (solid).
    (b) LLSs of $C_d^L(T, T_B)$ at $\mathrm{Re}_\alpha = 40$ (dashed) and $160$ (solid).
    Filled circles on the curves show positions of their maximum value.
    (c) Values of the exponent $\check{\gamma}$ as a function of $T_B/T_\eta$, which are measured by the maximum value of the LLSs at $\mathrm{Re}_\alpha = 40$ (red) and $160$ (green).
    The blue solid line shows $\ln (T_B/T_\eta) - 0.25$.
  }
  \label{fig:CL_small}
\end{figure}

In this subsection, we repeat what we have done in the previous subsection.
Therefore, we describe only the diffences.
In terms of the correlation along the diagonal line,
the exponent $\gamma$ is determend as we did in Fig. \ref{fig:CL_large}(b),
see Fig. \ref{fig:CL_small}(a) and (b).
In what follows we write the exponents with check in  the small $T_B$ condition.
In Fig. \ref{fig:CL_small}(c), we show the measured $\check{\gamma}$ as a function of $T_B/T_\eta$.
This choice of the variable yields a curve
independent of $\mathrm{Re}_\alpha$, which we fit with
\begin{equation}
  \check{\gamma}(T_B, T_\eta) = \ln\left(\frac{T_B}{T_\eta}\right) - \upsilon.
  \label{gamma_small}
\end{equation}
Here the constant $\upsilon$ is $0.25 \pm 0.02$ which is determined by a least-square method.
Therefore, at $T_B < T_\eta$, the scaling law of $C_d^L(T, T_B)$, can be,
\begin{equation}
  C_d^L(T, T_B)
  = G \varepsilon T \left(\frac{T_B}{T}\right)^{\check{\gamma}}
  = G  T^{1-\ln\left(\frac{T_B}{T_\eta}\right) + \upsilon}.
\end{equation}

\begin{figure}[htbp]
  \centering{
    \includegraphics[clip,scale=0.4]{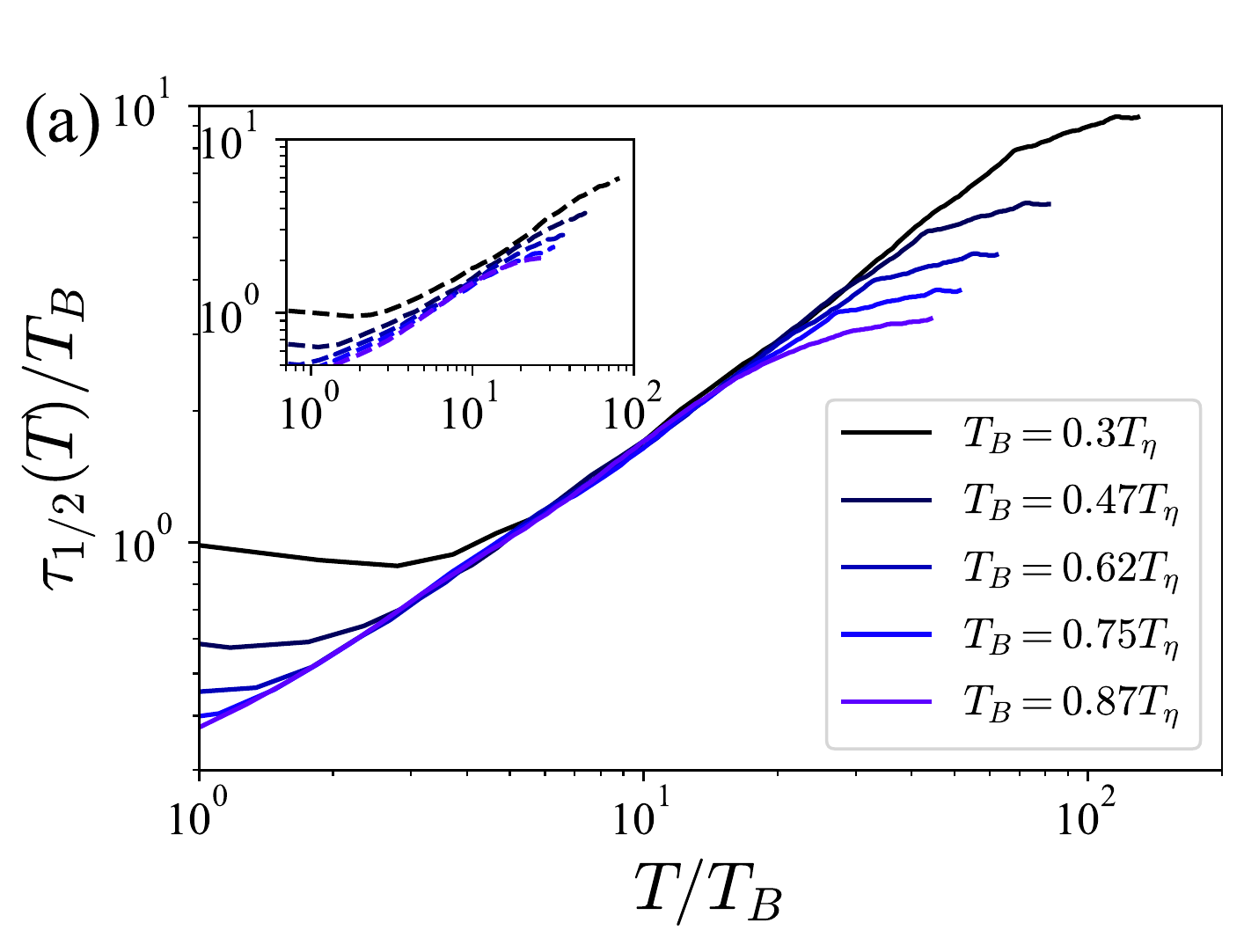}
    \includegraphics[clip,scale=0.4]{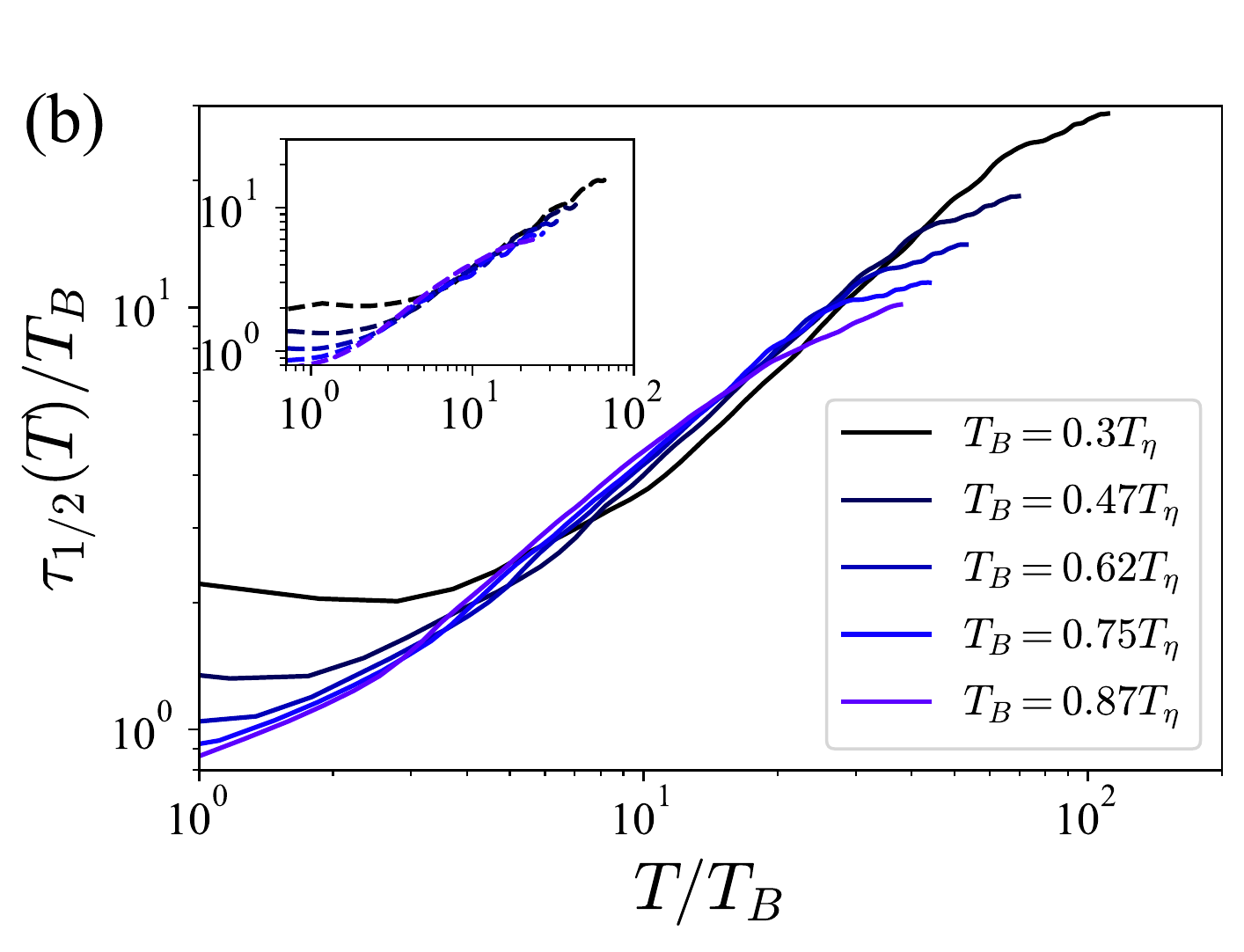}
    \includegraphics[clip,scale=0.4]{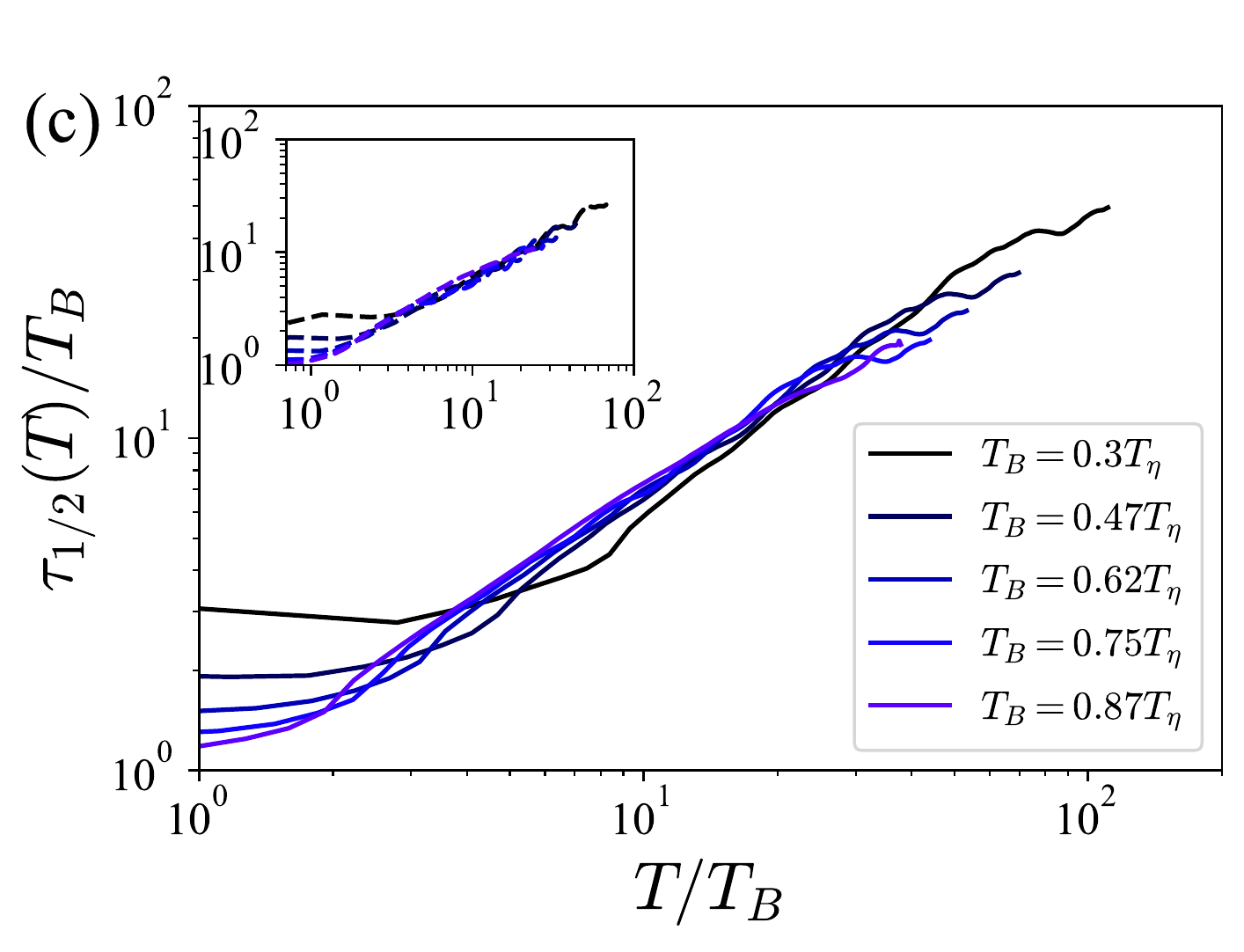}
  }
  \caption{
    Time evolution of the $n$-th decay time scale, $\tau_{1/n} (T)$
    for (a) $n=2$, (b) $n=8$, (c) $n=32$
    at $\mathrm{Re}_\alpha=160$.
    The insets show the same plots as the outsets but at $\mathrm{Re}_\alpha=80$.
  }
  \label{fig:tau_n_small}
\end{figure}

\begin{figure}[htbp]
  \centering{
    \includegraphics[clip,scale=0.4]{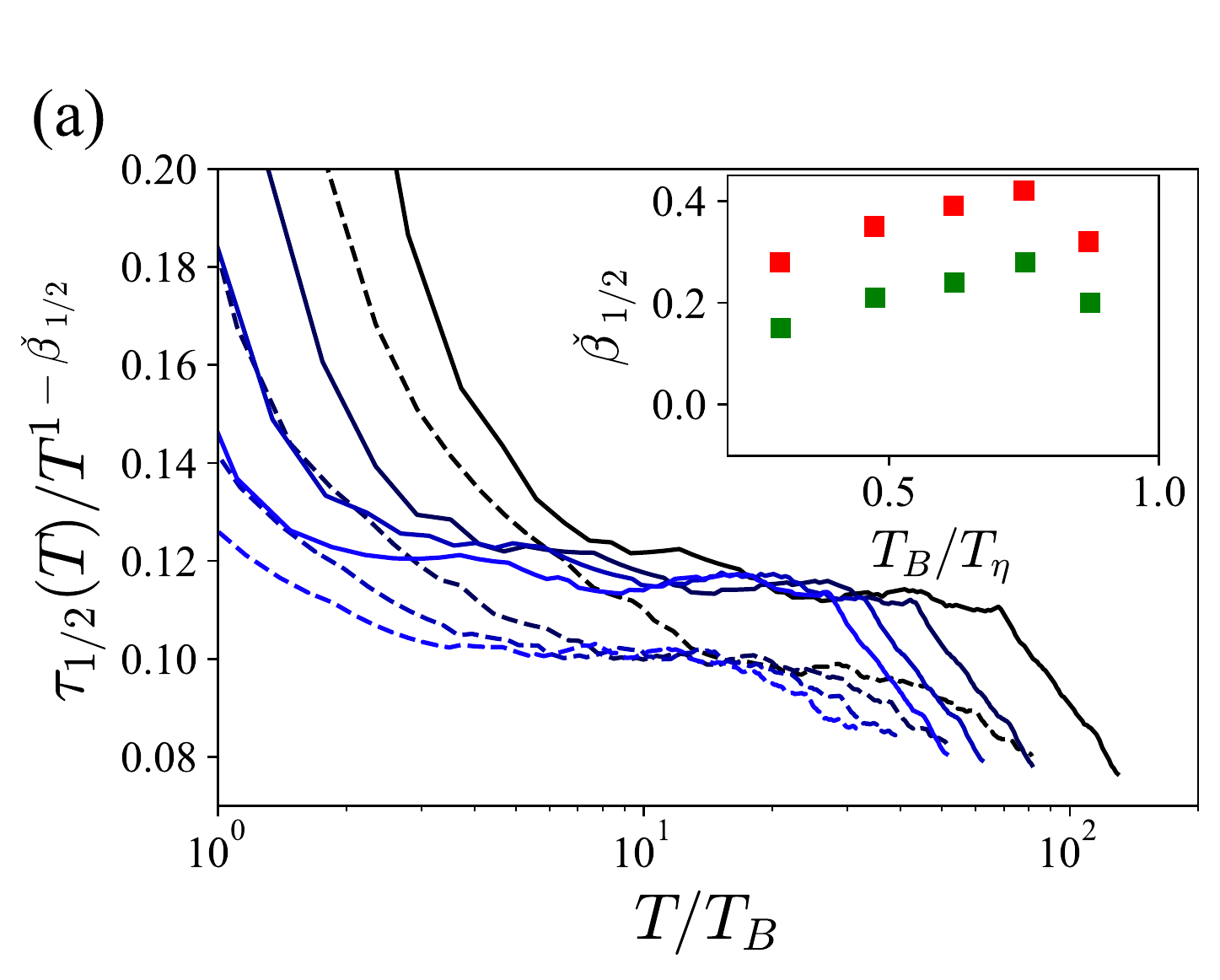}
    \includegraphics[clip,scale=0.4]{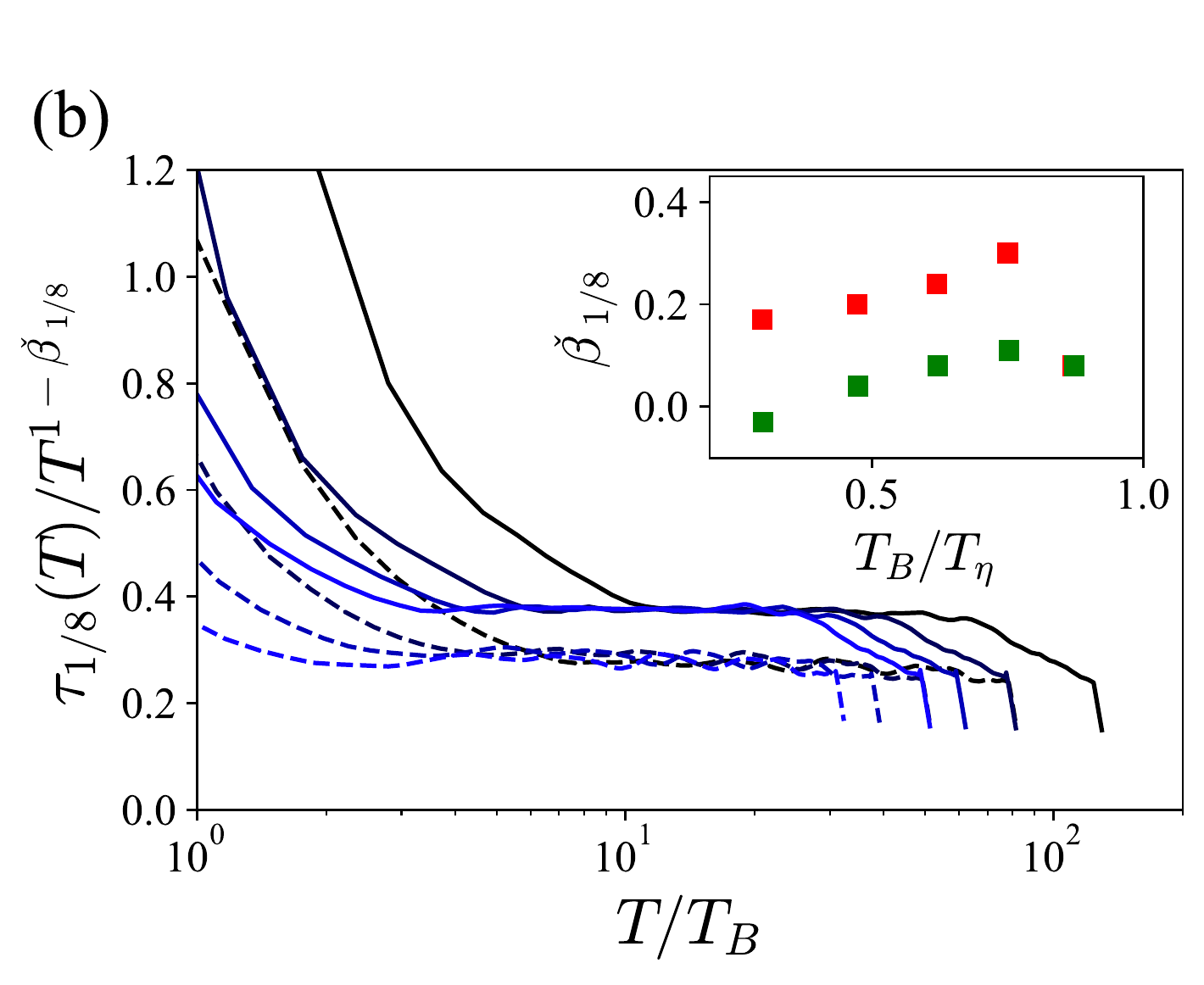}
    \includegraphics[clip,scale=0.4]{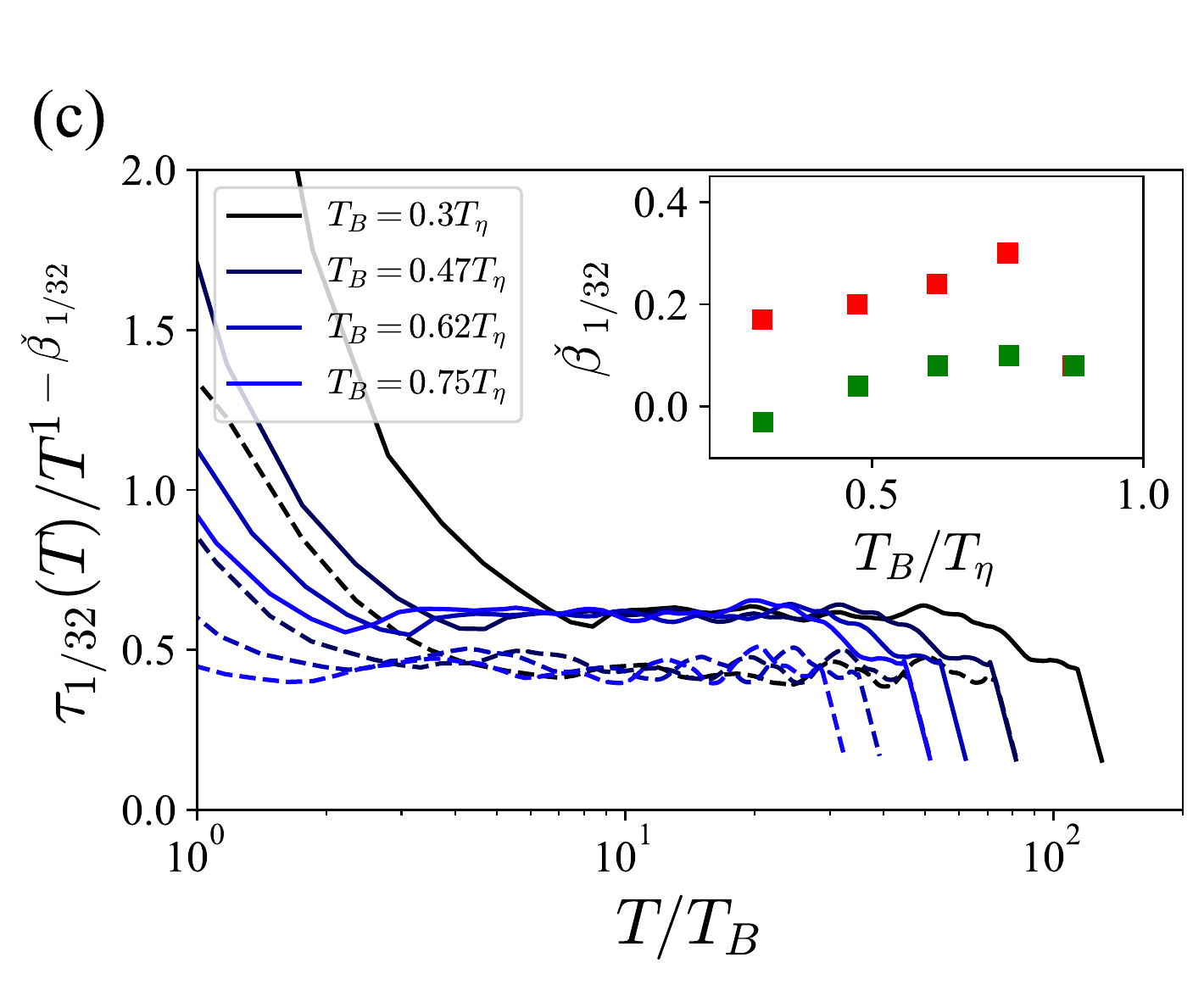}
  }
  \caption{
    Compensated graphs of $\tau_{1/n}(T)$ by $\check{\beta}_{1/n}$ for (a) $n=2$, (b) $n=8$, and (c) $n=32$
    at $\mathrm{Re}_\alpha = 80$ (dashed) and $\mathrm{Re}_\alpha = 160$ (solid).
    The Insets show $\beta_{1/n}$ as a function of $T_B/T_\eta$ at $\mathrm{Re}_\alpha = 80$ (red) and $\mathrm{Re}_\alpha = 160$ (green).
  }
  \label{fig:beta_small}
\end{figure}

Next, we focus on the other part of the correlation $C_p^L(T,\tau, T_B)$ defined in Eq.(\ref{defcd})
and the scaling exponent, $\check{\beta}$.
Figure \ref{fig:tau_n_small} shows $n$-th decay time, $\tau_{1/n}(T)$ and
Fig. \ref{fig:beta_small} shows its compensated plots by $T^{1-\check{\beta}_{1/n}}$.
Here, the exponent $\check{\beta}_{1/n}$ is selected in the same way as in the previous large $T_B$ case.
It should be noted that what we show in Fig.\ref{fig:beta_small} is not $D_{n,\check{\beta}_{1/n}}(T)$ defined in Eq.(\ref{defD}),
but $\tau_{1/n}(T)/T^{1-\check{\beta}_{1/n}}$.  
The plotted variable $\tau_{1/n}(T)/T^{1-\check{\beta}_{1/n}}$  is less
dependent of $T_B$  than $D_{n,\check{\beta}_{1/n}}(T)$,
which is consistent with the $T_B$ independent behavior of the $n$-th decay time
shown in Fig.\ref{fig:tau_n_small}.

Although the compensated data shown in Fig.\ref{fig:beta_small} oscillate getting
stronger for large $n$,
we observe a plateau region for each graph.
It is noticeable that the values of the plateaus depend on $Re_\alpha$.
This tendency is not present in the large $T_B$ case as shown
in Fig.\ref{fig:tau_n_large_compe} (ignoring data for the smallest $T_B$).
If the self-similarity given in Eq.(\ref{eq:CL_scaling_law})
is valid for $C_p^{L}(T, \tau, T_B)$, the values of the plateaus should
become independent of $Re_\alpha$. Therefore, Fig.\ref{fig:beta_small} suggests
two possibilities: one is that $C_p^{L}(T, \tau, T_B)$ is not self-similar;
the other is that $C_p^{L}(T, \tau, T_B)$ is self-similar but with yet another
time scale, $T_X$.

The second possibility is more likely, although numerical evidence
is marginally convincing as we will see.
With the hypothetical time scale $T_X$,
a similarity variable for $C_p^{L}(T, \tau, T_B)$
can be made as $\tau/[T_X^{\check\beta} T^{1-\check{\beta}}]$.
Therefore, the similarity function $g^L$ under the small $T_B$ conditions is likely the similar form
to Eq. (\ref{eq:gL_scaling}).
The difference is that we just replace $T_B$ in the similarity variable by $T_X$.
Moreover, as we discussed with Fig.\ref{fig:beta_small},
the levels of the plateaus of the vertical axis, which we denote $E_{n,\check{\beta}_{1/n}}(T)$
are independent of $T_B$.
Regarding change in the numerical values of the plateau levels as we vary $n$,
we observe
$E_{2, \check{\beta}_{1/2}} \simeq (1/3) E_{8, \check{\beta}_{1/8}} \simeq (1/5) E_{32, \check{\beta}_{1/32}}$,
yielding $\tau_{1/n}(T) =   \tau_{1/2}(T) \log_2 n$.
This implies that  $C_p^L(T, \tau, T_B)$
for $T_B < T_\eta$ decays exponentially in all the range of $\tau$.
Therefore, $g^L$ under the small $T_B$ condition is likely to have two self-similar forms such as
\begin{equation}
  g^L\left(\frac{\tau}{T_X^\beta T^{1 - \beta}} \right) =
  \begin{cases}
    \displaystyle \exp\left[{- \check{k}_1 \left(\frac{\tau}{T_X^{\check{\beta}_1} T^{1-\check{\beta}_1}}\right)}\right]  & \text{for} \quad \tau \lesssim T_B, \\
                                                                                                                          &                                     \\
    \displaystyle  \exp\left[{- \check{k}_2 \left(\frac{\tau}{T_X^{\check{\beta}_2} T^{1-\check{\beta}_2}}\right)}\right] & \text{for} \quad \tau > T_B,
  \end{cases}
  \label{eq:gL_small}
\end{equation}
where $\check{\beta}_1$ and $\check{\beta}_2$ are exponents for the two self-similar regimes of $\tau$
and $\check{k}_1$ and $\check{k}_2$ are constants.
We consider that the exponent $\check{\beta}_1$ for small $\tau$ is represented by $\beta_{1/2}$ and $\check{\beta}_2$ for large $\tau$ is by $\beta_{1/8} \simeq \beta_{1/32}$.
Equation (\ref{eq:gL_small}) is analogous to Eq.(\ref{eq:gL_scaling}) under the large $T_B$ condition.
We cannot find a simple form of $\check{\beta}_{1/n}$ unlike the large $T_B$ case.

\begin{figure}[htbp]
  \centering{
    \includegraphics[clip,scale=0.5]{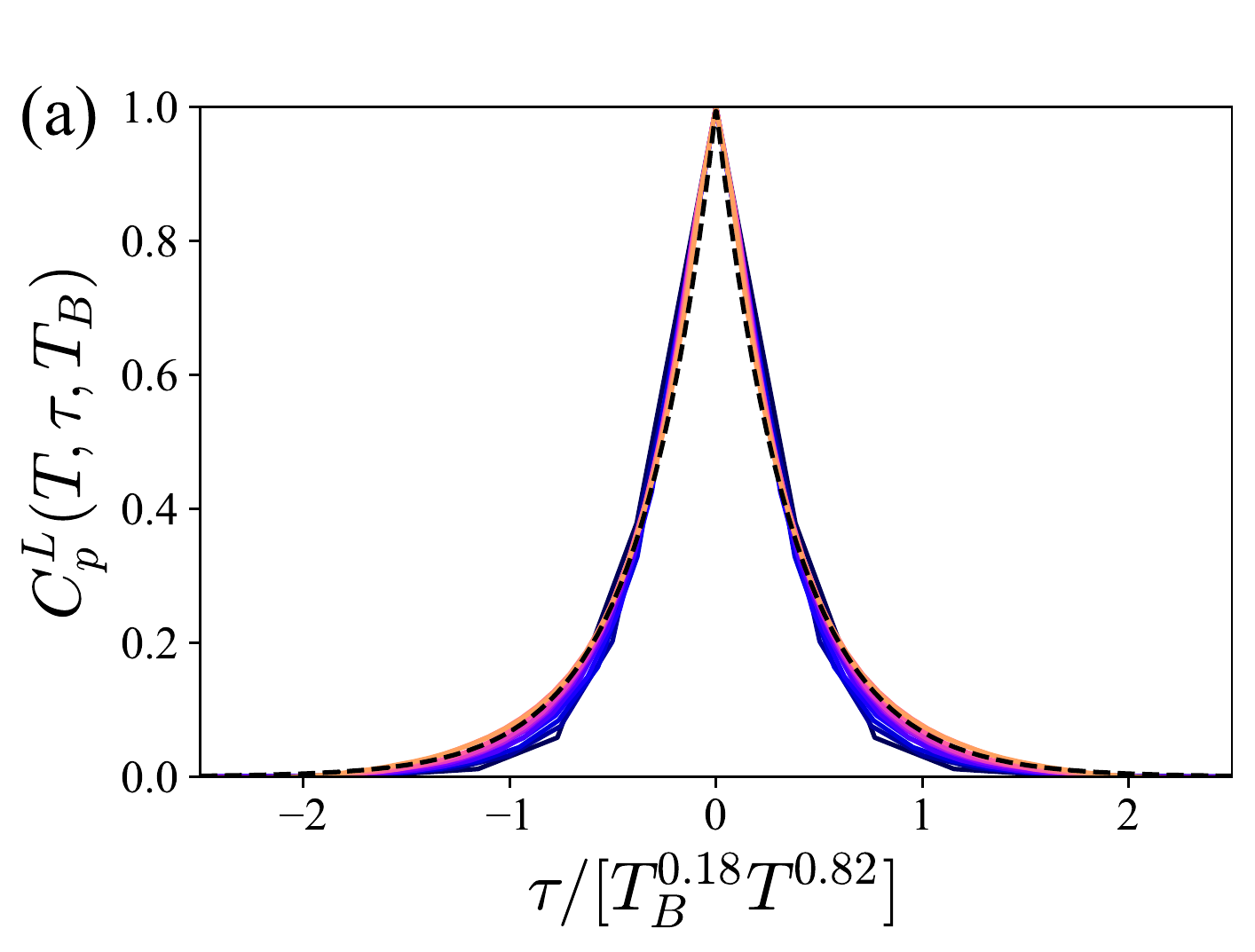}
    \includegraphics[clip,scale=0.5]{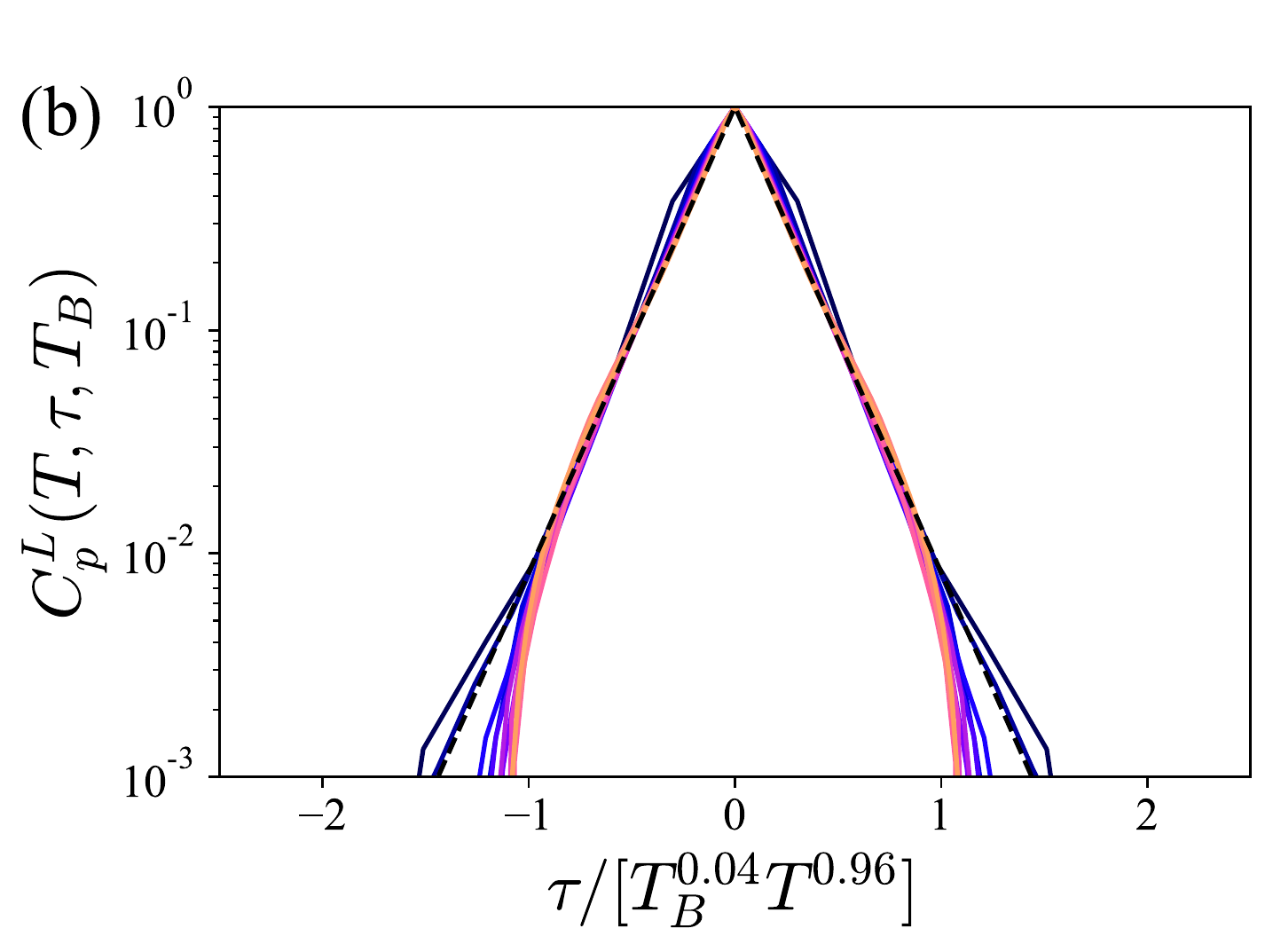}
  }
  \caption{
  Normalized correlation function, $C_p^L(T,\tau, \varepsilon)$, with $T_B=0.47T_\eta$
  as a function of rescaled variable (a) $\tau/[T_B^{1-\check{\beta}_{1/2}}~T^{\check{\beta}_{1/2}}]$
  and (b) $\tau/[T_B^{1-\check{\beta}_{1/32}}~T^{\check{\beta}_{1/{32}}}]$ under the small $T_B$ condition.
  Here the exponents are $\check{\beta}_{1/2} = 0.82$ and $\check{\beta}_{1/{32}} = 0.96$.
  Different curves have different $T$s in $T_B < T < 0.54 T_L$
  at $\mathrm{Re}_\alpha = 160$.
  The black dashed lines correspond to
  (a) $\exp(-2.7|\tau|/[T_B^{\check{\beta}_{1/2}}~T^{1-\beta_{1/2}}])$ and
  (b) $\exp(-4.8|\tau|/[T_B^{\check{\beta}_{1/32}}~T^{1-\beta_{1/32}}])$.
  }
  \label{fig:gL_norm_small}
\end{figure}

We show now that the two different scaling behaviors
given in Eq.(\ref{eq:gL_small}) are consistent to the DNS data.
Figure \ref{fig:gL_norm_small} shows $C_p^L(T,\tau, \varepsilon)$
as a functions of $\tau / T^{\check{\beta}_{1/n}}$ for $n=2$ and $n=32$.
More precisely, since we do not know $T_X$,
we use $T_B$ to non-dimensionalize $\tau / T^{\check{\beta}_{1/n}}$
in the horizontal axis of Fig.\ref{fig:gL_norm_small}.
All the curves in each panel collapse to one curve with a suitable rescaling of $\tau$.
The similarity function $g^L$ can be fitted with an exponential curve as shown in Fig. \ref{fig:gL_norm_small}.

Let us summarize the results in this section.
We have considered numerically scaling behavior of $C^L(r_0, T,\tau, \varepsilon)$
in comparison with the ansatz (\ref{eq:CL_scaling_law})
under two conditions, $T_B > T_\eta$ and $T_B < T_\eta$.
The DNS data of $C^L(r_0, T,\tau, \varepsilon)$ are consistent to
the ansatz for both conditions.
The difference between the two conditions is in the functional forms of
the exponents,
$\gamma\left(\frac{T_\eta}{T_B},\frac{T_L}{T_B}\right)$ (Eqs.(\ref{eq:gamma_large}) and (\ref{gamma_small})),
and
$\beta\left(\frac{T_\eta}{T_B},\frac{T_L}{T_B}\right)$ (Eq.(\ref{eq:beta_large})),
although $\beta$ for the latter case was not identified.
Furthermore, $\gamma$ and $\beta$ are probably continuous at $T_B = T_\eta$.
In particular, under the large $T_B$ condition, our DNS data indicate
that
$\gamma$ and $\beta$ do not approach zero as $T_B \to T_\eta$.
If this is not a finite Reynolds number effect, the non-zero limits
of the exponents imply that
the TTLVCF has exponents that deviate from the K41 dimensional analysis.
Consequently, the squared relative separation $\langle r^2(t) \rangle$
disagrees with the Richardson--Obukhov law $t^3$, even if we take $r_0 \to 0$
at infinite Reynolds number. We will discuss this point in the next section.

\section{Implications on the Richardson--Obukhov law \label{sec:richardson}}
Finally, we consider implications of the above scaling behaviors
of $C^L(r_0, T,\tau, \varepsilon)$ on the relative separations of particle pairs.
The second moment of the relative separation $r(t)$
can be reduced to
\begin{eqnarray}
  \langle r^2(t) \rangle
  &=& r_0^2
  + 2 \int_0^t \bm{r_0} \cdot \langle \delta \bm{v}(t_1) \rangle ~dt_1
  + \int_0^t \int_0^t \langle \delta \bm{v}(t_1) \delta \bm{v}(t_2) \rangle ~dt_1 dt_2 \nonumber \\
  &\sim&
  r_0^2 + 2\int_0^{t/2} ~dT \int_{0}^{2T}~ d\tau  ~ C^L(r_0, T,\tau,\varepsilon)
  + 2\int_{t/2}^t ~dT \int_{0}^{2(t-T)}~ d\tau  ~ C^L(r_0,T,\tau,\varepsilon)
  \label{eq:r2}
\end{eqnarray}
where the average time is $T = (t_1 + t_2) / 2$ and the relative time is $\tau = t_2 - t_1$.
We also assume  $\bm{r}_0 \cdot \langle \delta \bm{v}(t) \rangle = 0$ by taking
the direction of the initial separation $\bm{r}_0$ being randomly and isotropically distributed,
and use the symmetry of $C^L(r_0, T,\tau, \varepsilon)$ with respect to the diagonal line $t_1 = t_2$.

First, we consider the scaling law of $\langle r^2(t) \rangle$ under the large $T_B$ condition,
$T_\eta \ll T_B \ll T_L$, at sufficiently large Reynolds numbers.
Under this condition, $C^L(r_0, T,\tau, \varepsilon)$ has the self-similar forms (\ref{eq:CL_scaling_law}) and (\ref{Phig}).
The exponent $\gamma$ is given by Eq.(\ref{eq:gamma_large}).
The self-similar function $g^{L}$ and the other exponent $\beta$
take two different forms as given in Eqs.(\ref{eq:beta_large})--(\ref{eq:gL_scaling}),
depending on $\tau \lesssim T_B$ or $\tau \gg T_B$.
More precisely, taking into consideration the self-similarity,
$g^L$ may take the following forms,
\begin{equation}
  g^L\left(\frac{\tau}{T_B^{\beta} T^{1 - \beta} }\right)
  =
  \begin{cases}
    \displaystyle	g^L_1\left(\frac{\tau}{T_B^{\beta_1}T^{1-\beta_1}}\right) & \text{for} \quad \tau \leq c_1 T_B\left(\frac{T}{T_B}\right)^{1-\beta_1}, \\
                                                                            &                                                                           \\
    \displaystyle	g^L_2\left(\frac{\tau}{T_B^{\beta_2}T^{1-\beta_2}}\right) & \text{for} \quad \tau \geq c_2 T_B\left(\frac{T}{T_B}\right)^{1-\beta_2},
  \end{cases}
  \label{eq:gL_scaling_mod}
\end{equation}
where $c_1$ and $c_2$ are constants which determine the time to switch from $g^L_1$ to
$g^L_2$.
These constants are dependent on Reynolds number.
The transition time is around $T_B$, that is,
$c_1T_B(T/T_B)^{1-\beta_1} \sim T_B, ~ c_2 T_B(T/T_B)^{1-\beta_2} \sim T_B$ for $T_B \ll T \ll T_L$.
From this, we can estimate the values of $c_1$ and $c_2$:
\begin{equation}
  c_1 \sim \left(\frac{T_B}{T_L}\right)^{1-\beta_1}, \quad
  c_2 \sim \left(\frac{T_B}{T_L}\right)^{1-\beta_2}.
\end{equation}
Therefore, $c_1$ and $c_2$ are very small constants
at sufficiently large Reynolds numbers.
Now, we substitute the self-similar forms (\ref{eq:gL_scaling_mod}) and
calculate the integrals under the condition $t \gg T_B$, then we obtain,
\begin{align}
  \langle r^2(t) \rangle
  \sim & ~
  \left(\int_0^{c_1}g_1^L(x) dx\right) t^{3-\gamma-\beta_1}
  + \left(\int_{c_2}^\infty g_2^L(x) dx\right) t^{3-\gamma-\beta_2} \label{leading}                  \\
       & + \int_{T_B}^{t/2} dT~ T^{2-\gamma-\beta_2} \int_{\infty}^{2(T/T_B)^{\beta_2}} g^L_2(x) ~dx
  + \int_{t/2}^{t} dT~ T^{2-\gamma-\beta_2} \int_{\infty}^{\frac{2(t-T)}{T_B^{\beta_2}T^{1-\beta_2}}} g^L_2(x) ~dx. \label{subleading}
\end{align}
We should consider which term becomes dominant at large Reynolds numbers.
To calculate it in more detail, we use the functional forms $g^L_1(x) = e^{-k_1 x}$ and $g^L_2(x) = e^{-k_2 x}$
observed in Fig.\ref{fig:gL_norm_large}.
Thereby, we can calculate Eqs.(\ref{leading})-(\ref{subleading}) as follows:
\begin{align}
  \langle r^2(t) \rangle
  \sim & ~
  (1-e^{-c_1}) t^{3-\gamma-\beta_1}
  + e^{-c_2} t^{3-\gamma-\beta_2} \nonumber                                                      \\
       & - \Gamma\left(\frac{3-\gamma-\beta_2}{\beta_2},2k_2\left(t/2T_B\right)^{\beta_2}\right)
  - t^{3-\gamma-\beta_2}
  \int_0^{1/2} (1-z)^{3-\gamma-\beta_2} \exp\left[-\frac{2k_2t^{\beta_2}z}{T_B^{\beta_2}(1-z)^{1-\beta_2}}\right] dz \label{subleading2}.
\end{align}
Here, $\Gamma (a,x)$ is the upper incomplete gamma function defined by $\Gamma (a,x) = \int_x^\infty z^{a-1} e^{-z} dz$.

Now we consider conditions to recover the Richardson-Obukhov law, $\langle r^2(t) \rangle \propto t^3$.
It is known that the upper incomplete gamma function, $\Gamma(a,x)$, has the following asymptotic series:
$\Gamma(a,x) \sim x^{a-1}e^{-x}[1 + \frac{a-1}{x} + \frac{(a-1)(a-2)}{x^2} + \cdots]$ as $x \to \infty$ \cite{Abramowitz1964}.
With these asymptotic formulae, we have
$\Gamma ((3-\gamma-\beta_2)/\beta_2,2k_2(t/2T_B)^{\beta_2}) \propto t^{3-\gamma-\beta_2}\exp[-2k_2 (t/2T_B)^{\beta_2}]$,
as $t\to \infty$.
Here we assume $\beta_2 \not= 0$ at infinitely large Reynolds number,
i.e., $\omega_2\not=0$ in Eq.(\ref{eq:beta_large}).
Furthermore, the last term of Eq.(\ref{subleading2}) can be estimated as,
\begin{equation}
  t^{3-\gamma-\beta_2} \int_0^{1/2} (1-z)^{3-\gamma-\beta_2} \exp\left[-\frac{2k_2t^{\beta_2}z}{T_B^{\beta_2}(1-z)^{1-\beta_2}}\right]
  \leq C t^{3-\gamma-2\beta_2},
\end{equation}
where $C$ is a constant.
Therefore, considering $\beta_1 > \beta_2$ as obtained in DNS,
the dominant power-law scaling at large $t$ and large $Re_\alpha$ is given by
\begin{equation}
  \langle r^2(t) \rangle \sim t^{3-\gamma-\beta_2}.
  \label{eq:r2_scaling_infty}
\end{equation}
Here the power-law exponent of $\langle r^2(t) \rangle$ is related to those of the TTLVCF $C^L(r_0, T, \tau, \varepsilon)$.
In particular, it involves $\beta_2$, which implies that $\langle r^2(t) \rangle$ is affected
by the correlation $C^L(r_0, T, \tau, \varepsilon)$ far from the diagonal line.

We have found the empirical form of $\gamma$ as a function of $T_B, T_\eta$, and $T_L$, which
is given in Eq.(\ref{eq:gamma_large}). It suggests that $\gamma \to 0$ as $Re_\alpha \to \infty$.
The similar form of $\beta_2$ given in Eq.(\ref{eq:beta_large}) indicates that
$\beta_2 \to \omega_2^{0.4}$ as $Re_\alpha \to \infty$.
As we discussed in Sec.\ref{sec:dns}, with our DNS data we are not able to conclude whether $\omega_2$
vanishes or not. However, at the practically accessible Reynolds numbers, $\beta_2$
is not zero as indicated by Fig.\ref{fig:beta_large}.
Therefore, now including the constant factor,
the Richardson--Obukhov law is modified at finite Reynolds numbers to
\begin{equation}
  \langle r^2(t) \rangle = \frac{2G\varepsilon}{k_2(3-\gamma-\beta_2)} T_B^{\gamma+\beta_2} t^{3-\gamma-\beta_2}
  + \mathrm{(subleading ~terms)},
  \label{eq:mod_RO}
\end{equation}
where we take $c_2 \sim 0$.
Although $G$ and $k_2$ are slightly dependent on $T_\eta, T_L,$ and $T_B$,
they are estimated as $G\varepsilon \sim 1.5$ and $k_2 \sim 3.0$ from the DNS data.

Now let us consider the numerical value of the Richardson constant $g_R$ involved
in the Richardson--Obukhov law $\langle r^2(t) \rangle = g_R \varepsilon t^3$.
To evaluate $g_R$, we substitute the values of $G, \varepsilon$ and $k_2$
in Eq.(\ref{eq:mod_RO}) by assuming that they do not change
much at infinite Reynolds number. We also assume that the modified exponent
$3-\gamma-\beta_2$ approach $3$ at infinite Reynolds number, i.e.,
$\gamma \to 0$ and $\beta_2 \to 0$ as $Re_\alpha \to \infty$.
Then the Richardson constant is estimated as $g_R = 2G / k_2 \sim 2 \times 10^1$, which is distinctly different
from $g_R = 0.5$ and $3.8$ obtained in previous experimental and numerical studies, respectively\cite{Jullien1999,Boffetta2002c}.
This discrepancy of $g_R$ is not surprising since
the measurements in the previous studies were done under the small $T_B$ condition.

\begin{figure}[htbp]
  \includegraphics[scale=0.7,clip]{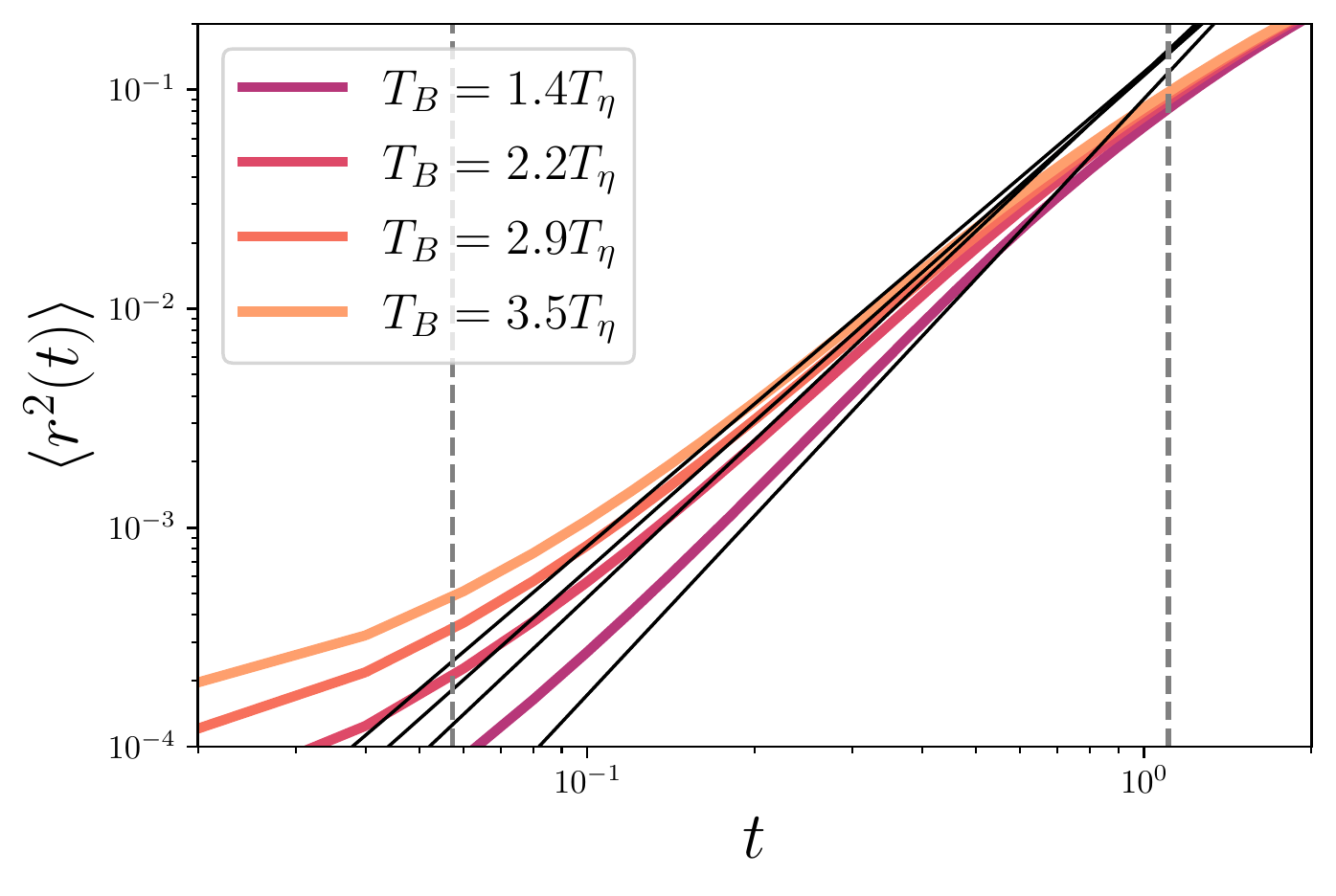}
  \caption{
    Mean-squared relative separation of particle pairs, $\langle r^2(t) \rangle$, for various initial separations at $\mathrm{Re}_\alpha = 160$ (colored lines) and the scaling law (\ref{eq:mod_RO}) (black lines).
    Two vertical dashed lines show the dissipation time scale $T_\eta$ (left) and the integral scale $T_L$ (right).
  }
  \label{fig:r2_scaling}
\end{figure}

Here, we validate the scaling law (\ref{eq:mod_RO}) by our DNS data.
Figure \ref{fig:r2_scaling} shows $\langle r^2(t) \rangle$ as a function of $t$ for various initial separations at $\mathrm{Re}_\alpha = 160$ with the scaling laws (\ref{eq:mod_RO}).
The scaling law (\ref{eq:mod_RO}) roughly coincides with the DNS data, especially at larger $T_B$.
Strictly speaking, the values of scaling exponents $3-\gamma-\beta_2$ in Eq. (\ref{eq:mod_RO}) are slightly larger than those of DNS data.
This is because we assumed that the self-similar function $g^L$ takes the forms (\ref{eq:gL_scaling_mod}) at finite Reynolds numbers.
Contrary to this assumption, self-similar function $g^L$ may take more complicated forms
at finite Reynolds numbers.
In other words, the scaling exponent $\beta$ may be not binary but continuous function of $\tau$.
In this view, the scaling exponent for $\langle r^2(t) \rangle$ is slightly smaller than
$3-\gamma-\beta_2$ because $\beta$ may be monotonically decreasing function of $\tau$.

Although the scaling laws (\ref{eq:mod_RO}) is good agreement with the DNS data for larger $T_B$,
this is not the case for smaller $T_B$ such as $T_B=1.4 T_\eta$.
This deviation does not mean failure of the scaling laws (\ref{eq:mod_RO})
because we assume the scale separation, $T_\eta \ll T_B \ll T_L$.
The scaling law (\ref{eq:mod_RO}) is still useful to understand the
qualitative tendency of $\langle r^2(t) \rangle$ at smaller $T_B$ than $T_\eta$ too.

We then consider the scaling law of $\langle r^2(t) \rangle$ under the small $T_B$ condition.
By doing an analogous calculation to that of the large $T_B$ condition,
we have
\begin{equation}
  \langle r^2(t) \rangle =
  \frac{2G\varepsilon}{\check{k}_2(3-\check{\gamma}-\check{\beta}_2)}
  T_X^{\check{\gamma}+\check{\beta}_2}
  t^{3-\check{\gamma}-\check{\beta}_2}
  + \mathrm{(subleading ~terms)},
  \label{eq:mod_RO_small}
\end{equation}
at large $t$ and large $Re_\alpha$.
This is similar to Eq.(\ref{eq:mod_RO}) for the large $T_B$ condition.
A crucial difference between Eqs.(\ref{eq:mod_RO_small}) and (\ref{eq:mod_RO}) is that
the exponent $\check{\gamma}$ is negative as seen from Eq.(\ref{gamma_small}).
This enables one to tune $T_B$ for given $T_\eta$ such that
$-\check{\gamma}(T_B, T_\eta) - \check{\beta}_2(T_\eta/T_B, T_L/T_B) = 0$
under the small $T_B$ condition (our DNS data suggest that $\check{\beta}_2$ is generally positive).
Consequently, we observe $\langle r^2(t) \propto t^3$, the same scaling exponent as the Richardson--Obukhov law.
In contrast, this sort of tuning leading to $t^3$ is not possible under the large $T_B$ condition
since $\gamma$ is always positive, see Eq.(\ref{eq:gamma_large}).

Indeed, it is known that $\langle r^2(t) \rangle \propto t^3$ can be observed even at moderate Reynolds number
by tuning the initial separation $r_0$ which satisfies the small $T_B$ condition $T_B \le T_\eta$.
See, for example, Refs. \cite{Jullien1999, Boffetta2002c, Kellay2002, Rivera2005,Rivera2016, Kishi2020} in
the 2D energy inverse-cascade turbulence.
Specifically, with the tuned initial separation $r_0$, $T_B$ is close to $T_\eta$.
In those circumstances, the equivalent of the Richardson constant can be given by
$\check{g}_R = 4G T_X^{\check{\gamma} + \check{\beta}_2}/(3 \check{k}_2)$ from Eq.(\ref{eq:mod_RO_small}).
It is noted that the value of $G$ in this range is strongly dependent on $T_B$ and $T_\eta$.
The Richardson constant measured in the previous experimental and numerical studies \cite{Jullien1999,Boffetta2002c}
with the tuned initial separation should be therefore compared to $\check{g}_R$.
However, we do not compare it quantitatively since we can not determine the accurate value of $T_X$.

Let us now argue that  the nature of the $t^3$ law with the tuned initial separation
is different from that of the Richardson--Obukhov law.
In general terms, by the Richardson--Obukhov $t^3$ law, it is understood that the $t^3$ law
holds irrespective of the value of the initial separation $r_0$, provided that
the inertial subrange is sufficiently wide. Strictly speaking, one should add a condition that $r_0$ is
inside the inertial subrange \cite{Batchelor1950}.
The large $T_B$ condition which we have considered conforms to the added condition.
From Eq.(\ref{eq:mod_RO}), the Richardson--Obukhov law corresponds to $\gamma = 0$ and $\beta_2 = 0$,
and the resultant $t^3$ law does not depend on $T_B$, or equivalently $r_0$.
For the sake of the argument, let us relax the added condition. Now we consider the small $T_B$ condition.
From Eq.(\ref{eq:mod_RO_small}), The $t^3$ law with the tuned initial
separation corresponds to $-\check{\gamma} - \check{\beta}_2 = 0$ and
the resultant $t^3$ law has the prefactor $T_X^{\check{\gamma} + \check{\beta}_2}$, which
potentially depends on $T_B$.
Therefore, the $t^3$ law observed at a given Reynolds number (however large)
by tuning the initial separation is different from the Richardson--Obukhov $t^3$ law.
The agreement of the power-law exponents is coincidental.
In this sense, the $t^3$ scaling of $\langle r^2(t) \rangle$ observed at moderate Reynolds numbers
is a different state from the complete similarity for $\langle r^2(t) \rangle$,
which is consistent with the dimensional analysis naively using the K41 phenomenology.

It is interesting that such a coincidence do not occur under the large $T_B$ condition.
Then, in this condition, can we say anything about observability of the bona fide
Richardson--Obukhov law? As far as our DNS data suggest, the exponents $\gamma$ and $\beta_2$
do not vanish under the large $T_B$ condition.
Consequently, the Richardson--Obukhov law is not observable at the current Reynolds numbers.
It should be noted that this is caused not by the intermittency effects, but by correlation of the Lagrangian velocity.
Extrapolation of the data suggests a possibility that $\beta_2$ does not vanish at infinite Reynolds number as we discussed.
This implies that the no matter how large the Reynolds number is, the Richardson--Obukhov law is not observable.

We have presented here a framework to study the Richardson--Obukhov law by way of the self-similarity
of the TTLVCF. It can be adapted to the 3D turbulence.
The $t^3$ law with the tuned initial separation is also known in
the 3D case, see, for example, \cite{OTT2000,Biferale2005,Sawford2008,Bitane2013b,D.Buaria2015}.
Our analysis in the 3D case will be reported elsewhere.


\section{Concluding remarks\label{sec:conclusion}}
We have investigated the two-time Lagrangian velocity correlation function (TTLVCF)
for particle pairs with the incomplete self-similarity and the DNS of the 2D energy inverse-cascade turbulence.
First, we have made the self-similar ansatz (\ref{eq:CL_scaling_law})
of the correlation function by using the idea of incomplete similarity.
The ansatz includes the Bachelor time,
the Kolmogorov dissipation length, and the integral length
as similarity variables, meaning that finite Reynolds number effects and
the initial separation dependence are encoded.
The ansatz is characterized by
the two scaling exponents, $\beta$ and $\gamma$, and
the one-variable function $g_L$.
The exponent $\gamma$ concerns
the equal-time correlation along the diagonal line through the origin
shown in Fig.\ref{fig:intro}.
The other exponent $\beta$ concerns how the correlation decreases
along the direction perpendicular to the diagonal line.
Since the ansatz is an example of the incomplete self-similarity,
the two exponents cannot be determined by dimensional analysis.

In order to verify the ansatz,
we have performed DNS of the 2D inverse energy-cascade turbulence and calculated
the TTLVCF by varying parameters such as
the Batchelor time $T_B$ and the dissipation time $T_\eta$. We split the DNS study
into two parts: the large and small $T_B$ conditions.
For both conditions, we showed that the ansatz describes the DNS results
reasonably well.  Then we measured
the values of the two exponents and the functional form of $g_L$ from the DNS data
which are in some cases too noisy to obtain reliable measurements.
The measurements indicated that the exponents depend on $T_B$ and $T_\eta$.
In theory, we assumed that they are independent.
This dependence of the exponents is empirically determined
as Eqs.(\ref{eq:gamma_large}), (\ref{eq:beta_large}), and (\ref{gamma_small}).
The function $g^L$ is determined as an exponential function.
Moreover, our results indicate that at finite Reynolds numbers,
the correlation in general has correction described
by non-zero $\gamma$, non-zero $\beta$ and the function $g^L$
to the K41 dimensional analysis for both large and small $T_B$ conditions.

We next considered the limit of these empirical relations at infinite Reynolds number.
The extrapolation of the relations obtained
at moderate Reynolds numbers was subject to uncertainty.
Thus, we could not determine the accurate values of the scaling exponents, $\gamma$ and $\beta$
at infinite Reynolds number.
Especially, we observed that $\beta$ may not approach zero,
which suggests a possibility that the TTLVCF
is not consistent with the K41 dimensional analysis
at infinite Reynolds number under the large $T_B$ condition.

Finally, we have considered the relationship between the scaling law of the TTLVCF
and the Richardson--Obukhov $t^3$ law for the second moment of relative separation via
the integral (\ref{eq:r2}).
With the asymptotic argument, we found that
the Richardson--Obukhov law is not recovered for finite $T_B$
at finite Reynolds numbers under the large $T_B$ condition.
Instead, the scaling law of the squared separation of particle pairs is modified as Eq.(\ref{eq:mod_RO}).
The modified scaling exponent, $3-\gamma-\beta_2$, is determined by the scaling behavior
of the TTLVCF on the diagonal line $\tau=0$ shown in Fig.\ref{fig:intro} and far from the diagonal line.
Namely, the scaling exponent is influenced by
correlations between two particles not only at a same time but also at quite different times.
Moreover, using the scaling law of the TTLVCF,
we explained why we, nevertheless, observe the $t^3$ scaling
at moderate Reynolds numbers with a special initial separation under the small $T_B$ condition
because $\gamma$ and $\beta_2$ take a negative and positive values, respectively.
Therefore, we concluded that the physics of this $t^3$-scaling behavior is different
from that of the Richardson--Obukhov law.

In this paper, we assumed that forcing effects are negligible.
The external forcing is limited to small scales for 2D.
In fact, the Eulerian statistics in the Fourier space
such as the energy spectrum or the energy flux is influenced by the forcing
only in the vicinity of the forcing scales\cite{XIAO2009,Boffetta2010, Mizuta2013}.
Hence the influence is considered as local.
This may be the reason why the empirically found functional forms of
the scaling exponents, $\beta$ and $\gamma$ depend only $T_\eta, T_L$ and $T_B$ given by
Eqs. (\ref{eq:beta_large}) and (\ref{eq:gamma_large}) under the large $T_B$ condition.
Strictly speaking, we can neglect the forcing effects
if correlation between the forcing and the Lagrangian velocity, $\langle f_i \delta v_j \rangle$,
rapidly decays in time. Here $f_i$ is the forcing increment between two Lagrangian particles and
$\delta v_j$ is the relative velocity between them.
We speculate that the cross correlation rapidly decays because the characteristic time scales of the forcing
and the velocity in the forcing scale are small.
%

Here, we discuss whether or not the scaling laws (\ref{Phig}) and
the modified Richardson--Obukhov law (\ref{eq:mod_RO}) can be applied to 3D turbulence.
First, the scaling laws (\ref{Phig}) could be suggested by only using the dimensinal analysis, which can
be applied to both 2D and 3D turbulence.
Namely, the special properties of 2D turbulence such as the inverse cascade of energy flux are not necessary.
This makes us convincing that this scaling laws (\ref{Phig}) can be implimeted to 3D turbulence.
On the other hand, more careful investigation is needed for the modified Richardson--Obukhov law (\ref{eq:mod_RO}).
This is because this is derived by using the DNS results such as Eq. (\ref{eq:gL_scaling_mod}).

Given the self-similar form Eqs.(\ref{eq:CL_scaling_law}) and (\ref{Phig}) of the TTLVCF,
one would like to ``derive'' it
from the Navier--Stokes equations using only plausible assumptions.
More precisely, we propose to use it as an input to a set of integro-differential
equations (closure equations) for the Lagrangian correlation function obtained by a closure approximation
such as direct-interaction approximation \cite{Kraichnan1959, Kraichnan1965, Kaneda1981,Kida1997}.
One standard procedure in the last step of the closures is to substitute certain
self-similar forms for the correlation function and the linear response function
to those in the closure equations and then to study consistency of
the self-similar forms with the closure equations. By input, we mean to input
the ansatz studied here into closure equations of, for example, a direct-interaction
approximation. This may give analytically functional forms of the scaling exponents,
$\beta(T_\eta/T_B, T_L/T_B)$ and $\gamma(T_\eta/T_B, T_L/T_B)$ and their limits
at infinite Reynolds number.

Closure approximations have been applied to study the Richardson--Obukhov $t^3$ law,
see e.g., \cite{Kraichnan1966, Ishihara2002}.
However, these studies have used one-time Lagrangian
velocity correlation function given by Eq.(\ref{ac}) in Introduction, which is different
from the TTLVCF $C^L(r_0, T,\tau, \varepsilon)$ studied here.
In fact, the TTLVCF is unexplored
with the Lagrangian renormalization approximation \cite{Kaneda1981} and
perhaps other closure approximations \footnote{Y. Kaneda (private communication)}.
Therefore, the results in this study play an important role to develop a new avenue of
closure theories.

Another approach can be to develop a stochastic model of
turbulent relative dispersion using the ansatz we have obtained here.
Recently, continuous time random walk (CTRW) models \cite{Thalabard2014,Bourgoin2015}
developed for the relative dispersion.
These models are constructed to be consistent with the Richardson--Obukhov law.
It is possible to modify these models to have the self-similar properties
of the TTLVCF obtained in this paper.
Building such a model corresponds to incorporating effects of time correlations \cite{Scatamacchia2012,Eyink2013b}
and finite propagation speed of the relative diffusion \cite{Ogasawara2006b,Kanatani2009}.
We will report a stochastic modeling based on the ansatz elsewhere.

%
%
%

\begin{acknowledgments}
  We are grateful for stimulating discussion with Yukio Kaneda.
  The numerical computations in this work were performed
  at the computer facility of the Yukawa Institute for Theoreical Physics at Kyoto University.
  This study was supported by the Research Institute for Mathematical Sciences
  at Kyoto University and by Kakenhi grant (A) No. 19H00641 from JSPS.
\end{acknowledgments}

\bibliography{paper}

\end{document}